\documentclass[twocolumn,amsmath,amssymb]{revtex4}
\usepackage{amsmath,amssymb,graphicx}
\usepackage{graphicx}
\usepackage{dcolumn}
\usepackage{bm}
\usepackage{epsfig}
\usepackage{here}
\usepackage{float}
\usepackage{amsmath}
\usepackage[usenames]{color}
\usepackage{amsmath}
\usepackage{epsfig}
\usepackage{color}

\newcommand{\bea}{\begin{eqnarray}}
\newcommand{\eea}{\end{eqnarray}}

\newcommand{\lblue}{\textcolor[rgb]{0.00,0.00,0.00}}
\newcommand{\dgreen}{\textcolor[rgb]{0.00,0.00,0.00}}

\begin{document}
%%%%%%%%%%%%%%%%%%%%%%%%%%%%%%%%%%%%%%%%%%%%%%%%%%%%%%%%%%%%%%%
\title{Exact formulations of relativistic electrodynamics and magnetohydrodynamics with helically coupled scalar field}
\author{Jai-chan Hwang${}^{1}$, Hyerim Noh${}^{2}$}
\address{${}^{1}$Particle Theory  and Cosmology Group,
         Center for Theoretical Physics of the Universe,
         Institute for Basic Science (IBS), Daejeon, 34051, Republic of Korea
         \\
         ${}^{2}$Theoretical Astrophysics Group, Korea Astronomy and Space Science Institute, Daejeon, Republic of Korea
         }

%%%%%%%%%%%%%%%%%%%%%%%%%%%%%%%%%%%%%%%%%%%%%%%%%%%%%%%%%%%%%%%
%\date{\today}

%%%%%%%%%%%%%%%%%%%%%%%%%%%%%%%%%%%%%%%%%%%%%%%%%%%%%%%%%%%%%%%
\begin{abstract}

We present the general relativistic electrodynamics and magnetohydrodynamics with a helically coupled scalar field. We consider three component system with the fluid, scalar field and electromagnetic fields with the helical coupling. We derive three exact formulations: the covariant formulation, the ADM formulation, and the fully nonlinear and exact perturbation formulation. We also derive the weak-gravity limit with fully relativistic fluid and fields. The latter two formulations are presented in the cosmological context.

\end{abstract}
%%%%%%%%%%%%%%%%%%%%%%%%%%%%%%%%%%%%%%%%%%%%%%%%%%%%%%%%%%%%%%%

%%%%%%%%%%%%%%%%%%%%%%%%%%%%%%%%%%%%%%%%%%%%%%%%%%%%%%%%%%%%%%%
\maketitle
\tableofcontents

%%%%%%%%%%%%%%%%%%%%%%%%%%%%%%%%%%%%%%%%%%%%%%%%%%%%%%%%%%%%%%%
%
%
%
%%%%%%%%%%%%%%%%%%%%%%%%%%%%%%%%%%%%%%%%%%%%%%%%%%%%%%%%%%%%%%%
\section{Introduction}

Magnetic fields are ubiquitous in celestial environments. The relativistic processes involving magnetic field are also widespread in astronomical objects. The scalar field appearing in high-energy physics is now main ingredients in cosmological processes, playing essential roles in driving inflation, dark matter and dark energy. For the scalar field, relativistic treatment is necessary. The combination of the magnetic field and the scalar field naturally leads to a helical electromagnetism \cite{Sikivie-1983} which disappears without the scalar field coupling.

The electromagnetic field helically coupled with the scalar field (the axion as an example) naturally leads to $\alpha$-dynamo term \cite{Krause-Radler-1980} causing exponential growth of magnetic field in the linear stage and to magnetic helicity generation which has important implications in enhancing the large-scale magnetic field through the inverse cascade, propagating the magnetic field to larger scales \cite{Frisch-1975, Brandenburg-Subramanian-2005}. Although the origin and evolution of the cosmic magnetic field are largely unknown in astrophysics and cosmology, the helically coupled scalar field may have important roles in understanding the subject, see \cite{Grasso-Rubinstein-2001, Widrow-2002, Giovannini-2004, Durrer-Neronov-2013, Subramanian-2016, Vachaspati-2021} for reviews.

Relativistic electrodynamics (ED) and magnetohydrodynamics (MHD) are important in many aspects of relativistic astrophysics and cosmology. Here, we present three exact relativistic formulations and the weak-gravity approximation of ED and MHD of the three component system including the helical coupling: the fluid (FL), the scalar field (SF) and the electromagnetic (EM) field. We consider a Lagrangian density
\bea
   & & L = {c^4 \over 16 \pi G} ( R - 2 \Lambda )
       - {1 \over 2} \phi^{;c} \phi_{,c} - V (\phi) + L_{\rm m}
   \nonumber \\
   & & \qquad
       - {1 \over 16 \pi} F_{ab} F^{ab}
       \lblue{- {g_{\phi \gamma} \over 4} f (\phi)
       F_{ab} F^{*ab}}
       + {1 \over c} J_{\rm em}^a A_a,
\eea
where $L_{\rm m}$ is the fluid part, $R$ the scalar curvature, $\Lambda$ the cosmological constant, $\phi$ the SF, $F_{ab}$ the EM tensor and the other symbols will be explained later. The $g_{\phi \gamma}$ term is the helical coupling.

In ED, the presence of helical-coupling causes extra charge and current densities and provides windows for direct detection of the axion particles by EM means \cite{Sikivie-1983, Axion-EM-coupling}. In MHD, the helical coupling causes the $\alpha$-dynamo term in MHD treatment: for expressions in the four formulations, see Sec.\ \ref{sec:Induction-eq}.

The covariant ($1+3$) and the ADM (Arnowitt-Deser-Misner, $3+1$) formulations are reformulations of Einstein's gravity without any restriction on the geometry and the energy-momentum contents. The fully nonlinear and exact perturbation (FNLE) formulation is based on nonlinear and exact perturbations in the Friedmann background. Here, for simplicity, we ignore the tensor-type perturbation and impose the spatial gauge conditions. The formulation is general without imposing the temporal gauge (hypersurface, slicing) condition.

The weak-gravity approximation is a limiting case of the FNLE formulation in the limit of weak gravity but with fully relativistic fluid and fields. The formulation is possible in the uniform-expansion gauge (maximal slicing in the flat spacetime background). By taking the nondynamic background, the equations of FNLE and weak-gravity formulations are valid for general energy-momentum configuration.

Previously, in \cite{Hwang-Noh-2022-axion-MHD-WG} we presented the weak-gravity and slow-motion limit of the system and studied gravitational and magnetic instability in the case of the massive scalar field with $f \propto \phi^2$ coupling. Here, we present the fully relativistic and nonlinear ED and MHD of the system without the restrictions used in \cite{Hwang-Noh-2022-axion-MHD-WG}. As we consider general potential $V$ and coupling $f$, here we do not handle the axion case with massive scalar field. For axion, together with $f \propto \phi^2$, the Klein and the Madelung transformations are available with proper non-relativistic limit. In the absence of the $g_{\phi \gamma}$-coupling, the covariant formulation and the post-Newtonian approximation of the axion were studied in \cite{Hwang-Noh-2022-axion-PN}. For a general potential $V(\phi)$, the SF does not have Newtonian limit, thus the post-Newtonian approximation is not available.

The complete sets of equations in the three exact formulations of ED are derived in Secs.\ \ref{sec:covariant-formulation}-\ref{sec:FNLE-formulation}. The weak-gravity approximation of the FNLE formulation of ED is derived in Sec.\ \ref{sec:WG-formulation}. MHDs in the four formulations are derived in Sec.\ \ref{sec:MHD}. In three appendices we derive the covariant, ADM and FNLE formulations for a general single-component fluid. Being a general fluid, the formulations is valid in the multi-component fluids and fields with the fluid quantities interpreted as collective ones. As the main purpose of this work is deriving the formulations, we will present some details needed for the derivation.

%%%%%%%%%%%%%%%%%%%%%%%%%%%%%%%%%%%%%%%%%%%%%%%%%%%%%%%%%%%%%%%
%
%
%
%%%%%%%%%%%%%%%%%%%%%%%%%%%%%%%%%%%%%%%%%%%%%%%%%%%%%%%%%%%%%%%
\section{Covariant (1+3) formulation}
                              \label{sec:covariant-formulation}

The covariant ($1+3$) formulation \cite{Ehlers-1993, Ellis-1971, Ellis-1973, Hawking-1966, Bertschinger-1995} is derived in the Appendix \ref{Appendix-cov} for a most general imperfect fluid. The fundamental set of equations are Eqs.\ (\ref{entropy})-(\ref{cov-Maxwell-4}). The formulation in the Appendix \ref{Appendix-cov}, although $u_a$ is used as the four-vector, is based on a generic time-like four-vector $U_a$. From now on, we will use $U_a$ for the generic four-vector, reserving $u_a$ for the fluid four-vector and $n_a$ for the normal four-vector. The $U_a$ can be the FL comoving ($u_a$), the collective-matter comoving, the SF comoving, the normal ($n_a$), etc.

In the presence of multiple component fluids and fields, we may regard the fluid quantities in the Appendix \ref{Appendix-cov} as the ones for the collective fluids and fields. Thus, our task in this section is to derive the conservation equations for the FL, the Maxwell equations for the EM field, the equation of motion for the SF, and the collective fluid quantities. Einstein's equations in (\ref{Raychaudhury-eq-cov})-(\ref{cov-Maxwell-4}) together with the collective fluid quantities provide the gravity part.

%%%%%%%%%%%%%%%%%%%%%%%%%%%%%%%%%%%%%%%%%%%%%%%%%%%%%%%%%%%%%%%
\subsection{Maxwell equations}

The EM tensor is decomposed as \cite{Ellis-1973}
\bea
   F_{ab}
       \equiv U_a E_b - U_b E_a - \eta_{abcd} U^c B^d,
   \label{Fab}
\eea
with $E_a U^a \equiv 0 \equiv B_a U^a$; $U_a$ is a generic time-like four-vector normalized with $U^c U_c \equiv -1$; $E_a$ and $B_a$ are based on the four-vector. In terms of the EM four-potential $A_a$, we have
\bea
   F_{ab} \equiv \nabla_a A_b - \nabla_b A_a,
\eea
where $\nabla_a$ is the covariant derivative. The dual tensor is
\bea
   & & F^{*ab}
       \equiv {1 \over 2} \eta^{abcd} F_{cd}
       = U^a B^b - U^b B^a + \eta^{abcd} U_c E_d,
   \nonumber \\
   & &
       F_{ab} = - {1 \over 2} \eta_{abcd} F^{*cd},
   \label{F*ab}
\eea
thus
\bea
   E_a \equiv F_{ab} U^b, \quad
       B_a \equiv F^*_{ab} U^b.
   \label{E-B-def}
\eea
We have two invariants
\bea
   & & F^{ab} F_{ab} = - 2 \left( E^2 - B^2 \right)
       = - F^{*ab} F^*_{ab},
   \nonumber \\
   & &
       F^{ab} F^*_{ab} = - 4 E^a B_a,
   \label{EM-invariants}
\eea
where $E^2 \equiv E^a E_a$; $F_{ab} F^{*ab}$ is parity-odd and leads to asymmetry between the two circular polarization states, thus helical.

For fluid quantities and the electric and magnetic field strengths based on the generic four-vector $U_a$, we can take either the normal-frame (which is closer to the laboratory-frame) with $U_a = n_a$ or the comoving frame. In the presence of the FL, the SF and the EM field, the comoving can be fluid-comoving, the scalar field-comoving, etc. The fluid quantities of the FL are often expressed in the fluid-comoving frame, and the EM fields are often expressed in the normal-frame. Expression of the SF is simplified in the SF-comoving frame. We reserve $U_a = u_a$ for the fluid-comoving frame (thus, $u_a$ is the fluid four-vector).

Variation with respect to $A_a$ gives the Maxwell equations
\bea
   & & F^{ab}_{\;\;\;\; ;b} = {4 \pi \over c} \left( J_{\rm em}^a
       \lblue{- c g_{\phi \gamma} f_{,b} F^{*ab}} \right),
   \nonumber \\
   & &
       F^{*ab}_{\;\;\;\;\;\; ;b} = 0
       \quad ({\rm or} \quad
       F_{[ab;c]} = 0),
   \label{Maxwell-cov}
\eea
with
\bea
   J_{\rm em}^a
       \equiv \varrho_{\rm em} c U^a + j^{a}, \quad
       j_a U^a \equiv 0, \quad
       J^a_{{\rm em};a} = 0,
   \label{current-def}
\eea
where semicolon indicates four-dimensional covariant derivative.
Thus
\bea
   \varrho_{\rm em} c
       = - J_{\rm em}^a U_a, \quad
       j^{a} = h^a_b J_{\rm em}^b,
\eea
where $\varrho_{\rm em}$ and $j^a$ are the charge and current densities, respectively, based on $U_a$; $h_{ab} \equiv g_{ab} + U_a U_b$ is the projection tensor. From these we can derive the four Maxwell equations \cite{Ellis-1973}
\bea
   & & E^a_{\;\; ;b} h^b_a
       = 4 \pi \left( \varrho_{\rm em}
       \lblue{- g_{\phi \gamma} f_{,a} B^a} \right)
       - 2 \omega^a B_a,
   \label{Maxwell-cov-1} \\
   & & h^a_b \widetilde {\dot E}{}^b
       = \left( \eta^a_{\;\;bcd} U^d \omega^c
       + \sigma^a_{\;\;b}
       - {2 \over 3} \delta^a_b \theta \right) E^b
   \nonumber \\
   & & \qquad
       + \eta^{abcd} U_d
       \left( a_b B_c - B_{b;c} \right)
   \nonumber \\
   & & \qquad
       - {4 \pi \over c} \left[ j^{a}
       \lblue{+ c g_{\phi \gamma} f_{,b} \left( U^b B^a
       - \eta^{abcd} U_c E_d \right)} \right],
   \label{Maxwell-cov-2} \\
   & & B^a_{\;\; ;b} h^b_a
       = 2 \omega^a E_a,
   \label{Maxwell-cov-3} \\
   & & h^a_b \widetilde {\dot B}{}^b
       = \left( \eta^a_{\;\;bcd} U^d \omega^c
       + \sigma^a_{\;\;b}
       - {2 \over 3} \delta^a_b \theta \right) B^b
   \nonumber \\
   & & \qquad
       - \eta^{abcd} U_d
       \left( a_b E_c - E_{b;c} \right),
   \label{Maxwell-cov-4}
\eea
where $\widetilde {\dot E}{}^a \equiv E^a_{\;\; ;b} U^b$. Equations (\ref{Maxwell-cov-1})-(\ref{Maxwell-cov-4}) follow, respectively, from
\bea
   & & U_a F^{ab}_{\;\;\;\; ;b} = \dots, \quad
       h^a_c F^{cb}_{\;\;\;\; ;b} = \dots,
   \nonumber \\
   & &
       U_a \eta^{abcd} F_{bc;d} = 0, \quad
       h^a_e \eta^{ebcd} F_{bc;d} = 0.
\eea
Apparently, the $g_{\phi \gamma}$-coupling can be interpreted as introducing extra charge and current densities \cite{Sikivie-1983, Wilczek-1987}. We have
\bea
   & & J_{\rm ax}^a
       = \lblue{- c g_{\phi \gamma} f_{,b} F^{*ab}}, \quad
       \varrho_{\rm ax}
       = \lblue{- g_{\phi \gamma} f_{,a} B^a},
   \nonumber \\
   & &
       j^a_{\rm ax}
       = \lblue{c g_{\phi \gamma} f_{,b} \left( U^b B^a
       - \eta^{abcd} U_c E_d \right)}.
\eea
These are axion (SF) induced electric charge and current densities.

We can derive
\bea
   & & \left( B^2 \right)^{\widetilde \cdot}
       = 2 \sigma_{ab} B^a B^b - {4 \over 3} \theta B^2
   \nonumber \\
   & & \qquad
       - 2 \eta^{abcd} U_d B_a
       \left( a_b E_c + E_{b;c} \right).
\eea
The current conservation $J_{{\rm em};a}^a = 0$ gives
\bea
   c \left( \widetilde {\dot \varrho}_{\rm em}
       + \theta \varrho_{\rm em} \right)
       = - j^a_{\;\; ;a}
       = - h^b_a j^a_{\;\; ;b} - j^a a_a,
\eea
which is also valid for $J^a_{\rm ax}$.

%%%%%%%%%%%%%%%%%%%%%%%%%%%%%%%%%%%%%%%%%%%%%%%%%%%%%%%%%%%%%%%
\subsection{Equation of motion}

Variation with respect to $\phi$ gives the equation of motion
\bea
   \Box \phi
       = V_{,\phi}
       \lblue{+ {g_{\phi \gamma} \over 4}
       f_{,\phi} F_{ab} F^{*ab}}.
   \label{EOM-cov}
\eea
Using $F_{ab} F^{*ab} = - 4 E^a B_a$ which is frame-invariant, we have
\bea
   & & - \Box \phi + V_{,\phi}
       = \widetilde {\ddot \phi}
       + \theta \widetilde {\dot \phi}
       + V_{,\phi}
       - h_a^b \left( h^{ac} \phi_{,c} \right)_{;b}
       - h_a^b \phi_{,b} a^a
   \nonumber \\
   & & \qquad
       = \lblue{ g_{\phi \gamma} f_{,\phi} E^a B_a},
   \label{EOM-cov-2}
\eea
based on the generic four-vector $U_a$.

%%%%%%%%%%%%%%%%%%%%%%%%%%%%%%%%%%%%%%%%%%%%%%%%%%%%%%%%%%%%%%%
\subsection{Fluid quantities}

In the presence of all three components, $T_{ab}$ can be separated into three components as $T_{ab} = T^{\rm FL}_{ab} + T^{\rm SF}_{ab} + T^{\rm EM}_{ab}$ with each component in Eqs.\ (\ref{Tab-FL-cov})-(\ref{Tab-EM-cov}). The fluid quantities in the Appendix \ref{Appendix-cov} can be regarded as the collective ones with $\mu = \mu^{\rm FL} + \mu^{\rm SF} + \mu^{\rm EM}$, etc; for individual fluid quantities, see Eq.\ (\ref{fluid-FL-cov}) for FL in the fluid-frame, and Eqs.\ (\ref{fluid-SF-cov}) and (\ref{fluid-EM-cov}) for the SF and the EM field, respectively, in the generic frame. Thus, the fluid quantities used in Einstein's equations in (\ref{entropy})-(\ref{cov-Maxwell-4}) are collective ones based on the generic frame. The energy-momentum conservation in Eqs.\ (\ref{cov-E-conserv-App}) and (\ref{cov-Mom-conserv-App}) also apply for the collective fluid quantities. In the following, unless mentioned otherwise, the fluid quantities $\mu$ etc. are in the fluid-comoving frame

In the multi-component case, it is convenient to introduce the individual conservation equations. The conservation equations of individual component are constrained by
\bea
   T^{ab}_{\;\;\;\; ;b}
       = ( T_{\rm FL}^{ab} + T_{\rm SF}^{ab} + T_{\rm EM}^{ab}
       )_{;b} = 0.
\eea
For the FL, we additionally have mass conservation equation in (\ref{cov-Mass-conserv-App}). For the EM field, we additionally have the Maxwell equations in Eq.\ (\ref{Maxwell-cov}); in the covariant form decomposed using $U_a$, see Eqs.\ (\ref{Maxwell-cov-1})-(\ref{Maxwell-cov-4}). For the SF, we have the equation of motion in Eq.\ (\ref{EOM-cov}); in the covariant form decomposed using $U_a$, see (\ref{EOM-cov-2}).

Variation with respect to $g_{ab}$ gives Einstein's equation and the energy-momentum tensor, with $\delta ( \sqrt{-g} L ) = {1 \over 2} T^{ab} \sqrt{-g} \delta g_{ab}$. We have $T_{ab} = T^{\rm FL}_{ab} + T^{\rm SF}_{ab} + T^{\rm EM}_{ab}$ with
\bea
   & & T^{\rm FL}_{ab} = \mu u_a u_b + p h_{ab}
       + q_a u_b + q_b u_a + \pi_{ab},
   \label{Tab-FL-cov} \\
   & & T^{\rm SF}_{ab} = \phi_{,a} \phi_{,b}
       - \left( {1 \over 2} \phi^{;c} \phi_{,c}
       + V \right) g_{ab},
   \label{Tab-SF-cov} \\
   & & T^{\rm EM}_{ab}
       = {1 \over 4 \pi} \left( F_{ac} F_b^{\;\;c}
       - {1 \over 4} g_{ab} F_{cd} F^{cd} \right),
   \label{Tab-EM-cov}
\eea
where $\mu = \mu^{\rm FL}$, etc. In $T_{ab}$, the fluid and the two fields are clearly separated. The $g_{\phi \gamma}$-coupling term does not contribute to the energy-momentum tensor, thus to the fluid quantities. The $g_{\phi \gamma}$-coupling occurs only in the Maxwell equations and the equation of motion. The EM coupling with the fluid appears in the energy and momentum conservation equations.

Fluid quantities of the FL in Eq.\ (\ref{Tab-FL-cov}) are defined based on the fluid four-vector $U_a = u_a$, with
\bea
   & & \mu = T^{\rm FL}_{ab} u^a u^b, \quad
       p = {1 \over 3} T^{\rm FL}_{ab} h^{ab}, \quad
       q_a = - T^{\rm FL}_{cd} u^c h^d_a,
   \nonumber \\
   & &
       \pi_{ab} = T^{\rm FL}_{cd} h^c_a h^d_b - p h_{ab},
   \label{fluid-FL-cov}
\eea
where we often do not distinguish the fluid quantities of the fluid by index FL. For a single component fluid, the fluid-frame condition implies vanishing flux $q_a \equiv 0$.

The fluid quantities based on the generic $U_a$-frame are similarly introduced as
\bea
   T_{ab} = \mu U_a U_b + p ( g_{ab} + U_a U_b )
       + q_a U_b + q_b U_a + \pi_{ab},
   \label{Tab-fluid-cov-U}
\eea
thus,
\bea
   & & \mu = T_{ab} U^a U^b, \quad
       p = {1 \over 3} T_{ab} h^{ab}, \quad
       q_a = - T_{cd} U^c h^d_a,
   \nonumber \\
   & &
       \pi_{ab} = T_{cd} h^c_a h^d_b - p h_{ab},
   \label{fluid-quantities-cov-U}
\eea
with the conditions in Eq.\ (\ref{q-pi-relations-App}); here, $\mu = \mu^{(U)}$ etc. and $h_{ab} \equiv g_{ab} + U_a U_b$; we do not distinguish $h^{(U)}_{ab}$, $h^{(u)}_{ab}$, etc. by superscript, which can be understood in the context. In the above relations, $T_{ab}$ can be $T^{\rm EM}_{ab}$, $T^{\rm SF}_{ab}$ or the collective one $T^{\rm tot}_{ab}$.

The energy-momentum tensors of the SF and the EM field in Eqs.\ (\ref{Tab-SF-cov}) and (\ref{Tab-EM-cov}) are introduced without reference to the four-vector; $T_{ab}$ is frame invariant. Using the generic four-vector $U_a$, we can expand
\bea
   & & T^{\rm SF}_{ab}
       = \left( {1 \over 2} \widetilde {\dot \phi}{}^2
       + V + {1 \over 2} h^{ab} \phi_{,a} \phi_{,b} \right)
       U_a U_b
   \nonumber \\
   & & \qquad
       + \left( {1 \over 2} \widetilde {\dot \phi}{}^2
       - V - {1 \over 2} h^{ab} \phi_{,a} \phi_{,b} \right) h_{ab}
       + h_a^c h_b^d \phi_{,c} \phi_{,d}
   \nonumber \\
   & & \qquad
       - \widetilde {\dot \phi}
       \left( U_a h_b^c \phi_{,c}
       + U_b h_a^c \phi_{,c} \right),
   \label{Tab-SF-cov-U} \\
   & & T^{\rm EM}_{ab}
       = {1 \over 4 \pi} \bigg[
       {1 \over 2} \left( E^2 + B^2 \right)
       \left( U_a U_b + h_{ab} \right)
       - E_a E_b
   \nonumber \\
   & & \qquad
       - B_a B_b
       + \left( U_a \eta_{bcde} + U_b \eta_{acde} \right)
       E^c B^d U^e \bigg],
   \label{Tab-EM-cov-U}
\eea
where we used $\phi_{,a} = h_a^b \phi_{,b} - \widetilde {\dot \phi} U_a$ with $\widetilde {\dot \phi} \equiv \phi_{,c} U^c$ for the SF, and Eq.\ (\ref{Fab}) for the EM field.

For the scalar field, from Eqs.\ (\ref{Tab-SF-cov}) and (\ref{fluid-quantities-cov-U}), we have
\bea
   & & \mu^{\rm SF} = {1 \over 2} \widetilde {\dot \phi}{}^2
       + V + {1 \over 2} h^{ab} \phi_{,a} \phi_{,b},
   \nonumber \\
   & &
       p^{\rm SF} = {1 \over 2} \widetilde {\dot \phi}{}^2
       - V - {1 \over 6} h^{ab} \phi_{,a} \phi_{,b}, \quad
       q^{\rm SF}_a = - \widetilde {\dot \phi} h^b_a \phi_{,b},
   \nonumber \\
   & &
       \pi^{\rm SF}_{ab} = h_a^c \phi_{,c} h_b^d \phi_{,d}
       - {1 \over 3} h_{ab} h^{cd} \phi_{,c} \phi_{,d},
   \label{fluid-SF-cov}
\eea
using the generic four-vector $U_a$; thus, here $\mu^{\rm SF} = \mu^{\rm SF}_{(U)}$, etc. The SF-comoving frame is defined as $q^{\rm SF}_a \equiv 0$. In this frame we have $h_a^b \phi_{,b} = 0$, thus
\bea
   \mu^{\rm SF} = {1 \over 2} \widetilde {\dot \phi}{}^2
       + V, \quad
       p^{\rm SF} = {1 \over 2} \widetilde {\dot \phi}{}^2
       - V, \quad
       \pi^{\rm SF}_{ab} = 0,
   \label{fluid-SF-cov-comoving-frame}
\eea
and Eq.\ (\ref{EOM-cov-2}) simplifies to
\bea
   - \Box \phi + V_{,\phi}
       = \widetilde {\ddot \phi}
       + \theta \widetilde {\dot \phi}
       + V_{,\phi}
       = \lblue{ g_{\phi \gamma} f_{,\phi} E^a B_a},
   \label{EOM-cov-CF}
\eea
where the four-vector is defined as $U_a \widetilde {\dot \phi} \equiv - \phi_{,a}$ or $\widetilde {\dot \phi} h^b_a \phi_{,b} = 0$ where $U_a$ is the SF-comoving frame; for the axion with time-average involved, the latter case applies and we have to use Eqs.\ (\ref{EOM-cov-2}) and (\ref{fluid-SF-cov}), \cite{Hwang-Noh-2022-axion-PN}.

For the EM field, from Eqs.\ (\ref{Tab-EM-cov}) and (\ref{fluid-quantities-cov-U}), we have
\bea
   & & \mu^{\rm EM} = {1 \over 8 \pi} \left( E^2 + B^2 \right)
       = 3 p^{\rm EM},
   \nonumber \\
   & &
       q^{\rm EM}_a
       = {1 \over 4 \pi} \eta_{abcd} E^b B^c U^d,
   \nonumber \\
   & &
       \pi^{\rm EM}_{ab}
       = - {1 \over 4 \pi} \left[ E_a E_b + B_a B_b
       - {1 \over 3} h_{ab}
       \left( E^2 + B^2 \right) \right],
   \label{fluid-EM-cov}
\eea
using the general four-vector $U_a$; thus, here $\mu^{\rm SF} = \mu^{\rm SF}_{(U)}$, etc. The EM field is often evaluated in the fluid-comoving frame $u_a$ and the normal frame $n_a$. Later we will introduce notations distinguishing the EM field in the two frames, see Eq.\ (\ref{e-E}).

%%%%%%%%%%%%%%%%%%%%%%%%%%%%%%%%%%%%%%%%%%%%%%%%%%%%%%%%%%%%%%%
\subsection{Fluid conservation equations}

For the fluid, the energy-momentum tensor and fluid quantities are given in Eqs.\ (\ref{Tab-FL-cov}) and (\ref{fluid-FL-cov}). From Eq.\ (\ref{Tab-FL-cov}) we have
\bea
   & & T^{ab}_{{\rm FL}; b} u_a
       = - \left[ \widetilde {\dot {\mu}}
       + \left( \mu + p \right) \theta
       + \pi^{ab} \sigma_{ab}
       + q^a_{\;\;;a} + q^a a_a \right],
   \nonumber \\
   \label{E-conserv-cov-u} \\
   & & T^{cb}_{{\rm FL}; b} h^a_c
       = \left( \mu + p \right) a_a
       + h^b_a \left( p_{,b} + \pi^c_{b;c}
       + \widetilde {\dot {q}}_b \right)
   \nonumber \\
   & & \qquad
       + \left( \omega_{ab} + \sigma_{ab}
       + {4 \over 3} \theta h_{ab} \right) q^b.
   \label{Mom-conserv-cov-u}
\eea
As the contraction with $u_a$ indicates, here we are using the fluid four-vector, and all the kinematic quantities $\theta$, etc. and the projection tensor are associated with $u_a$. We can instead contract Eq.\ (\ref{Tab-FL-cov}) using the arbitrary frame four-vector $U_a$ and $h^{(U)}_{ac} \equiv g_{ac} + U_a U_c$, and the results are mixture of the above two conservation equations; when $U_a = n_a$, we have the ADM energy and momentum conservation equations.

For the EM field, from Eq.\ (\ref{Tab-EM-cov}), using Eq.\ (\ref{Maxwell-cov}), we have
\bea
   T^{ab}_{{\rm EM}; b}
       = \dgreen{- {1 \over c} F^{ab} J_b^{\rm em}}
       \lblue{+ g_{\phi \gamma}
       F^{ab} F^*_{bc} f^{;c}}.
\eea
Thus, using Eqs.\ (\ref{Fab}) and (\ref{current-def}), we have
\bea
   & & T^{ab}_{{\rm EM}; b} U_a
       = \dgreen{{1 \over c} E^a j_a}
       \lblue{+ g_{\phi \gamma}
       E^a B_a \widetilde {\dot f}},
   \nonumber \\
   & &
       T^{cb}_{{\rm EM}; b} h^a_c
       = \dgreen{- \varrho_{\rm em} E^a
       - {1 \over c} \eta^{abcd} j_b B_c U_d}
   \nonumber \\
   & & \qquad
       \lblue{+ g_{\phi \gamma} E^c B_c h^{ab} f_{,b}}.
\eea

For the scalar field, from Eq.\ (\ref{Tab-SF-cov}), using Eq.\ (\ref{EOM-cov}), we have
\bea
   T^{ab}_{{\rm SF}; b} = \phi^{;a} \left( \Box \phi
       - V_{,\phi} \right)
       = \lblue{- g_{\phi \gamma} f^{;a} E^b B_b}.
\eea
Thus,
\bea
   & & T^{ab}_{{\rm SF}; b} U_a
       = \lblue{- g_{\phi \gamma}
       E^a B_a \widetilde {\dot f}},
   \nonumber \\
   & &
       T^{cb}_{{\rm SF}; b} h^a_c
       = \lblue{- g_{\phi \gamma} E^b B_b h^{ac} f_{,c}}.
\eea
Combining the two field contributions, we have
\bea
   & & \left( T^{ab}_{\rm SF} + T^{ab}_{\rm EM} \right)_{;b} U_a
       = \dgreen{{1 \over c} E^a j_a},
   \\
   & & \left( T^{cb}_{\rm SF} + T^{cb}_{\rm EM} \right)_{;b}
       h^a_c
       = \dgreen{- \varrho_{\rm em} E^a
       - {1 \over c} \eta^{abcd} j_b B_c U_d},
\eea
where the $g_{\phi \gamma}$ interaction terms have canceled. The collective energy and momentum conservation equations are
\bea
   & & \left( T^{ab}_{\rm FL} + T^{ab}_{\rm SF}
       + T^{ab}_{\rm EM} \right)_{;b} U_a = 0,
   \nonumber \\
   & &
       \left( T^{cb}_{\rm FL} + T^{ab}_{\rm SF}
       + T^{cb}_{\rm EM} \right)_{;b} h^a_c = 0.
\eea

In the following we indicate the electric and magnetic fields in the fluid-comoving and normal-frames, respectively, as
\bea
   & & e_a \equiv E^{(u)}_a, \quad
       b_a \equiv B^{(u)}_a; \quad
       E_a \equiv E^{(n)}_a, \quad
       B_a \equiv B^{(n)}_a,
   \nonumber \\
   & &
       \varrho_{\rm em} \equiv \varrho_{\rm em}^{(n)}, \quad
       j_a \equiv j_a^{(n)}.
   \label{e-E}
\eea

In the fluid-comoving ($u_a$) frame, the fluid conservation equations give
\bea
   & & \widetilde {\dot {\mu}}
       + \left( \mu + p \right) \theta
       + \pi^{ab} \sigma_{ab}
       + q^a_{\;\;;a} + q^a a_a
       = \dgreen{{1 \over c} e^a j^{(u)}_a}
   \nonumber \\
   & & \qquad
       = \dgreen{{1 \over c} \gamma E^a j_a
       - \varrho_{\rm em} E^a u_a
       - {1 \over c} \eta^{abcd} j_a
       u_b n_c B_d},
   \label{E-conserv-cov} \\
   & & \left( \mu + p \right) a_a
       + h^b_a \left( p_{,b} + \pi^c_{b;c}
       + \widetilde {\dot {q}}_b \right)
   \nonumber \\
   & & \qquad
       + \left( \omega_{ab} + \sigma_{ab}
       + {4 \over 3} \theta h_{ab} \right) q^b
   \nonumber \\
   & & \qquad
       = \dgreen{\varrho_{\rm em}^{(u)} e_a
       + {1 \over c} \eta_{abcd} j^b_{(u)} b^c u^d}
   \nonumber \\
   & & \qquad
       = \dgreen{\varrho_{\rm em}
       ( E_a + u_a E_b u^b)
       + {1 \over c} ( n_a - \gamma u_a ) j^b E_b}
   \nonumber \\
   & & \qquad
       \dgreen{+ {1 \over c} \eta_{abcd} \left[
       j^b ( \gamma B^c + n^c B_e u^e ) u^d
       + u^b n^c B^d j_e u^e
       \right]}.
   \nonumber \\
   \label{Mom-conserv-cov}
\eea
The $g_{\phi \gamma}$ interaction term does not affect the fluid conservation equations. In the second steps we expressed the EM contributions using the normal-frame quantities. The relations follow from Eqs.\ (\ref{E-B-def}) and (\ref{current-def}) as
\bea
   & & e_a = \gamma E_a + n_a E_b u^b
       - \eta_{abcd} u^b n^c B^d,
   \nonumber \\
   & & b_a = \gamma B_a + n_a B_b u^b
       + \eta_{abcd} u^b n^c E^d,
   \nonumber \\
   & & \varrho^{(u)}_{\rm em}
       = \gamma \varrho_{\rm em}
       - {1 \over c} j_a u^a,
   \nonumber \\
   & & j^{(u)}_a = j_a
       + u_a j_b u^b
       + \varrho_{\rm em} c
       (n_a - \gamma u_a),
   \label{rho-j-u-n}
\eea
where we set $\varrho_{\rm em} \equiv \varrho^{(n)}_{\rm em}$ and $j_a \equiv j^{(n)}_a$.
%We note that Eq.\ (\ref{E-conserv-cov-u}), (\ref{Mom-conserv-cov-u}), (\ref{E-conserv-cov}) and (\ref{Mom-conserv-cov}) are already based on the fluid four-vector $u_a$; thus the EM parts of the above two equations are based on $u_a$ as well. In the general $U_a$ we should go back to Eq.\ (\ref{Tab-conserv-cov}), see below Eq.\ (\ref{Mom-conserv-cov-u}).

Mass conservation follows from $J^a_{\;\; ;a} = 0$ with $J^a \equiv \overline \varrho u^a$, thus
\bea
   \widetilde {\dot {\overline \varrho}}
       + \theta \overline \varrho = 0.
   \label{Mass-conserv-cov}
\eea
We have
\bea
   \mu \equiv \varrho c^2, \quad
       \varrho \equiv \overline \varrho
       \left( 1 + {\Pi / c^2} \right),
   \label{mu-varrho}
\eea
where $\overline \varrho$ is the mass density, and $\overline \varrho \Pi$ is the internal energy density; $\Pi = \varepsilon$ in the Appendix.

%For the fluid quantities in the fluid-comoving frame we can choose $U_a = u_a$ in Eqs.\ (\ref{E-conserv-cov}) and (\ref{Mom-conserv-cov}), so that the EM contributions in the two equations become \dgreen{$(1/c) e^a j^{(u)}_a$} and \dgreen{$\varrho^{(u)}_{\rm em} e^a + (1/c) \eta^{abcd} j^{(u)}_b b_c u_d$}, respectively. In MHD, we have $e_a \equiv 0$, thus the EM term in Eq.\ (\ref{E-conserv-cov}) vanishes. How can we recover the Joule heating term, then? Nothing must be wrong, as we simply have $u^a T^b_{a;b} = 0$ in the covariant energy conservation and $n^a T^b_{a;b} = 0$ in the ADM energy conservation. \lred{[CHECK???]}

Here we summarize the complete set of equations of the covariant formulation. The gravity (with the metric and kinematic variable associated with the generic four-vector $U_a$) is determined by Eqs.\ (\ref{Raychaudhury-eq-cov})-(\ref{cov-constr-3}); the fluid quantities are collective ones including the FL, SF and EM field in Eqs.\ (\ref{fluid-FL-cov}), (\ref{fluid-SF-cov}), and (\ref{fluid-EM-cov}) like $\mu = \mu^{\rm FL} + \mu^{\rm SF} + \mu^{\rm EM}$, etc.; as the fluid quantities of the FL is defined in the fluid-comoving frame, it is convenient to consider the collective fluid quantities in the fluid-comoving frame as well. The Maxwell equations are given in Eqs.\ (\ref{Maxwell-cov-1})-(\ref{Maxwell-cov-4}). The charge density and current density are provided externally. The scalar field equation is in Eq.\ (\ref{EOM-cov-2}). Three fluid conservation equations (the energy, momentum and mass conservations determining $\mu$, $u_a$ and $\overline \varrho$, respectively) are given in Eqs.\ (\ref{E-conserv-cov}), (\ref{Mom-conserv-cov}) and (\ref{Mass-conserv-cov}); here, the fluid quantities are for the fluid only given in Eq.\ (\ref{fluid-FL-cov}). Equation of state may provide $\Pi$, $p$, $q_a$ and $\pi_{ab}$. For the SF and EM field we still have a freedom to take the four-vector: like the fluid four-vector $u_a$, or the normal four-vector $n_a$, etc.

%%%%%%%%%%%%%%%%%%%%%%%%%%%%%%%%%%%%%%%%%%%%%%%%%%%%%%%%%%%%%%%
%
%
%
%%%%%%%%%%%%%%%%%%%%%%%%%%%%%%%%%%%%%%%%%%%%%%%%%%%%%%%%%%%%%%%
\section{ADM (3+1) formulation}
                                   \label{sec:ADM-formulation}

The ADM equations are derived in the Appendix \ref{Appendix-ADM}. From this section, we indicate the spacetime covariant quantities by overtildes like $\widetilde u_a$, $\widetilde E_a$, etc.

The ADM metric and its inverse are \cite{ADM, Bardeen-1980}
\bea
   & & \widetilde g_{00} \equiv - N^2 + N^i N_i, \quad
       \widetilde g_{0i} \equiv N_i, \quad
       \widetilde g_{ij} \equiv h_{ij},
   \nonumber \\
   & & \widetilde g^{00} = - N^{-2}, \quad
       \widetilde g^{0i} = N^{-2} N^i,
   \nonumber \\
   & &
       \widetilde g^{ij} = h^{ij} - N^{-2} N^i N^j,
   \label{ADM-metric-def}
\eea
where $h^{ij}$ is an inverse of the three-space intrinsic metric $h_{ij}$
\bea
   h^{ik} h_{jk} \equiv \delta^i_j,
\eea
and the index of $N_i$ is raised and lowered by $h_{ij}$ and its inverse metric $h^{ij}$. Thus,
\bea
   h_{ij} = \widetilde g_{ij}, \quad
       N_i = \widetilde g_{0i}, \quad
       N = 1/\sqrt{ - \widetilde g^{00} }.
\eea
We have
\bea
   \sqrt{ - \widetilde g} = N \sqrt{h}, \quad
       \widetilde g \equiv {\rm det} ( \widetilde g_{ab} ), \quad
       h \equiv {\rm det} (h_{ij}).
\eea
The connection and curvature in terms of the ADM metric are presented in the Appendix \ref{Appendix-ADM}.
The ADM fluid quantities are
\bea
   & & E \equiv \widetilde n_a \widetilde n_b \widetilde T^{ab},
       \quad
       J_i \equiv - \widetilde n_b \widetilde T^b_i, \quad
       S_{ij} \equiv \widetilde T_{ij},
   \nonumber \\
   & &
       S \equiv h^{ij} S_{ij}, \quad
       \overline{S}_{ij} \equiv S_{ij}
       - {1\over 3} h_{ij} S,
   \label{ADM-fluid-def}
\eea
where indices of $J_i$ and $S_{ij}$ are raised and lowered by $h_{ij}$ as the metric and its inverse. The fundamental set of ADM equations for a general imperfect fluid are Eqs.\ (\ref{Mom-constraint-ADM})-(\ref{ADM-Mom-conserv-App}). In the multiple component fluids and fields, the fluid quantities in these equations and above can be regarded as the collective ones. In our three component system we have $E = E^{\rm FL} + E^{\rm SF} + E^{\rm EM}$, etc., and all we additionally need are conservation equations for the FL, the Maxwell equations for the EM field, equation of motion for the SF, and the collective fluid quantities all in the ADM notation. We will derive these in this section.

For the normal four-vector $\widetilde n_a$, we have
\bea
   \widetilde n_i \equiv 0, \quad
       \widetilde n_0 = - N, \quad
       \widetilde n^i = - {N^i \over N}, \quad
       \widetilde n^0 = {1 \over N}.
   \label{n_a-ADM}
\eea
For the fluid velocity vector we have a couple of alternative expressions of often used in the literature. These are
\bea
   & & V^i \equiv {\widetilde h^{(n)i}_{\;\;\;\;\; b}
       \widetilde u^b \over - \widetilde n_c \widetilde u^c}
       = {1 \over N} \left( {\widetilde u^i \over \widetilde u^0}
       + N^i \right)
       = {1 \over N} \left( \overline V^i + N^i \right),
   \nonumber \\
   & &
       \overline V^i \equiv {\widetilde u^i \over \widetilde u^0}
       = {d x^i \over d x^0}
       = N V^i - N^i,
   \label{V-bar-V}
\eea
where indices of $V^i$ and $\overline V^i$  are raised and lowered by $h_{ij}$ and its inverse; $V_i$ is the fluid three-velocity measured by the Eulerian observer with $\widetilde n_a$ \cite{Banyuls-1997}, and $\overline V_i$ is the fluid coordinate three-velocity \cite{Wilson-Mathews-2003, Baumgarte-Shapiro-2010}. We use $V_i$ in this work, and the relation to $\overline V_i$ is simple as given above. Thus, our fluid four-vector is
\bea
   & & \widetilde u_i = \gamma V_i, \quad
       \widetilde u_0 = \gamma \left( N_i V^i - N \right),
   \nonumber \\
   & &
       \widetilde u^i \equiv \gamma \left( V^i - {1 \over N} N^i \right) \equiv {\gamma \over N} \overline V^i, \quad
       \widetilde u^0 = {1 \over N} \gamma,
   \label{u_a-ADM}
\eea
with the Lorentz factor
\bea
   \gamma \equiv - \widetilde n_c \widetilde u^c
       = N \widetilde u^0
       = {1 \over \sqrt{1 - V^k V_k}}.
\eea
%The fluid velocity will be used in the fluid quantities of the FL, see Eq.\ (\ref{ADM-fluid-FL}).

%%%%%%%%%%%%%%%%%%%%%%%%%%%%%%%%%%%%%%%%%%%%%%%%%%%%%%%%%%%%%%%
\subsection{Maxwell equations}

We introduce the EM field in the ADM notation. From $\widetilde b_a \widetilde u^a \equiv 0$, using Eq.\ (\ref{u_a-ADM}), we have
\bea
   & &
       \widetilde b_i \equiv b_i, \quad
       \widetilde b_0 = b_i \left( N^i - N V^i \right),
   \nonumber \\
   & &
       \widetilde b^i = b^i - {N^i \over N} b_j V^j, \quad
       \widetilde b^0 = {1 \over N} b_i V^i,
\eea
and similarly for $\widetilde e_a$ and $\widetilde j^{(u)}_a$ with $\widetilde e_i \equiv e_i$ and $\widetilde j^{(u)}_i \equiv j^{(u)}_i$; indices of $e_i$, $b_i$, and $j^{(u)}_i$ are raised and lowered by $h_{ij}$ and its inverse.
Similarly, from $\widetilde B_a \widetilde n^a \equiv 0$, using Eq.\ (\ref{n_a-ADM}), we have
\bea
   \widetilde B_i \equiv B_i, \quad
       \widetilde B_0 = B_i N^i, \quad
       \widetilde B^i = B^i, \quad
       \widetilde B^0 = 0,
\eea
and similarly for $\widetilde E_a$ and $\widetilde j_a$ with $\widetilde E_i \equiv E_i$ and $\widetilde j_i \equiv j_i$; indices of $E_i$, $B_i$, and $j_i$ are raised and lowered by $h_{ij}$ and its inverse. In the two frames we have
\bea
   \widetilde J_a
       = \varrho^{{\rm em}}_{(u)} c \widetilde u_a
       + \widetilde j^{(u)}_a
       = \varrho^{{\rm em}}_{(n)} c \widetilde n_a
       + \widetilde j^{(n)}_a,
   \label{j-relation-ADM}
\eea
thus
\bea
   & & \varrho_{\rm em} = \gamma \varrho^{(u)}_{\rm em}
       + {1 \over c}
       V^i j_i^{(u)}, \quad
       \varrho^{(u)}_{\rm em} = \gamma \varrho_{\rm em}
       - \gamma {1 \over c}
       V^i j_i,
   \nonumber \\
   & &
       j_i = j_i^{(u)}
       + c \varrho^{(u)}_{\rm em} \gamma V_i,
   \nonumber \\
   & &
       j^{(u)}_i
       = j_i
       + \gamma^2 V_i V_j j^j
       - c \varrho_{\rm em} \gamma^2 V_i,
   \label{j-u-n-relation-ADM}
\eea
where we set $\varrho_{\rm em} \equiv \varrho^{(n)}_{\rm em}$ and $j_i \equiv j^{(n)}_i$.

Now, we derive relations of EM field between the normal and comoving frames. From $\widetilde b_a \equiv \widetilde F^*_{ab} \widetilde u^b$ and $\widetilde e_a \equiv \widetilde F_{ab} \widetilde u^b$, evaluating the EM tensor in the normal-frame, and using Eqs.\ (\ref{Fab}), (\ref{F*ab}), (\ref{u_a-ADM}) and (\ref{eta-ADM}), we have
\bea
   b_i = \gamma \left( B_i
       - \overline \eta_{ijk} V^j E^k \right), \quad
       e_i = \gamma \left( E_i
       + \overline \eta_{ijk} V^j B^k \right).
   \label{b-B-relation-ADM}
\eea
Similarly, from $\widetilde B_a \equiv \widetilde F^*_{ab} \widetilde n^b$ and $\widetilde E_a \equiv \widetilde F_{ab} \widetilde n^b$, evaluating the EM tensor in the fluid-frame, and using Eqs.\ (\ref{Fab}), (\ref{F*ab}), (\ref{n_a-ADM}) and (\ref{eta-ADM}), we have
\bea
   & & B_i = \gamma \left( b_i
       - V_i V_j b^j
       + \overline \eta_{ijk} V^j e^k \right),
   \nonumber \\
   & &
       E_i = \gamma \left( e_i
       - V_i V_j e^j
       - \overline \eta_{ijk} V^j b^k \right).
   \label{B-b-relation-ADM}
\eea

The EM tensor becomes
\begin{widetext}
\bea
   & & \widetilde F_{ij} = \overline \eta_{ijk} B^k,\quad
       \widetilde F_{0i} = - N E_i - \left( {\bf N} \times {\bf B} \right)_i; \quad
       \widetilde F^{ij} = - {N^i \over N} E^j + {N^j \over N} E^i
       + \overline \eta^{ijk} B_k,\quad
       \widetilde F^{0i} = {1 \over N} E^i;
   \nonumber \\
   & & \widetilde F^i_{\;\;\;j} = - {N^i \over N} E_j
       + \overline \eta^i_{\;\;jk} B^k, \quad
       \widetilde F^i_{\;\;\;0} = - {N^i \over N} E_j N^j
       + N E^i + \overline \eta^{ijk} N_j B_k, \quad
       \widetilde F^0_{\;\;\;i} = {1 \over N} E_i, \quad
       \widetilde F^0_{\;\;\;0} = {1 \over N} E_i N^i,
   \label{Fab-ADM}
\eea
where the vector product is associated with $\overline \eta_{ijk}$ and $h_{ij}$; bold face letters indicate the three-space vectors, ${\bf N} = N^i$ and ${\bf B} = B^i$, etc., with the location of the index depending on the context. In a matrix form, we have
\bea
   \widetilde F_{ab}
       =
       \left(
       \begin{array}{rrrr}
           0 & -NE_x - ({\bf N} \times {\bf B})_x & -NE_y - ({\bf N} \times {\bf B})_y & -NE_z - ({\bf N} \times {\bf B})_z \\
         NE_x + ({\bf N} \times {\bf B})_x &    0 &  \overline \eta_{xyz} B^z & -\overline \eta_{xyz} B^y \\
         NE_y + ({\bf N} \times {\bf B})_y & -\overline \eta_{xyz} B^z &    0 & \overline \eta_{xyz} B^x \\
         NE_z + ({\bf N} \times {\bf B})_z &  \overline \eta_{xyz} B^y & -\overline \eta_{xyz} B^x &    0
       \end{array}
       \right).
   \label{Fab-matrix-ADM}
\eea

In the normal-frame, the Maxwell equations in Eqs.\ (\ref{Maxwell-cov-1})-(\ref{Maxwell-cov-4}), using Eqs.\ (\ref{n_a-ADM}) and (\ref{connection-ADM}), become
\bea
   & & E^i_{\;\; :i} = 4 \pi \left( \varrho_{\rm em}
       \lblue{- g_{\phi \gamma} B^i f_{,i}} \right),
   \label{Maxwell-ADM-1} \\
   & & E^i_{\;\;,0}
       = N K E^i + E^i_{\;\; :j} N^j - E^j N^i_{\;\; :j}
       + \overline \eta^{ijk} ( N B_k )_{:j}
       - {4 \pi \over c} N \left[ j^i
       \lblue{+ c g_{\phi \gamma} \left( B^i \widehat {\dot f}
       - \overline \eta^{ijk} E_j f_{,k} \right)} \right],
   \label{Maxwell-ADM-2} \\
   & & B^i_{\;\; :i} = 0,
   \label{Maxwell-ADM-3} \\
   & & B^i_{\;\;,0}
       = N K B^i + B^i_{\;\; :j} N^j - B^j N^i_{\;\; :j}
       - \overline \eta^{ijk} ( N E_k )_{:j},
   \label{Maxwell-ADM-4}
\eea
where
\bea
   \widehat {\dot \phi} \equiv \phi_{,a} \widetilde n^a
       = {1 \over N}
       \left( \partial_0 - N^i \nabla_i \right) \phi.
   \label{widehat}
\eea
The colon and $\nabla_i$ indicate the covariant derivative associated with the intrinsic metric $h_{ij}$. %In the absence of the helical coupling, these were presented in Eqs.\ (5.144)-(5.147) of \cite{Baumgarte-Shapiro-2010}.

In the vector notation, using Eq.\ (\ref{extrinsic-curvature-trace}), we have
\bea
   & & \nabla \cdot {\bf E}
       = 4 \pi \left( \varrho_{\rm em}
       \lblue{- g_{\phi \gamma} {\bf B} \cdot \nabla f} \right),
   \label{Maxwell-ADM-1-vector} \\
   & & {1 \over \sqrt{h}} ( \sqrt{h} {\bf E} )_{,0}
       = - \nabla \times ( {\bf N} \times {\bf E} )
       + {\bf N} \nabla \cdot {\bf E}
       + \nabla \times ( N {\bf B} )
       - {4 \pi \over c} N \left[ {\bf j}
       \lblue{+ c g_{\phi \gamma}
       \left( {\bf B} \widehat {\dot f}
       - {\bf E} \times \nabla f \right)} \right],
   \label{Maxwell-ADM-2-vector} \\
   & & \nabla \cdot {\bf B} = 0,
   \label{Maxwell-ADM-3-vector} \\
   & & {1 \over \sqrt{h}} ( \sqrt{h} {\bf B} )_{,0}
       = - \nabla \times ( {\bf N} \times {\bf B} )
       - \nabla \times ( N {\bf E} ),
   \label{Maxwell-ADM-4-vector}
\eea
where the curl and divergence operators are associated with $\overline \eta_{ijk}$ and $h_{ij}$.

%%%%%%%%%%%%%%%%%%%%%%%%%%%%%%%%%%%%%%%%%%%%%%%%%%%%%%%%%%%%%%%
\subsection{Equation of motion}

From Eq.\ (\ref{EOM-cov}), we have
\bea
   - \Box \phi + V_{,\phi}
       = \widehat {\ddot \phi} - K \widehat {\dot \phi}
       - {1 \over N} \left( N \phi^{:i} \right)_{:i}
       + V_{,\phi}
       = \lblue{g_{\phi \gamma} f_{,\phi} E^i B_i}.
   \label{EOM-ADM}
\eea

%%%%%%%%%%%%%%%%%%%%%%%%%%%%%%%%%%%%%%%%%%%%%%%%%%%%%%%%%%%%%%%
\subsection{Fluid quantities}

The FL parts of ADM fluid quantities can be derived from Eqs.\ (\ref{ADM-fluid-def}), using the following ADM notation for the fluid part. The fluid quantities based on the fluid four-vector $\widetilde u_a$ are introduced as
\bea
   \widetilde \mu = \mu, \quad
       \widetilde p = p, \quad
       \widetilde q_i \equiv Q_i, \quad
       \widetilde \pi_{ij} \equiv \Pi_{ij},
   \label{fluid-quantities-ADM}
\eea
where indices of $Q_i$ and $\Pi_{ij}$ are raised and lowered using $h_{ij}$ and its inverse. The ADM fluid quantities can be expressed in terms of the fluid quantities in the comoving frame, using Eq.\ (\ref{u_a-ADM}), as
\bea
   & & E^{\rm FL} = \mu
       + \left( \mu + p \right) \left( \gamma^2 - 1 \right)
       + 2 \gamma V^i Q_i
       + V^i V^j \Pi_{ij},
   \nonumber \\
   & & S^{\rm FL} = 3 p
       + \left( \mu + p \right) \left( \gamma^2 - 1 \right)
       + 2 \gamma V^i Q_i
       + V^i V^j \Pi_{ij},
   \nonumber \\
   & & J^{\rm FL}_i = \left( \mu + p \right) \gamma^2 V_i
       + \gamma \left( Q_i
       + V_i V^j Q_j \right)
       + V^j \Pi_{ij},
   \nonumber \\
   & & S^{\rm FL}_{ij}
       = \left( \mu + p \right) \gamma^2 V_i V_j
       + p h_{ij} + \Pi_{ij}
       + \gamma \left( Q_i V_j + Q_j V_i \right),
   \nonumber \\
   & & \overline S^{\rm FL}_{ij}
       = \Pi_{ij}
       - {1 \over 3} h_{ij} \Pi^k_k
       + \left( \mu + p \right)
       \left[ \gamma^2 V_i V_j
       - {1 \over 3} h_{ij} \left( \gamma^2 - 1 \right) \right]
       + \gamma \left( Q_i V_j + Q_j V_i
       - {2 \over 3} h_{ij} Q^k V_k \right),
   \label{ADM-fluid-FL}
\eea
where $\Pi^i_i = V^i V^j \Pi_{ij}$. The ADM fluid quantities, being defined as in Eq.\ (\ref{ADM-fluid-def}), are frame independent. For a single component fluid, by following the fluid, we may set $Q_i \equiv 0$.

The SF parts of ADM fluid quantities can be derived from Eqs.\ (\ref{ADM-fluid-def}) and (\ref{Tab-SF-cov})
\bea
   & & E^{\rm SF} = {1 \over 2} \widehat {\dot \phi}{}^2
       + V + {1 \over 2} \phi^{:k} \phi_{,k}, \quad
       J^{\rm SF}_i = - \phi_{,i} \widehat {\dot \phi}, \quad
       S^{\rm SF}_{ij} = \phi_{,i} \phi_{,j}
       + \left( {1 \over 2} \widehat {\dot \phi}{}^2
       - V
       - {1 \over 2} \phi^{:k} \phi_{,k} \right) h_{ij},
   \nonumber \\
   & & S^{\rm SF} = 3 \left( {1 \over 2} \widehat {\dot \phi}{}^2
       - V
       - {1 \over 6} \phi^{:k} \phi_{,k} \right), \quad
       \overline S^{\rm SF}_{ij}
       = \phi_{,i} \phi_{,j}
       - {1 \over 3} \phi^{:k} \phi_{,k} h_{ij}.
   \label{ADM-fluid-SF}
\eea

The EM parts of ADM fluid quantities can be derived from Eqs.\ (\ref{ADM-fluid-def}) and (\ref{Tab-EM-cov-U}) in the normal-frame
\bea
   & & E^{\rm EM}
       = {1 \over 8 \pi} \left( E^2 + B^2 \right), \quad
       J^{\rm EM}_i
       = {1 \over 4 \pi} ( {\bf E} \times {\bf B} )_i, \quad
       S^{\rm EM}_{ij}
       = - {1 \over 4 \pi} \left[ E_i E_j + B_i B_j
       - {1 \over 2} h_{ij} \left( E^2 + B^2 \right) \right],
   \nonumber \\
   & & S^{\rm EM}
       = {1 \over 8 \pi} \left( E^2 + B^2 \right), \quad
       \overline S^{\rm EM}_{ij}
       = - {1 \over 4 \pi} \left[ E_i E_j + B_i B_j
       - {1 \over 3} h_{ij} \left( E^2 + B^2 \right) \right].
   \label{ADM-fluid-EM}
\eea
%These were presented in Eqs.\ (5.119)-(5.122) of \cite{Baumgarte-Shapiro-2010}.
For the EM parts of the fluid quantities in the fluid-frame, thus evaluating Eq.\ (\ref{fluid-EM-cov}) in the $u_a$-frame, and using Eq.\ (\ref{b-B-relation-ADM}), we have
\bea
   & & \mu^{\rm EM} = 3 p^{\rm EM}
       = {\gamma^2 \over 8 \pi} \left[
       \left( 2 - {1 \over \gamma^2} \right)
       \left( E^2 + B^2 \right)
       + 4 \overline \eta_{ijk} E^i V^j B^k
       - 2 ( E_i V^i )^2
       - 2 ( B_i V^i )^2 \right],
   \nonumber \\
   & & Q^{\rm EM}_i = {1 \over 4 \pi}
       \left\{ \overline \eta_{ijk} E^j B^k
       + E_i E_j V^j + B_i B_j V^j
       - \gamma^2 ( E^2 + B^2 ) V_i
       - 2 \gamma^2 V_i \overline \eta_{jk\ell} E^j V^k B^\ell
       + \gamma^2 V_i \left[ ( E_j V^j )^2
       + ( B_j V^j )^2 \right] \right\},
   \nonumber \\
   & & \Pi^{\rm EM}_{ij}
       = - {1 \over 4 \pi} \bigg\{
       E_i E_j + B_i B_j
       + ( E_i \overline \eta_{jk\ell}
       + E_j \overline \eta_{ik\ell} ) V^k B^\ell
       - ( B_i \overline \eta_{jk\ell}
       + B_j \overline \eta_{ik\ell} ) V^k E^\ell
       + \overline \eta_{ik\ell} \overline \eta_{jmn}
       V^k V^m ( E^\ell E^n + B^\ell B^n )
   \nonumber \\
   & & \qquad
       - {1 \over 3} \gamma^2 ( h_{ij} + \gamma^2 V_i V_j )
       \left[ \left( 2 - {1 \over \gamma^2} \right)
       \left( E^2 + B^2 \right)
       + 4 \overline \eta_{k\ell m} E^k V^\ell B^m
       - 2 ( E_k V^k )^2
       - 2 ( B_k V^k )^2 \right] \bigg\}.
\eea
Evaluating these in the normal-frame, we have
\bea
   \mu^{{\rm EM}(n)} = E^{\rm EM}
       = S^{\rm EM} = 3 p^{{\rm EM}(n)}, \quad
       Q^{{\rm EM}(n)}_i = J^{\rm EM}_i, \quad
       \Pi^{{\rm EM}(n)}_{ij} = \overline S^{\rm EM}_{ij}.
\eea

%%%%%%%%%%%%%%%%%%%%%%%%%%%%%%%%%%%%%%%%%%%%%%%%%%%%%%%%%%%%%%%
\subsection{Fluid conservation equations}

In the presence of three components, the conservation equations in Eqs.\ (\ref{ADM-E-conserv-App}) and (\ref{ADM-Mom-conserv-App}) are valid for collective fluid quantities, like $E = E^{\rm FL} + E^{\rm EM} + E^{\rm SF}$, etc. These are also consistent with the Maxwell equation and equation of motion for the EM field and the SF. Using the ADM fluid quantities for the two fields in Eqs.\ (\ref{ADM-fluid-EM}) and (\ref{ADM-fluid-SF}), we can show, for the EM part
\bea
   E^{\rm EM}_{,0} N^{-1} + \dots
       = \dgreen{- {1 \over c} E^i j_i}
       \lblue{- g_{\phi \gamma} E^i B_i \widehat {\dot f}}, \quad
       J^{\rm EM}_{i,0} N^{-1} + \dots
       = \dgreen{- \varrho_{\rm em} E_i
       - {1 \over c} \overline \eta_{ijk} j^j B^k}
       \lblue{+ g_{\phi \gamma} E^j B_j f_{,i}},
\eea
where we used the Maxwell equations in Eqs.\ (\ref{Maxwell-ADM-1})-(\ref{Maxwell-ADM-4}); ellipsis ($\dots$) indicates that the left-hand-side is the same as the original ADM conservation equations in Eqs.\ (\ref{ADM-E-conserv-App}) and (\ref{ADM-Mom-conserv-App}) with the fluid quantities replaced for the individual component, here the EM part. For the SF part, we have
\bea
   E^{\rm SF}_{,0} N^{-1} + \dots
       = \lblue{g_{\phi \gamma} E^i B_i \widehat {\dot f}}, \quad
       J^{\rm SF}_{i,0} N^{-1} + \dots
       = \lblue{- g_{\phi \gamma} E^j B_j f_{,i}},
\eea
where we used the equation of motion in Eq.\ (\ref{EOM-ADM}) in the first relation. Thus, we have
\bea
   ( E^{\rm EM} + E^{\rm SF} )_{,0} N^{-1} + \dots
       = \dgreen{- {1 \over c} E^i j_i}, \quad
       ( J^{\rm EM}_i + J^{\rm SF}_i )_{,0} N^{-1} + \dots
       = \dgreen{- \varrho_{\rm em} E_i
       - {1 \over c} \overline \eta_{ijk} j^j B^k},
\eea
where the $g_{\phi \gamma}$ interaction terms have canceled. Therefore, the fluid parts of the ADM conservation equations, with $E = E^{\rm FL}$, etc., give
\bea
   & & E_{,0} N^{-1} - E_{,i} N^i N^{-1}
       - K \left( E + {1 \over 3} S \right)
       - \overline{S}^{ij} \overline{K}_{ij}
       + N^{-2} \left( N^2 J^i \right)_{:i}
       = \dgreen{{1 \over c} E^i j_i},
   \label{ADM-E-conserv-FL} \\
   & & J_{i,0} N^{-1} - J_{i:j} N^j N^{-1}
       - J_j N^j_{\;\;:i} N^{-1} - K J_i + E N_{,i} N^{-1}
       + S^j_{i:j} + S_i^j N_{,j} N^{-1}
       = \dgreen{\varrho_{\rm em} E_i
       + {1 \over c} \overline \eta_{ijk} j^j B^k},
   \label{ADM-Mom-conserv-FL}
\eea
with no $g_{\phi \gamma}$ interaction terms appearing in the conservation equations for the fluid; in these two equations $E = E^{\rm FL}$, etc. These can be written in conservative forms as
\bea
   & & \left( \sqrt{h} E \right)_{,0}
       + \left[ \sqrt{h} \left( N J^i - N^i E \right) \right]_{,i}
       = \sqrt{h} \left( N S^{ij} K_{ij}
       - N_{,i} J^i \right)
       \dgreen{+ {1 \over c} \sqrt{h} N E^i j_i},
   \label{ADM-E-conserv-FL-2} \\
   & & \left( \sqrt{h} J_i \right)_{,0}
       + \left[ \sqrt{h} \left( N S^j_i
       - N^j J_i \right) \right]_{,j}
       = \sqrt{h} \left( {1 \over 2} N S^{jk} h_{jk,i}
       + J_j N^j_{\;\;,i}
       - E N_{,i} \right)
       \dgreen{+ \sqrt{h} N \left( \varrho_{\rm em} E_i
       + {1 \over c} \overline \eta_{ijk} j^j B^k \right)}.
   \label{ADM-Mom-conserv-FL-2}
\eea
\end{widetext}
The continuity equation, $(\overline \varrho \widetilde u^c)_{;c} \equiv 0$, gives
\bea
   \left( \sqrt{h} D \right)_{,0}
       + \left[ \sqrt{h} D
       \left( N V^i - N^i \right) \right]_{,i} = 0,
   \label{ADM-mass-conserv-FL}
\eea
where we define
\bea
   D \equiv \gamma \overline \varrho, \quad
       \mu \equiv \varrho c^2, \quad
       \varrho \equiv \overline \varrho
       \left( 1 + {\Pi \over c^2} \right).
\eea
Using
\bea
   {\cal E} \equiv E - D c^2,
\eea
the energy conservation equation can be written as
\bea
   & & \left( \sqrt{h} {\cal E} \right)_{,0}
       + \left[ \sqrt{h} \left( N J^i
       - D c^2 N V^i
       - N^i {\cal E} \right) \right]_{,i}
   \nonumber \\
   & & \qquad
       = \sqrt{h} \left( N S^{ij} K_{ij}
       - N_{,i} J^i \right)
       \dgreen{+ {1 \over c} \sqrt{h} N E^i j_i},
   \label{ADM-E-conserv-FL-3}
\eea
where ${\cal E} = {\cal E}^{\rm FL}$, etc.; the same equations without the EM parts are valid for ${\cal E} = {\cal E}^{\rm tot}$, etc.

Here we summarize the complete set of equations of the ADM formulation. The gravity (with the metric variables $N$, $N_i$ and $h_{ij}$) is determined by Eqs.\ (\ref{Mom-constraint-ADM})-(\ref{ADM-prop-tracefree}). The fluid quantities are collective ones including the FL, SF and EM field in Eqs.\ (\ref{ADM-fluid-FL})-(\ref{ADM-fluid-EM}). The Maxwell equations are given in Eqs.\ (\ref{Maxwell-ADM-1-vector})-(\ref{Maxwell-ADM-4-vector}). The charge density and current density are provided externally, for example, by Ohm's law. The scalar field equation (determining $\phi$) is in Eq.\ (\ref{EOM-ADM}). Three fluid conservation equations (the energy, momentum and mass conservations determining $E$, $J_i$ and $D$, respectively) are given in Eqs.\ (\ref{ADM-E-conserv-FL})-(\ref{ADM-mass-conserv-FL}) and (\ref{ADM-E-conserv-FL-3}); here, the fluid quantities are for the fluid only given in Eq.\ (\ref{ADM-fluid-FL}). Equation of state may provide $S$ and $S_{ij}$.

%%%%%%%%%%%%%%%%%%%%%%%%%%%%%%%%%%%%%%%%%%%%%%%%%%%%%%%%%%%%%%%
%
%
%
%%%%%%%%%%%%%%%%%%%%%%%%%%%%%%%%%%%%%%%%%%%%%%%%%%%%%%%%%%%%%%%
\section{FNLE formulation}
                                    \label{sec:FNLE-formulation}

The FNLE equations are derived in the Appendix \ref{Appendix-FNLE}. From this section, we indicate the ADM fluid quantities of the FL and EM field by overlines like $\overline Q_i$, $\overline \Pi_{ij}$, $\overline E_i$, $\overline B_i$, $\overline j_i$, etc., and indicate the covariant quantities by overtildes like $\widetilde u_a$, $\widetilde E_a$, etc. Different notations are inevitable as, for example, even for the magnetic field $B_i$, we have $\widetilde B_i$, $\overline B_i$, and a plain $B_i$. The first one indicates the $i$-component of a spacetime covariant four-vector $\widetilde B_a$, the index of the second one is associated with the intrinsic metric, $h_{ij}$, and the index of the final one is raised and lowered by $\delta_{ij}$ and its inverse; the latter two do not have the temporal index. To minimize the notational complication, we omitted the overtildes in Sec.\ \ref{sec:covariant-formulation} and Appendix \ref{Appendix-cov} and overlines in Sec.\ \ref{sec:ADM-formulation} and Appendix \ref{Appendix-ADM}.

The FNLE metric is \cite{Hwang-Noh-2013}
\bea
   & & \widetilde g_{00}
       = - a^2 \left( 1 + 2 \alpha \right), \quad
       \widetilde g_{0i} = - a \chi_i,
   \nonumber \\
   & &
       \widetilde g_{ij} = a^2 \left( 1 + 2 \varphi \right) \delta_{ij}.
   \label{metric-FNLE}
\eea
The index $0$ indicates the conformal time $\eta$ with $a d \eta \equiv c dt$ and the spatial index of $\chi_i$ is raised and lowered by $\delta_{ij}$ and its inverse $\delta^{ij}$. In the FNLE formulation, we introduced the metric (of vector and tensor-type perturbations) and fluid variables (like velocity and stress) with spatial indices associated with the comoving part of three-space metric of the Friedmann background \cite{Noh-2014, Hwang-Noh-Park-2016}; for a flat Friedmann background, the three-space metric becomes $\delta_{ij}$. This is a basic assumption used for the formulation. The FNLE equations are exact but not covariant. Here we consider a flat Friedmann background.

In our metric convention we have only five degrees of freedom; we {\it ignored} the transverse-tracefree tensor-type perturbation (losing two physical degrees of freedom), and imposed a spatial gauge condition (three gauge degrees of freedom) without losing any generality or convenience; these make the spatial part of the metric tensor simple as above. We still have not imposed the temporal gauge condition. After imposing the temporal gauge condition, we have four physical degrees of freedom, two for scalar-type perturbation and two for vector-type perturbation; we may decompose the vector variables into the longitudinal and transverse parts as $\chi_i \equiv \chi_{,i} + \chi_i^{(v)}$ and $v_i \equiv - v_{,i} + v^{(v)}_i$ with $\chi^{(v)i}_{\;\;\;\;\;\;, i} \equiv 0$ and $v^{(v)i}_{\;\;\;\;\;\;, i} \equiv 0$ where $v_i$ is defined in Eq.\ (\ref{u_a-FNLE}); such a decomposition is possible even to fully nonlinear order while couplings occur in equations from the nonlinear order. Extension of the formulation including the tensor mode and without imposing the spatial gauge condition is presented in \cite{Gong-2017}.

As the temporal gauge condition we can impose $\alpha \equiv 0$ (synchronous gauge), $\varphi \equiv 0$ (uniform-curvature gauge), $\chi \equiv 0$ (zero-shear gauge), $\kappa \equiv 0$ (uniform-expansion gauge) with $\kappa$ defined in Eq.\ (\ref{extrinsic-curvature}), $v \equiv 0$ (comoving gauge). Except for synchronous gauge, where we have remnant gauge mode even after imposing the condition, the other fundamental gauge conditions, together with the spatial gauge condition we already imposed, completely fix (temporal and spatial) gauge degrees of freedom. The remaining variables after imposing such gauge conditions are free from the gauge degrees of freedom and have unique gauge invariant combination. Thus, the variables can be regarded as equivalently gauge-invariant \cite{Bardeen-1988}. These comments on the gauge issue apply to fully nonlinear orders in perturbation \cite{Noh-Hwang-2004, Hwang-Noh-2013}.

The fundamental set of FNLE equations for a general imperfect fluid are Eqs.\ (\ref{eq1})-(\ref{eq6-FL-FNLE-App}). In the multiple component fluids and fields, the fluid quantities in these equations can be regarded as the collective ones, like $E = E^{\rm FL} + E^{\rm EM} + E^{\rm SF}$, etc., and all we additionally need are conservation equations for the FL, the Maxwell equations for the EM field, equation of motion for the SF, and the collective fluid quantities all in the FNLE notation. We will derive these in this section.

The normal four-vector is
\bea
   & & \widetilde n_i \equiv 0, \quad
       \widetilde n_0 = - a {\cal N},
   \nonumber \\
   & &
       \widetilde n^i
       = {\chi^i \over a^2 {\cal N} (1 + 2 \varphi)}, \quad
       \widetilde n^0 = {1 \over a {\cal N}},
   \label{n_a-FNLE}
\eea
where ${\cal N}$ is defined in Eq.\ (\ref{N-definition}).
The fluid four-vector is
\bea
   & &
       \widetilde u_i \equiv a \gamma {v_i \over c}, \quad
       \widetilde u_0 = - \gamma \left( a {\cal N}
       + {\chi^i \over 1 + 2 \varphi} {v_i \over c} \right),
   \nonumber \\
   & &
       \widetilde u^i
       = {\gamma \over a (1 + 2 \varphi)}
       \left( {v^i \over c} + {\chi^i \over a {\cal N}} \right), \quad
       \widetilde u^0 = {1 \over a {\cal N}} \gamma,
   \label{u_a-FNLE}
\eea
with the Lorentz factor
\bea
   \gamma
       \equiv {1 \over \sqrt{ 1 - {1 \over 1 + 2 \varphi}
       {v^2 \over c^2}}},
\eea
where $v^2 \equiv v^k v_k$. Compared with the ADM fluid velocities in Eqs.\ (\ref{V-bar-V}) we have
\bea
   V_i \equiv a {v_i \over c}, \quad
       \overline V_i \equiv a^2 {\cal N} {\overline v_i \over c}, \quad
       \overline v^i = v^i + {c \chi^i \over a {\cal N}},
   \label{V_i-v_i}
\eea
where indices of $v_i$ and $\overline v_i$ are raised and lowered by $\delta_{ij}$ and its inverse; $\overline v_i$ is used in the post-Newtonian study of \cite{Chandrasekhar-1965}.

%%%%%%%%%%%%%%%%%%%%%%%%%%%%%%%%%%%%%%%%%%%%%%%%%%%%%%%%%%%%%%%
\subsection{Maxwell equations}

We introduce the EM field in the FNLE notation. From $\widetilde b_a \widetilde u^a = 0$, we have
\bea
   & & \widetilde b_i \equiv a b_i, \quad
       \widetilde b_0 = - {a {\cal N} \over 1 + 2 \varphi}
       \left( {v^i \over c} + {\chi^i \over a {\cal N}} \right) b_i,
   \nonumber \\
   & &
       \widetilde b^i = {1 \over a (1 + 2 \varphi)} \left( b^i
       + {\chi^i \over a {\cal N} ( 1 + 2 \varphi )}
       {v^j \over c} b_j \right),
   \nonumber \\
   & &
       \widetilde b^0 = {1 \over a {\cal N} (1 + 2 \varphi)}
       {v^i \over c} b_i,
\eea
and similarly for $\widetilde e_a$ and $\widetilde j^{(u)}_a$ with $\widetilde e_i \equiv a e_i$ and $\widetilde j^{(u)}_i \equiv a j^{(u)}_i$. Indices of $b_i$, $e_i$ and $j^{(u)}_i$ are raised and lowered by the metric $\delta_{ij}$ and its inverse $\delta^{ij}$. Similarly, from $\widetilde B_a \widetilde n^a = 0$, we have
\bea
   & & \widetilde B_i \equiv a B_i, \quad
       \widetilde B_0 = - {\chi^i \over 1 + 2 \varphi} B_i,
   \nonumber \\
   & &
       \widetilde B^i = {1 \over a (1 + 2 \varphi)} B^i, \quad
       \widetilde B^0 = 0,
   \label{B-cov-FNLE}
\eea
and similarly for $\widetilde E_a$ and $\widetilde j^{(n)}_i$ with $\widetilde E_i \equiv a E_i$ and $\widetilde j^{(n)}_i \equiv a j^{(n)}_i$. Indices of $B_i$, $E_i$ and $j^{(n)}_i$ are raised and lowered by the metric $\delta_{ij}$ and its inverse $\delta^{ij}$. In the two frames we have
\bea
   \widetilde J_a
       = \varrho^{{\rm em}}_{(u)} c \widetilde u_a
       + \widetilde j^{(u)}_a
       = \varrho^{{\rm em}}_{(n)} c \widetilde n_a
       + \widetilde j^{(n)}_a,
   \label{j-relation}
\eea
thus
\bea
   & & \varrho_{\rm em} = \gamma \varrho^{(u)}_{\rm em}
       + {1 \over 1 + 2 \varphi} {1 \over c}
       {v^i \over c} j_i^{(u)},
   \nonumber \\
   & &
       \varrho^{(u)}_{\rm em} = \gamma \varrho_{\rm em}
       - {\gamma \over 1 + 2 \varphi} {1 \over c}
       {v^i \over c} j_i, \quad
       j_i = j_i^{(u)} + \varrho^{(u)}_{\rm em} \gamma v_i,
   \nonumber \\
   & &
       j^{(u)}_i
       = j_i
       + {\gamma^2 \over 1 + 2 \varphi} {v_i v_j \over c^2} j^j
       - \gamma^2 \varrho_{\rm em} v_i,
   \label{j-u-n-relation}
\eea
where we set $\varrho_{\rm em} \equiv \varrho^{(n)}_{\rm em}$ and $j_i \equiv j^{(n)}_i$.

Compared with the ADM notation, we have
\bea
   \overline B_i \equiv a B_i, \quad
       \overline B^i = {B^i \over a ( 1 + 2 \varphi )},
   \label{B-ADM-FNLE}
\eea
and similarly for $\overline E_i$, $\overline j_i$, $\overline b_i$, etc. We have
\bea
   \widetilde B^2 = \overline B^2
       = {B^2 \over 1 + 2 \varphi},
   \label{B-relations}
\eea
where $\widetilde B^2 \equiv \widetilde B^c \widetilde B_c$, $\overline B^2 \equiv \overline B^i \overline B_i$, and $B^2 \equiv B^i B_i$.

We derive relations of EM field between the two frames. From $\widetilde b_a \equiv \widetilde F^*_{ab} \widetilde u^b$ and $\widetilde e_a \equiv \widetilde F_{ab} \widetilde u^b$, and evaluating the EM tensor in the normal-frame, we have
\bea
   & & b_i = \gamma \left( B_i
       - {1 \over \sqrt{1 + 2 \varphi}} \eta_{ijk} {v^j \over c} E^k \right),
   \nonumber \\
   & &
       e_i = \gamma \left( E_i
       + {1 \over \sqrt{1 + 2 \varphi}} \eta_{ijk} {v^j \over c} B^k \right),
   \label{b-B-relation}
\eea
and from $\widetilde B_a \equiv \widetilde F^*_{ab} \widetilde n^b$ and $\widetilde E_a \equiv \widetilde F_{ab} \widetilde n^b$, and evaluating the EM tensor in the fluid-frame, we have
\bea
   & & B_i = \gamma \left( b_i
       - {1 \over 1 + 2 \varphi} {v_i v_j \over c^2} b^j
       + {1 \over \sqrt{1 + 2 \varphi}} \eta_{ijk} {v^j \over c} e^k \right),
   \nonumber \\
   & &
       E_i = \gamma \left( e_i
       - {1 \over 1 + 2 \varphi} {v_i v_j \over c^2} e^j
       - {1 \over \sqrt{1 + 2 \varphi}} \eta_{ijk} {v^j \over c} b^k \right).
   \nonumber \\
   \label{B-b-relation}
\eea
These also follow from Eqs.\ (\ref{b-B-relation-ADM}) and (\ref{B-b-relation-ADM}), using Eqs.\ (\ref{V_i-v_i}), (\ref{eta-FNLE}) and (\ref{B-ADM-FNLE}).

In the FNLE formulation the Maxwell equations in Eqs.\ (\ref{Maxwell-ADM-1})-(\ref{Maxwell-ADM-4}) become
\begin{widetext}
\bea
   & & {1 \over a ( 1 + 2 \varphi )^{3/2}}
       \left( \sqrt{1 + 2 \varphi} E^i \right)_{,i}
       = 4 \pi \left( \varrho_{\rm em}
       \lblue{- g_{\phi \gamma} {B^i \nabla_i f
       \over a (1 + 2 \varphi)}} \right),
   \label{Maxwell-FNL-1-1} \\
   & & {a \over c} {\partial \over \partial t}
       \left( {E_i \over a (1 + 2 \varphi)} \right)
       = - {{\cal N} \over 1 + 2 \varphi} {3 H - \kappa \over c} E_i
       + {1 \over a^2 (1 + 2 \varphi)}
       \bigg[ - \left( {E_i \over 1 + 2 \varphi} \right)_{,j} \chi^j
       + \left( {\chi_i \over 1 + 2 \varphi} \right)_{,j} E^j \bigg]
   \nonumber \\
   & & \qquad
       + {1 \over a (1 + 2 \varphi)^{3/2}} \eta_{ijk}
       \nabla^j \left( {\cal N} B^k \right)
       - {4 \pi \over c} {{\cal N} \over 1 + 2 \varphi} \left\{ j_i
       \lblue{+ g_{\phi \gamma} \left[ {1 \over {\cal N}} B^i
       \left( \dot f + {c \chi^j \nabla_j f \over a^2( 1 + 2 \varphi)}
       \right)
       - {c \eta^{ijk} E_j \nabla_k f \over a \sqrt{1 + 2 \varphi}} \right]}
       \right\},
   \label{Maxwell-FNL-2-1} \\
   & & \left( \sqrt{1 + 2 \varphi} B^i \right)_{,i} = 0,
   \label{Maxwell-FNL-3-1} \\
   & & {a \over c} {\partial \over \partial t}
       \left( {B_i \over a (1 + 2 \varphi)} \right)
       = - {{\cal N} \over 1 + 2 \varphi} {3 H - \kappa \over c} B_i
       + {1 \over a^2 (1 + 2 \varphi)}
       \bigg[ - \left( {B_i \over 1 + 2 \varphi} \right)_{,j} \chi^j
       + \left( {\chi_i \over 1 + 2 \varphi} \right)_{,j} B^j \bigg]
   \nonumber \\
   & & \qquad
       - {1 \over a (1 + 2 \varphi)^{3/2}} \eta_{ijk}
       \nabla^j \left( {\cal N} E^k \right).
   \label{Maxwell-FNL-4-1}
\eea
Using Eq.\ (\ref{eq1}), Eqs.\ (\ref{Maxwell-FNL-2-1}) and (\ref{Maxwell-FNL-4-1}) simplify, and in the vector notation we have
\bea
   & & {\nabla \cdot ( \sqrt{1 + 2 \varphi} {\bf E} )
       \over a ( 1 + 2 \varphi )^{3/2}}
       = 4 \pi \left( \varrho_{\rm em}
       \lblue{- g_{\phi \gamma} {{\bf B} \cdot \nabla f
       \over a (1 + 2 \varphi)}} \right),
   \label{Maxwell-FNL-1} \\
   & & {1 \over c} {\partial \over \partial t}
       \left( a^2 \sqrt{1 + 2 \varphi} {\bf E} \right)
       = \nabla \times \left( {1 \over \sqrt{1 + 2 \varphi}}
       \vec{\chi} \times {\bf E} \right)
       + a \nabla \times \left( {\cal N} {\bf B} \right)
       - \vec{\chi}
       {\nabla \cdot ( \sqrt{1 + 2 \varphi} {\bf E} )
       \over 1 + 2 \varphi}
   \nonumber \\
   & & \qquad
       - {4 \pi \over c} a^2 {\cal N} \sqrt{1 + 2 \varphi}
       \left\{ {\bf j}
       \lblue{+ g_{\phi \gamma} \left[ {1 \over {\cal N}} {\bf B}
       \left( \dot f + {c \vec{\chi} \cdot \nabla f \over a^2( 1 + 2 \varphi)} \right)
       - {c {\bf E} \times \nabla f \over a \sqrt{1 + 2 \varphi}}
       \right]}
       \right\},
   \label{Maxwell-FNL-2} \\
   & & \nabla \cdot \left( \sqrt{1 + 2 \varphi} {\bf B} \right)
       = 0,
   \label{Maxwell-FNL-3} \\
   & & {1 \over c} {\partial \over \partial t}
       \left( a^2 \sqrt{1 + 2 \varphi} {\bf B} \right)
       = \nabla \times \left( {1 \over \sqrt{1 + 2 \varphi}}
       \vec{\chi} \times {\bf B} \right)
       - a \nabla \times \left( {\cal N} {\bf E} \right).
   \label{Maxwell-FNL-4}
\eea
The $g_{\phi \gamma}$ coupling causes extra charge density and current densities \cite{Wilczek-1987}.

Equation (\ref{Maxwell-FNL-3}) shows divergence of $\sqrt{1 + 2 \varphi} {\bf B}$ vanishes instead of $\nabla \cdot {\bf B}$. This is partly because of our somewhat arbitrary definition of $B_i$ in Eq.\ (\ref{B-cov-FNLE}) which leads to relations in Eq.\ (\ref{B-relations}).
%Thus, if we define $\widetilde B_i \equiv a \sqrt{1 + 2 \varphi} B^*_i$, thus $B_i = \sqrt{1 + 2 \varphi} B_i^*$, and similarly for $E_i$, we have $\widetilde B^2 = \overline B^2 = B^{*2}$ but $\nabla \cdot [ (1 + 2 \varphi) {\bf B}^* ] = 0$.
The other factor is caused because we are using divergence operator of flat space while we are in a curved space.

%%%%%%%%%%%%%%%%%%%%%%%%%%%%%%%%%%%%%%%%%%%%%%%%%%%%%%%%%%%%%%%
\subsection{Equation of motion}

From Eq.\ (\ref{EOM-ADM}) we have
\bea
   - \Box \phi + V_{,\phi}
       = \widehat {\ddot \phi}
       + {1 \over c} \left( 3 H - \kappa \right)
       \widehat {\dot \phi}
       - {( {\cal N} \sqrt{1 + 2 \varphi}
       \phi^{,i} )_{,i} \over a^2 {\cal N} (1 + 2 \varphi)^{3/2}}
       + V_{,\phi}
       = \lblue{g_{\phi \gamma} f_{,\phi}
       {E^i B_i \over 1 + 2 \varphi}},
\eea
where
\bea
   \widehat {\dot \phi} \equiv \phi_{,a} \widetilde n^a
       = {1 \over {\cal N}}
       \left( {1 \over c} {\partial \over \partial t}
       + {\chi^i \over a(1 + 2 \varphi)} \nabla_i \right) \phi.
   \label{widehat-FNLE}
\eea
Thus
\bea
   & & \ddot \phi
       + \bigg( 3 H {\cal N}
       - {\cal N} \kappa
       - {\dot {\cal N} \over {\cal N}}
       - {c \chi^i {\cal N}_{,i} \over
       a^2 {\cal N} ( 1 + 2 \varphi )} \bigg) \dot \phi
       + {2 c \chi^i \over
       a^2 ( 1 + 2 \varphi )} \dot \phi_{,i}
       - {c^2 \over a^2 (1 + 2 \varphi)}
       \left( {\cal N}^2 \delta^{ij}
       - {\chi^i \chi^j \over a^2 (1 + 2 \varphi)} \right) \phi_{,ij}
   \nonumber \\
   & & \qquad
       + \bigg[ - {{\cal N}^2 c^2 \over a^2 (1 + 2 \varphi)}
       \left( {{\cal N}^{,i} \over {\cal N}}
       + {\varphi^{,i} \over 1 + 2 \varphi} \right)
       + \bigg( 3 H {\cal N}
       - {\cal N} \kappa
       - {\dot {\cal N} \over {\cal N}}
       - {c \chi^k {\cal N}_{,k} \over
       a^2 {\cal N} c^2 ( 1 + 2 \varphi )} \bigg)
       {c \chi^i \over a^2 (1 + 2 \varphi)}
   \nonumber \\
   & & \qquad
       + \left( {c \chi^i \over a^2 (1 + 2 \varphi)}
       \right)^{\displaystyle\cdot}
       + {c \chi^k \over a^4 (1 + 2 \varphi)}
       \left( {c \chi^i \over 1 + 2 \varphi} \right)_{,k}
       \bigg] \phi_{,i}
       + c^2 {\cal N}^2 V_{,\phi}
       = \lblue{
       {\cal N}^2 c^2 g_{\phi \gamma} f_{,\phi}
       {E^i B_i \over 1 + 2 \varphi}}.
   \label{EOM-FNLE}
\eea
{\it Ignoring} the metric perturbations, thus in the Friedmann background, we have
\bea
   \ddot \phi + 3 H \dot \phi
       - c^2 {\Delta \over a^2} \phi
       + c^2 V_{,\phi}
       = \lblue{c^2 g_{\phi \gamma} f_{,\phi}
       {\bf E} \cdot {\bf B}},
   \label{EOM-Friedmann}
\eea
where $F_{ab} \widetilde F^{ab} = - 4 {\bf E} \cdot {\bf B}$ is related to the time derivative of the magnetic helicity, $\int_V {\bf A} \cdot {\bf B} d^3 x$ with ${\bf B} \equiv \nabla \times {\bf A}$ \cite{Blackman-2015}.

%%%%%%%%%%%%%%%%%%%%%%%%%%%%%%%%%%%%%%%%%%%%%%%%%%%%%%%%%%%%%%%
\subsection{Fluid quantities}
                               \label{sec:FNLE-fluid-quantities}

The fluid quantities in the FNLE formulation, $E$, $S$, $m_i$ and $m_{ij}$, are defined using the ADM fluid quantities in Eq.\ (\ref{ADM-fluid-FNLE}). For the fluid part, from Eq.\ (\ref{ADM-fluid-FL}), we have
\bea
   & & E^{\rm FL} = \mu + \left( \mu + p \right)
       \left( \gamma^2 - 1 \right)
       + {2 \gamma \over 1 + 2 \varphi} {v^i \over c}
       {Q_i \over c}
       + {1 \over (1 + 2 \varphi)^2} \Pi_{ij}
       {v^i v^j \over c^2},
   \nonumber \\
   & & S^{\rm FL} = 3 p + \left( \mu + p \right)
       \left( \gamma^2 - 1 \right)
       + {2 \gamma \over 1 + 2 \varphi} {v^i \over c}
       {Q_i \over c}
       + {1 \over (1 + 2 \varphi)^2} \Pi_{ij}
       {v^i v^j \over c^2},
   \nonumber \\
   & & J^{\rm FL}_i
       = a \left[ \left( \mu + p \right) \gamma^2
       {v_i \over c}
       + \gamma \left( \delta^j_i
       + {1 \over 1 + 2 \varphi} {v^j v_i \over c^2} \right)
       {Q_j \over c}
       + {1 \over 1 + 2 \varphi} \Pi_i^j {v_j \over c}
       \right]
       \equiv a c m^{\rm FL}_i,
   \nonumber \\
   & & S^{\rm FL}_{ij}
       = a^2 \left[ \left( 1 + 2 \varphi \right) p
       \delta_{ij}
       + \left( \mu + p \right) \gamma^2 {v_i v_j \over c^2}
       + {\gamma \over c^2}
       \left( Q_i v_j + Q_j v_i \right)
       + \Pi_{ij} \right]
       \equiv a^2 (1 + 2 \varphi) m^{\rm FL}_{ij},
   \label{ADM-fluid-FL-FNLE}
\eea
where we introduced
\bea
   \overline Q_i \equiv a {Q_i \over c}, \quad
       \overline Q^i = {Q^i \over a c ( 1 + 2 \varphi)}, \quad
       \overline \Pi_{ij} \equiv a^2 \Pi_{ij}, \quad
       \overline \Pi^i_j = {\Pi^i_j \over 1 + 2 \varphi}, \quad
       \overline \Pi^{ij}
       = {\Pi^{ij} \over a^2 ( 1 + 2 \varphi )^2},
   \label{Q-Pi-2}
\eea
with the indices of $Q_i$ and $\Pi_{ij}$ raised and lowered by $\delta_{ij}$ and its inverse; $\mu$, $p$, $Q_i$ and $\Pi_{ij}$ are fluid quantities in the fluid-frame $\widetilde u_a$. We have
\bea
   \Pi^k_k = {1 \over 1 + 2 \varphi} \Pi_{ij}
       {v^i v^j \over c^2},
\eea
and $\overline m^{\rm FL}_{ij} \equiv m^{\rm FL}_{ij} - {1 \over 3} \delta_{ij} m^{{\rm FL}k}_{\;\;\;\;\; k}$ with $m^{{\rm FL}k}_{\;\;\;\;\; k} = S^{\rm FL}$.

For the SF part, from Eq.\ (\ref{ADM-fluid-SF}), we have
\bea
   & & E^{\rm SF} = {1 \over 2}
       \left( {1 \over {\cal N} c} \dot \phi
       + {\chi^i \phi_{,i} \over a^2 {\cal N} ( 1 + 2 \varphi)}
       \right)^2 + V
       + {1 \over 2} {\phi^{,i} \phi_{,i} \over
       a^2 ( 1 + 2 \varphi )},
   \nonumber \\
   & &
       S^{\rm SF} = {3 \over 2}
       \left( {1 \over {\cal N} c} \dot \phi
       + {\chi^i \phi_{,i} \over a^2 {\cal N} ( 1 + 2 \varphi)}
       \right)^2 -3 V
       - {1 \over 2} {\phi^{,i} \phi_{,i} \over
       a^2 ( 1 + 2 \varphi )},
   \nonumber \\
   & &
       J^{\rm SF}_i = - \phi_{,i}
       \left( {1 \over {\cal N} c} \dot \phi
       + {\chi^j \phi_{,j} \over a^2 {\cal N} ( 1 + 2 \varphi)}
       \right)
       \equiv a c m^{\rm SF}_i,
   \nonumber \\
   & &
       S^{\rm SF}_{ij} = \phi_{,i} \phi_{,j}
       + \bigg[ {1 \over 2}
       \left( {1 \over {\cal N} c} \dot \phi
       + {\chi^k \phi_{,k} \over a^2 {\cal N} ( 1 + 2 \varphi)}
       \right)^2 - V
       - {1 \over 2} {\phi^{,k} \phi_{,k} \over
       a^2 ( 1 + 2 \varphi )} \bigg]
       a^2 (1 + 2 \varphi ) \delta_{ij}
       \equiv a^2 (1 + 2 \varphi) m^{\rm SF}_{ij},
   \nonumber \\
   & &
       \overline S^{\rm EM}_{ij} = \phi_{,i} \phi_{,j}
       - {1 \over 3} \phi^{,k} \phi_{,k} \delta_{ij}
       \equiv a^2 (1 + 2 \varphi) \overline m^{\rm SF}_{ij}.
   \label{ADM-fluid-SF-FNLE}
\eea

For the EM part, from Eq.\ (\ref{ADM-fluid-EM}), we have
\bea
   & & E^{\rm EM} = S^{\rm EM}
       = {1 \over 8 \pi} {1 \over 1 + 2 \varphi} \left( E^2 + B^2 \right), \quad
       J^{\rm EM}_i
       = {1 \over 4 \pi} {a \over \sqrt{1 + 2 \varphi}}
       \eta_{ijk} E^j B^k
       \equiv a c m^{\rm EM}_i,
   \nonumber \\
   & &
       S^{\rm EM}_{ij}
       = - {1 \over 4 \pi} a^2 \left[ E_i E_j + B_i B_j
       - {1 \over 2} \delta_{ij} \left( E^2 + B^2 \right) \right]
       \equiv a^2 (1 + 2 \varphi) m^{\rm EM}_{ij},
   \nonumber \\
   & &
       \overline S^{\rm EM}_{ij}
       = - {1 \over 4 \pi} a^2 \left[ E_i E_j + B_i B_j
       - {1 \over 3} \delta_{ij} \left( E^2 + B^2 \right) \right]
       \equiv a^2 (1 + 2 \varphi) \overline m^{\rm EM}_{ij}.
   \label{ADM-fluid-EM-FNLE}
\eea
We can derive covariant fluid quantities of the EM field. From Eq.\ (\ref{fluid-EM-cov}) evaluated in the fluid-frame, and transforming to the fields variables to the normal-frame ones using Eq.\ (\ref{b-B-relation}), we have
\bea
   & & \mu^{{\rm EM}}
       = 3 p^{{\rm EM}}
   \nonumber \\
   & & \qquad
       = {1 \over 8 \pi} {\gamma^2 \over 1 + 2 \varphi}
       \bigg\{ \left( 2 - {1 \over \gamma^2} \right)
       \left( E^2 + B^2 \right)
       + {4 \over \sqrt{1 + 2 \varphi}} \eta_{ijk}
       E^i {v^j \over c} B^k
       - {2 \over 1 + 2 \varphi}
       \bigg[ \left( {v^i \over c} E_i \right)^2
       + \left( {v^i \over c} B_i \right)^2 \bigg] \bigg\},
   \nonumber \\
   & & {1 \over c} Q_i^{{\rm EM}}
       = {1 \over 4 \pi} {\gamma \over \sqrt{1 + 2 \varphi}}
       \bigg\{ \eta_{ijk} E^j B^k
       + {1 \over \sqrt{1 + 2 \varphi}}
       \left[ E_i {v^j \over c} E_j
       + B_i {v^j \over c} B_j
       - \gamma^2 {v_i \over c} \left( E^2 + B^2 \right) \right]
   \nonumber \\
   & & \qquad
       - {2 \gamma^2 \over 1 + 2 \varphi} {v_i \over c}
       \eta_{jk\ell} E^j {v^k \over c} B^\ell
       + {\gamma^2 \over (1 + 2 \varphi)^{3/2}} {v_i \over c}
       \bigg[ \left( {v^j \over c} E_j \right)^2
       + \left( {v^j \over c} B_j \right)^2 \bigg]
       \bigg\},
   \nonumber \\
   & & \Pi_{ij}^{{\rm EM}}
       = - {\gamma^2 \over 4 \pi} \bigg\{
       E_i E_j + B_i B_j
       + {1 \over \sqrt{1 + 2 \varphi}}
       \left( E_i \eta_{jk\ell} + E_j \eta_{ik\ell} \right)
       {v^k \over c} B^\ell
       - {1 \over \sqrt{1 + 2 \varphi}}
       \left( B_i \eta_{jk\ell} + B_j \eta_{ik\ell} \right)
       {v^k \over c} E^\ell
   \nonumber \\
   & & \qquad
       + {1 \over 1 + 2 \varphi} \eta_{ik\ell} \eta_{jmn}
       {v^k v^m \over c^2}
       ( E^\ell E^n + B^\ell B^n )
       - {1 \over 3} \left( \delta_{ij}
       + {\gamma^2 \over 1 + 2 \varphi} {v_i v_j \over c^2} \right)
   \nonumber \\
   & & \qquad
       \times
        \bigg\{ \left( 2 - {1 \over \gamma^2} \right)
       \left( E^2 + B^2 \right)
       + {4 \over \sqrt{1 + 2 \varphi}} \eta_{k\ell m}
       E^k {v^\ell \over c} B^m
       - {2 \over 1 + 2 \varphi}
       \bigg[ \left( {v^k \over c} E_k \right)^2
       + \left( {v^k \over c} B_k \right)^2 \bigg] \bigg\}
       \bigg\}.
   \label{cov-fluid-quantities-EM-u-2}
\eea

Einstein equation parts of the FNLE equations are presented in Eqs.\ (\ref{eq1})-(\ref{eq5}) where we have $E = E^{{\rm FL}} + E^{{\rm SF}} + E^{{\rm EM}}$ and similarly for $S$, $m_i$, $m_{ij}$ and $\overline m_{ij}$.

%%%%%%%%%%%%%%%%%%%%%%%%%%%%%%%%%%%%%%%%%%%%%%%%%%%%%%%%%%%%%%%
\subsection{Fluid conservation equations}

The fluid parts of the ADM conservation equations in conservative forms, Eqs.\ (\ref{ADM-E-conserv-FL-2}) and (\ref{ADM-Mom-conserv-FL-2}), give
\bea
   & &
       {1 \over a^3} \left[
       a^3 \left( 1 + 2 \varphi \right)^{3/2} E
       \right]^{\displaystyle\cdot}
       + {1 \over a} \left[
       \sqrt{1 + 2 \varphi}
       \left( {\cal N} c^2 m^i
       + {c \over a} \chi^i E \right) \right]_{,i}
       =
       - {1 \over a} \sqrt{1 + 2 \varphi}
       {\cal N}_{,i} c^2 m^i
   \nonumber \\
   & & \qquad
       - \sqrt{1 + 2 \varphi} m^{ij}
       \left[ \left( {\dot a \over a}
       + \dot \varphi + 2 {\dot a \over a} \varphi \right)
       \delta_{ij}
       + {c \over a^2} \chi_{i,j}
       - {c \over a^2 (1 + 2 \varphi)}
       \left( 2 \chi_i \varphi_{,j}
       - \delta_{ij} \chi^k \varphi_{,k} \right) \right]
   \nonumber \\
   & & \qquad
       \dgreen{+ \sqrt{1 + 2 \varphi}
       {\cal N} {\bf E} \cdot {\bf j}},
   \label{eq6-FL-FNLE} \\
   & & {1 \over a^4} \left[ a^4
       \left( 1 + 2 \varphi \right)^{3/2} m_i
       \right]^{\displaystyle\cdot}
       + {1 \over a} \left[
       \left( 1 + 2 \varphi \right)^{3/2}
       \left( {\cal N} m_i^j
       + {c \chi^j m_i \over a (1 + 2 \varphi)} \right)
       \right]_{,j}
   \nonumber \\
   & & \qquad
       = {1 \over a} \left( 1 + 2 \varphi \right)^{3/2} \bigg[
       {{\cal N} \varphi_{,i} \over 1 + 2 \varphi} S
       - {\cal N}_{,i} E
       - {c \over a} \left( {\chi^j \over 1 + 2 \varphi} \right)_{,i} m_j \bigg]
       \dgreen{+ \left( 1 + 2 \varphi \right)^{3/2} {\cal N}
       \left( \varrho_{\rm em} E_i
       + {( {\bf j} \times {\bf B})_i \over c \sqrt{1 + 2 \varphi}}
       \right)}.
   \label{eq7-FL-FNLE}
\eea
The mass conservation in Eq.\ (\ref{ADM-mass-conserv-FL}) gives
\bea
   {1 \over a^3} \left[ a^3
       \left( 1 + 2 \varphi \right)^{3/2}
       D \right]^{\displaystyle{\cdot}}
       + {1 \over a} \left[ \sqrt{1 + 2 \varphi}
       \left( {\cal N} v^i + {c \over a} \chi^i \right)
       D \right]_{,i} = 0,
   \label{eq0-FL-FNLE}
\eea
where $D \equiv \overline \varrho \gamma$. In another conservative form, using ${\cal E} \equiv E - D c^2$, we have
\bea
   & & {1 \over a^3} \left[
       a^3 \left( 1 + 2 \varphi \right)^{3/2} {\cal E}
       \right]^{\displaystyle\cdot}
       + {1 \over a} \left\{ \sqrt{1 + 2 \varphi}
       \left[ {\cal N} c^2 ( m^i - D v^i )
       + {c \over a} \chi^i {\cal E} \right] \right\}_{,i}
       = - {c^2 \over a} \sqrt{1 + 2 \varphi} {\cal N}_{,i} m^i
   \nonumber \\
   & & \qquad
       - \sqrt{1 + 2 \varphi}
       \left[ \left( H + \dot \varphi + 2 H \varphi \right) S
       - {c \over a^2}
       {\chi^i \varphi_{,i} \over 1 + 2 \varphi} S
       + {c \over a^2} m^{ij} \left( \chi_{i,j}
       - {2 \chi_i \varphi_{,j} \over 1 + 2 \varphi} \right)
       \right]
       \dgreen{+ \sqrt{1 + 2 \varphi} {\cal N} E^i j_i}.
   \label{eq6-FL-FNLE-3}
\eea
In Eqs.\ (\ref{eq6-FL-FNLE})-(\ref{eq6-FL-FNLE-3}) we have $E = E^{\rm FL}$, etc.; the same equations without the EM parts are valid for $E = E^{\rm tot}$, etc.

The covariant equations (\ref{E-conserv-cov}) and (\ref{Mom-conserv-cov}) lead to another form of conservation equations. Ignoring $q_a$ and $\pi_{ab}$, for simplicity, we have
\bea
   & & \dot \mu
       + {1 \over a (1 + 2 \varphi)}
       \left( {\cal N} v^i + {c \over a} \chi^i \right) \mu_{,i}
   \nonumber \\
   & & \qquad
       + (\mu + p)
       \bigg[
       {{\cal N} \over \gamma} ( 3 H - \kappa )
       + {\dot \gamma \over \gamma}
       + {{\cal N} \over \gamma a}
       \left( {\gamma v^i \over 1 + 2 \varphi} \right)_{,i}
       + {c \chi^i \gamma_{,i}
       \over a^2 ( 1 + 2 \varphi ) \gamma}
       + {{\cal N} \over a (1 + 2 \varphi)}
       \left( {{\cal N}_{,i} \over {\cal N}}
       + {3 \varphi_{,i} \over 1 + 2 \varphi} \right) v^i
       \bigg]
   \nonumber \\
   & & \qquad
       = \dgreen{ {{\cal N} \over \gamma} \widetilde e^a \widetilde j^{(u)}_a
       = {{\cal N} \over (1 + 2 \varphi) \gamma}
       e^i \left( j^{(u)}_i
       - {v_i v^j \over c^2 (1 + 2 \varphi)} j^{(u)}_j \right)
       = {{\cal N} \over 1 + 2 \varphi}
       \left( E^i + {1 \over \sqrt{1 + 2 \varphi}}
       \eta^{ijk} {v_j \over c} B_k \right)
       \left( j_i - \varrho_{\rm em} v_i \right),}
   \label{eq6-FL-FNLE-cov} \\
   & & {\mu + p \over c^2} \bigg[
       {1 \over a \gamma} \left( a \gamma v_i
       \right)^{\displaystyle{\cdot}}
       + {c^2 \over a} {\cal N}_{,i}
       + {c \over a^2} \left( {\chi^j \over 1 + 2 \varphi}
       \right)_{,i} v_j
       + { (\gamma v_i)_{,j} \over a (1 + 2 \varphi) \gamma}
       \left( {\cal N} v^j + {c \over a} \chi^j \right)
       + {{\cal N} \over a \gamma^2} p_{,i}
   \nonumber \\
   & & \qquad
       + {1 \over c^2} \dot p v_i
       + {v_i p_{,j} \over a (1 + 2 \varphi) c^2}
       \left( {\cal N} v^j + {c \over a} \chi^j \right)
       \bigg]
       = \dgreen{{{\cal N} \over a \gamma^2}
       \left( \varrho_{\rm em}^{(u)} \widetilde e_i
       + {1 \over c} \widetilde \eta_{ibcd}
       \widetilde j^b_{(u)} \widetilde b^c \widetilde u^d
       \right)}
   \nonumber \\
   & & \qquad
       = \dgreen{{{\cal N} \over \gamma^2}
       \left[ \varrho^{(u)}_{\rm em} e_i
       + {1 \over c \gamma \sqrt{1 + 2 \varphi}}
       \left( \eta_{ijk}
       + {\gamma^2 \over 1 + 2 \varphi} {v_i v^\ell \over c^2}
       \eta_{jk\ell} \right) j^j_{(u)} b^k \right]}
   \nonumber \\
   & & \qquad
       = \dgreen{{{\cal N} \over \gamma^2}
       \left[ \varrho_{\rm em} E_i
       - {\gamma^2 v_i \over (1 + 2 \varphi) c^2}
       E_j \left( j^j - \varrho_{\rm em} v^j \right)
       + {1 \over c \sqrt{1 + 2 \varphi}}
       \left( \eta_{ijk}
       + {\gamma^2 \over 1 + 2 \varphi} {v_i v^\ell \over c^2}
       \eta_{jk\ell} \right) j^j B^k \right].}
   \label{eq7-FL-FNLE-cov}
\eea

\end{widetext}
In the Friedmann background, thus ignoring metric perturbations, we have
\bea
   & &
       {1 \over a^3} \left( a^3 E \right)^{\displaystyle\cdot}
       + {\dot a \over a} S
       + {c^2 \over a} m^i_{\;,i}
       = \dgreen{{\bf E} \cdot {\bf j}},
   \label{eq6-FL-Friedmann} \\
   & & {1 \over a^4} \left( a^4 m_i
       \right)^{\displaystyle\cdot}
       + {1 \over a} m^j_{i,j}
       = \dgreen{\left[ \varrho_{\rm em} E_i
       + {1 \over a c} ( {\bf j} \times {\bf B} )_i
       \right]},
   \label{eq7-FL-Friedmann} \\
   & & {1 \over a^3} \left( a^3
       D \right)^{\displaystyle{\cdot}}
       + {1 \over a} \left(
       D v^i \right)_{,i} = 0,
   \label{eq0-FL-Friedmann} \\
   & &
       {1 \over a^3} \left( a^3 {\cal E} \right)^{\displaystyle\cdot}
       + {\dot a \over a} S
       + {c^2 \over a} ( m^i - D v^i)_{\;,i}
       = \dgreen{{\bf E} \cdot {\bf j}},
\eea
where $E = E^{\rm FL}$, etc.; the same equations without the EM parts are valid for $E = E^{\rm tot}$, etc. The above equations are valid for general fluid with $Q_i$ and $\Pi_{ij}$.

Here we summarize the complete set of equations of the FNLE formulation. The gravity (with the metric variables $\alpha$, $\varphi$, $\chi_i$ and $\kappa$) is determined by Eqs.\ (\ref{eq1})-(\ref{eq5}). The fluid quantities are collective ones including the FL, SF and EM field in Eqs.\ (\ref{ADM-fluid-FL-FNLE}), (\ref{ADM-fluid-SF-FNLE}), and (\ref{ADM-fluid-EM-FNLE}). The Maxwell equations (determining ${\bf E}$ and ${\bf B}$) are given in Eqs.\ (\ref{Maxwell-FNL-1})-(\ref{Maxwell-FNL-4}). The charge density and current density are provided externally, for example, by Ohm's law. The scalar field equation (determining $\phi$) is in Eq.\ (\ref{EOM-FNLE}). Three fluid conservation equations (the energy, momentum and mass conservations determining $\mu$, $v_i$ and $\overline \varrho$, respectively) are given in Eqs.\ (\ref{eq6-FL-FNLE})-(\ref{eq7-FL-FNLE-cov}); here, the fluid quantities are for the fluid only given in Eq.\ (\ref{ADM-fluid-FL-FNLE}). Equation of state may provide $p$ and $\Pi_{ij}$. In this work we do not consider the radiation process, and we may ignore $Q_i$.
We have not imposed temporal gauge condition, and the available gauge conditions are summarized below Eq.\ (\ref{metric-FNLE}).

%%%%%%%%%%%%%%%%%%%%%%%%%%%%%%%%%%%%%%%%%%%%%%%%%%%%%%%%%%%%%%%
%
%
%
%%%%%%%%%%%%%%%%%%%%%%%%%%%%%%%%%%%%%%%%%%%%%%%%%%%%%%%%%%%%%%%
\section{Weak-gravity approximation}
                                     \label{sec:WG-formulation}

We can formulate the weak-gravity approximation by taking the weak gravity limit in the FNLE formulation. For this purpose it is important to take a suitable temporal gauge condition. The weak-gravity formulation was derived in \cite{Hwang-Noh-2016, Noh-Hwang-Bucher-2019} for a single fluid and ordinary MHD in non-expanding medium. Here we extend to the expanding background with the SF and EM field.

We set
\bea
   \alpha \equiv {\Phi \over c^2}, \quad
       \varphi \equiv - {\Psi \over c^2}, \quad
       \chi_i \equiv a {P_i \over c^3}.
\eea
The weak gravity approximation is a limit with four conditions
\bea
   & & {\Phi \over c^2} \ll 1, \quad
      - {\Psi \over c^2} \ll 1,
   \nonumber \\
   & &
      \gamma^2 {t_\ell^2 \over t_g^2} \ll 1, \quad
      {G \varrho a^2 \over c^2 \Delta}
      \sim {a^2 H^2 \over c^2 \Delta}
      \sim {\ell^2 \over \ell_H^2} \ll 1,
\eea
where the third one is the action-at-a-distance condition; $t_\ell \equiv \ell/c$ with $\ell$ a characteristic scale, and $t_g \equiv 1/\sqrt{G \varrho}$ the gravitational time-scale. The fourth condition is the sub-horizon limit with $\ell_H \equiv c t_H$, $t_H \equiv 1/H$ and $H \equiv \dot a/a$ the Hubble-Lema\^itre paramater.

A consistent weak gravity formulation, with the matter and field parts kept fully relativistic and nonlinear, is possible in uniform-expansion gauge condition setting $\kappa \equiv 0$; $\kappa$ is defined in Eq.\ (\ref{eq1}), and in non-expanding background it is the same as the maximal slicing setting the trace of extrinsic curvature equals to zero as the hypersurface condition. An alternative temporal gauge condition available is zero-shear gauge setting the longitudinal part of $\chi_i$ equals to zero, but this leads to an inconsistency with the exact result known in spherical geometry \cite{Tolman-1939, Oppenheimer-Volkoff-1939}, by missing a pressure term in the Poisson's equation \cite{Hwang-Noh-2016}. For a consistent weak gravity formulation with fully relativistic matter and field, it is essential to keep $\chi_i$ term properly.

Using the weak gravity condition and imposing uniform-expansion gauge, Eq.\ (\ref{eq3}) gives
\bea
   {1 \over a^2}
       \left( {1 \over 2} \Delta \chi_i
       + {1 \over 6} \chi^k_{\;\;,ki} \right)
       =
       - {8 \pi G \over c^3} a m_i,
   \label{eq3-WG}
\eea
where $m_i$ can be read from Eqs.\ (\ref{ADM-fluid-FL-FNLE}), (\ref{ADM-fluid-SF-FNLE}) and (\ref{ADM-fluid-EM-FNLE}). Decomposing $\chi_i$ into the longitudinal and transverse parts, we have
\bea
   & & {1 \over a} \chi
       = - {12 \pi G \over c^3} a^2 \Delta^{-2} m^i_{\;\;,i},
   \label{eq3-WG-scalar} \\
   & & {1 \over a} \chi^{(v)}_i
       = - {16 \pi G \over c^3} a^2 \Delta^{-1}
       \left( m_i - \nabla_i \Delta^{-1} m^j_{\;\;,j} \right).
   \label{eq3-WG-vector}
\eea
{\it Assuming} $m_i \sim \varrho \gamma^2 v_i$, i.e., assuming the other contributions from FL, SF and EM field are smaller than or comparable to this term, considering the action-at-a-distance condition, we have
\bea
   {1 \over a} \chi_{,i}, \;\;
       {1 \over a} \chi^{(v)}_i \ll {v_i \over c}.
\eea

Fluid conservation equations in Eqs.\ (\ref{eq6-FL-FNLE})-(\ref{eq6-FL-FNLE-3}) become
\bea
   & & \dot E + {\dot a \over a} ( 3 E + S )
       + {c^2 \over a} m^i_{\;\;,i}
       = - {1 \over a} ( 2 \Phi - \Psi )_{,i} m^i
   \nonumber \\
   & & \qquad
       \dgreen{+ {\bf E} \cdot {\bf j}},
   \label{eq6-fluid-WG} \\
   & & {1 \over a^4} ( a^4 m_i )^{\displaystyle{\cdot}}
       + {1 \over a} m^j_{i,j}
       = - {1 \over a c^2} ( \Phi_{,i} E + \Psi_{,i} S )
   \nonumber \\
   & & \qquad
       \dgreen{+ \varrho_{\rm em} E_i
       + {1 \over c} ({\bf j} \times {\bf B} )_i},
   \label{eq7-fluid-WG} \\
   & & {1 \over a^3} ( a^3 D )^{\displaystyle{\cdot}}
       + {1 \over a} ( D v^i )_{,i} = 0,
   \\
   & & \dot {\cal E} + {\dot a \over a} ( 3 {\cal E} + S )
       + {c^2 \over a} (m^i - D v^i)_{,i}
       = - {1 \over a} ( 2 \Phi - \Psi )_{,i} m^i
   \nonumber \\
   & & \qquad
       \dgreen{+ {\bf E} \cdot {\bf j}},
   \label{eq6-cal-fluid-WG}
\eea
where
\bea
   & & E = \varrho c^2 + \left( \varrho c^2 + p \right)
       \left( \gamma^2 - 1 \right)
       + 2 \gamma {v^i Q_i \over c^2}
       + \Pi_{ij} {v^i v^j \over c^2},
   \nonumber \\
   & & S = 3 p + \left( \varrho c^2 + p \right)
       \left( \gamma^2 - 1 \right)
       + 2 \gamma {v^i Q_i \over c^2}
       + \Pi_{ij} {v^i v^j \over c^2},
   \nonumber \\
   & & m_i
       = \left( \varrho + {p \over c^2} \right) \gamma^2 v_i
       + \gamma \left( \delta^j_i
       + {1 \over 1 + 2 \varphi} {v^j v_i \over c^2} \right)
       {Q_j \over c^2}
   \nonumber \\
   & & \qquad
       + \Pi_i^j {v_j \over c^2},
   \nonumber \\
   & & m_{ij}
       = p \delta_{ij}
       + \left( \varrho + {p \over c^2} \right) \gamma^2 v_i v_j
       + {\gamma \over c^2}
       \left( Q_i v_j + Q_j v_i \right)
   \nonumber \\
   & & \qquad
       + \Pi_{ij},
   \nonumber \\
   & & D \equiv \gamma \overline \varrho, \quad
       {\cal E} \equiv E - D c^2 \quad
       \varrho \equiv \overline \varrho
       \left( 1 + {\Pi \over c^2} \right).
   \label{ADM-fluid-FL-WG}
\eea
In Eqs.\ (\ref{eq6-fluid-WG})-(\ref{ADM-fluid-FL-WG}) the fluid quantities are only for the fluid with $E = E^{\rm FL}$, etc. We note that the potential terms in Eqs.\ (\ref{eq6-fluid-WG}) and (\ref{eq6-cal-fluid-WG}) are negligible compared with the $m^i_{\;\;,i}$ term by the weak-gravity condition. Similarly, $\Psi_{,i} S$ term in Eq.\ (\ref{eq7-fluid-WG}) is also negligible compared with $m^j_{i,j}$ term by the same condition.

Maxwell's equations in Eqs.\ (\ref{Maxwell-FNL-1})-(\ref{Maxwell-FNL-4}) become
\bea
   & & \nabla \cdot {\bf E}
       = 4 \pi a \left( \varrho_{\rm em}
       \lblue{- g_{\phi \gamma} {1 \over a}
       {\bf B} \cdot \nabla f} \right),
   \label{Maxwell-WG-1} \\
   & & {1 \over c} {\partial \over \partial t}
       \left( a^2 {\bf E} \right)
       =
       a \nabla \times {\bf B}
   \nonumber \\
   & & \qquad
       - {4 \pi a^2 \over c} \left[ {\bf j}
       \lblue{+ g_{\phi \gamma} \left( {\bf B} \dot f
       - {c \over a} {\bf E} \times \nabla f \right)}
       \right],
   \label{Maxwell-WG-2} \\
   & & \nabla \cdot {\bf B} = 0,
   \label{Maxwell-WG-3} \\
   & & {1 \over c} {\partial \over \partial t}
       \left( a^2 {\bf B} \right)
       = - a \nabla \times {\bf E}.
   \label{Maxwell-WG-4}
\eea
Thus, all metric perturbations disappear, and these are the same as the Maxwell equations in the Friedmann background.

The equation of motion in Eq.\ (\ref{EOM-FNLE}) gives
\bea
   \ddot \phi + 3 H \dot \phi
       - c^2 {\Delta \over a^2} \phi
       + ( c^2 + 2 \Phi ) V_{,\phi}
       = \lblue{c^2 g_{\phi \gamma} f_{,\phi}
       {\bf E} \cdot {\bf B}},
   \label{EOM-WG}
\eea
where we kept $\Phi$ term as the mass term of the SF has $c^2$ order in the potential, $V = {m^2 c^2 \over 2 \hbar^2} \phi^2$.

The remaining ones are Einstein's equation determining gravity in Eqs.\ (\ref{eq1})-(\ref{eq5}). Equation (\ref{eq3}) is already used above and the other ones give
\bea
   & & {\Delta \over a^2} \Psi
       = {4 \pi G \over c^2} (E - E_b),
   \label{eq2-WG} \\
   & & {\Delta \over a^2} \Phi
       = {4 \pi G \over c^2} (E + S - E_b - S_b),
   \label{eq4-WG} \\
   & & {c \over a^3} \left\{ a \left[ 2 ( m^i_{\;\;,j}
       + m_j^{\;\;,i} )
       - ( \nabla^i \nabla_j + \delta^i_j \Delta )
       \Delta^{-1} m^k_{\;\;,k} \right]
       \right\}^{\displaystyle{\cdot}}
   \nonumber \\
   & & \qquad
       = {1 \over a^2} \left(
       \nabla^i \nabla_j - {1 \over 3} \delta^i_j \Delta \right)
       ( \Phi - \Psi )
       + {8 \pi G \over c^2} \overline m^i_j,
   \label{eq5-WG} \\
   & & 3 \dot \Psi + 3 {\dot a \over a} \Phi
       = - 12 \pi G a \Delta^{-1} m^i_{\;\;,i},
   \label{eq1-WG}
\eea
where the subindex $b$ indicates the background order, and we used Eqs.\ (\ref{eq3-WG-scalar}) and (\ref{eq3-WG-vector}). The fluid quantities in these equations are the collective one with $E = E^{\rm FL} + E^{\rm SF} + E^{\rm EM}$, etc., thus,
\bea
   & & E = E^{\rm FL} + {1 \over 2 c^2} \dot \phi^2 + V
       + {1 \over 2 a^2} \phi^{,i} \phi_{,i}
       \dgreen{+ {1 \over 8 \pi} ( E^2 + B^2 )},
   \nonumber \\
   & & S = S^{\rm FL} + {3 \over 2 c^2} \dot \phi^2 -3 V
       - {1 \over 2 a^2} \phi^{,i} \phi_{,i}
       \dgreen{+ {1 \over 8 \pi} ( E^2 + B^2 )},
   \nonumber \\
   & & m_i = m_i^{\rm FL}
       - {1 \over a c^2} \dot \phi \phi_{,i}
       \dgreen{+ {1 \over 4 \pi c} ({\bf E} \times {\bf B})_i},
\eea
where $E^{\rm FL}$, $S^{\rm FL}$ and $m_i^{\rm FL}$ are given in Eq.\ (\ref{ADM-fluid-FL-WG}).

To {\it prove} the consistency of the weak-gravity formulation we can show the validity of remaining two equations in Eqs.\ (\ref{eq5-WG}) and (\ref{eq1-WG}). By taking $\nabla_i \nabla^j$ in Eq.\ (\ref{eq5-WG}) and using Eqs.\ (\ref{eq2-WG}), (\ref{eq4-WG}) and the momentum conservation equation for collective fluid [the same as Eq.\ (\ref{eq7-fluid-WG}) without the EM terms], we can show its validity; $\Phi_{,i} E$ and $\Psi_{,i} S$ terms are negligible by the sub-horizon and the weak-gravity conditions, respectively. Finally, Eq.\ (\ref{eq1-WG}) is naturally valid by using Eqs.\ (\ref{eq3-WG-scalar}), (\ref{eq2-WG}), (\ref{eq4-WG}) and the energy conservation equation for collective fluid [the same as Eq.\ (\ref{eq6-fluid-WG}) without the EM term]; $(2 \Phi - \Psi)_{,i} m^i$ term is negligible by the weak-gravity condition. This {\it proves} the consistency of the weak-gravity approximation with fully relativistic consideration of the fluid and fields.

Equations (\ref{eq6-fluid-WG})-(\ref{eq4-WG}) are the complete set in the weak-gravity limit.

%%%%%%%%%%%%%%%%%%%%%%%%%%%%%%%%%%%%%%%%%%%%%%%%%%%%%%%%%%%%%%%
%
%
%
%%%%%%%%%%%%%%%%%%%%%%%%%%%%%%%%%%%%%%%%%%%%%%%%%%%%%%%%%%%%%%%
\section{MHD}
                                        \label{sec:MHD}

In this section, we consider MHD approximations of the four formulations. MHD treats the evolution of magnetic field ${\bf B}$ coupled with hydrodynamic variables $\varrho$ and ${\bf v}$. Here, we additionally have the helically coupled scalar field. The Maxwell equations are replaced to the induction equation which is Faraday's equation with the electric field determined in terms of the electric current by using certain forms of Ohm's law, and the electric current determined using Amp$\grave{\rm e}$re's law. Thus, only the Maxwell equations and the EM contribution to the fluid quantities are modified. As modifications in the fluid quantity are simple (replacing the electric field using Ohm's law and Amp$\grave{\rm e}$re's law), here we consider the Maxwell equations only.

Although the conducting fluid is often composed of the mixture of charged fluids with electrons and ions, MHD handles the case as a single fluid approximation with Ohm's law. MHD usually {\it takes} as Ohm's law a simple relation between the current and the electric field in the comoving frame of the fluid \cite{Weyl-1922, %Tolman-1934,
Eckart-1940, Jackson-1975}
\bea
   \widetilde j_a^{(u)} = \sigma \widetilde e_a,
   \label{Ohms-cov}
\eea
where $\sigma$ is the isotropic electrical conductivity. Whether this simple prescription is valid in relativistic regime, in particular, for relativistic velocity, is an open issue; here, we take this relation as one of the MHD assumption. MHD approximation further {\it assumes} large conductivity with \cite{Somov-1994}
\bea
   4 \pi \sigma \gg 1/T,
   \label{MHD-condition}
\eea
where $T$ the characteristic time scale involved in the system with $\partial /(\partial t) \sim 1/T$. The {\it ideal} MHD considers $\sigma \rightarrow \infty$ limit with finite current, thus $\widetilde e_a = 0$. Keeping finite $\sigma$ gives the {\it resistive} MHD. For ideal MHD, the Joule heating term disappears in the covariant energy conservation in Eq.\ (\ref{eq6-FL-FNLE-cov}), whereas the term survives in the conservative forms in Eqs.\ (\ref{eq6-FL-FNLE}) and (\ref{eq6-FL-FNLE-3}).

%%%%%%%%%%%%%%%%%%%%%%%%%%%%%%%%%%%%%%%%%%%%%%%%%%%%%%%%%%%%%%%
\subsection{Covariant formulation}

Using Eq.\ (\ref{rho-j-u-n}), Ohm's law in Eq.\ (\ref{Ohms-cov}) gives
\bea
   & & j_a + u_a j_b u^b
       + \varrho_{\rm em} c (n_a - \gamma u_a)
   \nonumber \\
   & & \qquad
       = \sigma \left( \gamma E_a + n_a E_b u^b
       - \eta_{abcd} u^b n^c B^d \right).
\eea
In this subsection we omitted overtilde indicating the spacetime covariant quantities. In the ideal MHD limit, we have
\bea
   \left( \gamma \delta^b_a + n_a u^b \right) E_b
       = \eta_{abcd} u^b n^c B^d.
\eea
Using the MHD condition in Eq.\ (\ref{MHD-condition}) in Gauss' law in Eq.\ (\ref{Maxwell-cov-1}), the convective current term, $\varrho_{\rm em} c (n_a - \gamma u_a)$, is negligible; we use $E^a_{\;\; ;b} h^b_a \sim E_a/L$ and $c (n_a - \gamma u_a) \sim L/T$, and assume the axion contribution is of the similar order as $\varrho_{\rm em}$; $L$ is the characteristic length scale. Thus, Ohm's law in MHD approximation is
\bea
   j_a + u_a j_b u^b
       = \sigma \left( \gamma E_a + n_a E_b u^b
       - \eta_{abcd} u^b n^c B^d \right).
   \label{Ohm-cov-MHD-1}
\eea
Now, we can set up the covariant MHD formulation.

Ohm's law in Eq.\ (\ref{Ohm-cov-MHD-1}) determines $E_a$
\bea
   \gamma E_a
       = \eta_{abcd} u^b n^c B^d
       + {1 \over \sigma} \left[ j_a
       + ( u_a - \gamma n_a) j_b u^b \right].
   \label{E-MHD-cov}
\eea
Gauss' law in Eq.\ (\ref{Maxwell-cov-1}) determines $\varrho_{\rm em}$
\bea
   \varrho_{\rm em}
       = {1 \over 4 \pi} E^a_{\;\; ;b} h^b_a
       \lblue{+ g_{\phi \gamma} f_{,a} B^a},
\eea
where $h_{ab}$ is associated with $n_a$ and $\omega_{ab} = 0$ for the normal-frame. Amp$\grave{\rm e}$re's law in Eq.\ (\ref{Maxwell-cov-2}) determines $j^a$
\bea
   & & j^{a}
       = {c \over 4 \pi} \bigg[ \left( \sigma^a_{\;\;b}
       - {2 \over 3} \delta^a_b \theta \right) E^b
       + \eta^{abcd} n_d
       \left( a_b B_c - B_{b;c} \right)
   \nonumber \\
   & & \qquad
       - h^a_b \widetilde {\dot E}{}^b \bigg]
       \lblue{- c g_{\phi \gamma} f_{,b} \left( n^b B^a
       - \eta^{abcd} n_c E_d \right)},
   \label{j-MHD-cov}
\eea
where we kept the displacement current term ($\widetilde {\dot E}{}^b$-term). Combining Eqs.\ (\ref{Ohm-cov-MHD-1}) and (\ref{j-MHD-cov}), the condition in Eq.\ (\ref{MHD-condition}) does imply that the displacement current can be ignored compared with the conductive current. Thus, the displacement current disappears in the induction equation using Ohm's law and Amp$\grave{\rm e}$re's law. However, the current term also appears in the fluid conservation equations without a direct combination with Ohm's law. Thus, we keep the displacement current term in Amp$\grave{\rm e}$re's law in Eqs.\ (\ref{j-MHD-cov}), (\ref{j-MHD-ADM}), and (\ref{j-MHD-FNLE}).

Faraday's law in Eq.\ (\ref{Maxwell-cov-4}) gives induction equation for $B^a$
\bea
   h^a_b \widetilde {\dot B}{}^b
       = \left( \sigma^a_{\;\;b}
       - {2 \over 3} \delta^a_b \theta \right) B^b
       - \eta^{abcd} n_d
       \left( a_b E_c - E_{b;c} \right),
   \label{Faraday-MHD-cov}
\eea
and the no-monopole condition in Eq.\ (\ref{Maxwell-cov-3}) gives
\bea
   B^a_{\;\; ;b} h^b_a
       = 0.
   \label{no-monopole-MHD-cov}
\eea
Equations (\ref{E-MHD-cov})-(\ref{no-monopole-MHD-cov}) provide Maxwell equations determining the magnetic field in MHD approximation. The EM variables $E_a$, $\varrho_{\rm em}$ and $j_a$ are determined by Eqs.\ (\ref{E-MHD-cov})-(\ref{j-MHD-cov}). We note that $E_a$ and $j_a$ are coupled in Eqs.\ (\ref{E-MHD-cov}) and (\ref{j-MHD-cov}). The other equations determining the gravity, fluid and the scalar field are the {\it same} as summarized below Eq.\ (\ref{mu-varrho}).

%%%%%%%%%%%%%%%%%%%%%%%%%%%%%%%%%%%%%%%%%%%%%%%%%%%%%%%%%%%%%%%
\subsection{ADM formulation}

Ohm's law in Eq.\ (\ref{Ohms-cov}), using Eqs.\ (\ref{j-u-n-relation-ADM}) and (\ref{b-B-relation-ADM}), gives
\bea
   \overline {\bf j}
       + \gamma^2 {\bf V} {\bf V} \cdot \overline {\bf j}
       - \varrho_{\rm em} \gamma^2 c {\bf V}
       = \sigma \gamma \left( \overline {\bf E}
       + {\bf V} \times \overline {\bf B} \right),
   \label{Ohms-ADM}
\eea
where the curl operation is associated with $\overline \eta_{ijk}$. Using the MHD condition in Eq.\ (\ref{MHD-condition}) in Gauss' law in Eq.\ (\ref{Maxwell-ADM-1-vector}), the convective current term, $c \varrho_{\rm em} \gamma^2 {\bf V}$, is negligible; we use $\nabla \cdot \overline {\bf E} \sim \overline {\bf E}/L$ and $c {\bf V} \sim L/T$, and assume the axion contribution is of the similar order as $\varrho_{\rm em}$. Thus, Ohm's law in MHD approximation gives
\bea
   \overline {\bf j}
       + \gamma^2 {\bf V} {\bf V} \cdot \overline {\bf j}
       = \sigma \gamma \left( \overline {\bf E}
       + {\bf V} \times \overline {\bf B} \right).
   \label{Ohms-ADM-SM}
\eea
For an ideal MHD we have
\bea
   \overline {\bf E}
       = - {\bf V} \times \overline {\bf B}.
\eea
Now, we can set up the ADM MHD formulation.

Ohm's law determines $\overline {\bf E}$ as
\bea
   \overline {\bf E}
       = - {\bf V} \times \overline {\bf B}
       + {1 \over \sigma \gamma} \left( \overline {\bf j}
       + \gamma^2 {\bf V} {\bf V} \cdot \overline {\bf j} \right).
   \label{E-MHD-ADM}
\eea
Gauss' law in Eq.\ (\ref{Maxwell-ADM-1-vector}) determines $\varrho_{\rm em}$
\bea
   \varrho_{\rm em}
       = {1 \over 4 \pi} \overline \nabla \cdot \overline {\bf E}
       \lblue{+ g_{\phi \gamma} \overline {\bf B} \cdot
       \overline \nabla f}.
\eea
Amp$\grave{\rm e}$re's law in Eq.\ (\ref{Maxwell-ADM-2-vector}) determines $\overline {\bf j}$
\bea
   & & \overline {\bf j} = {c \over 4 \pi N}
       \bigg[
       - \overline \nabla \times ( {\bf N}
       \times \overline {\bf E} )
       + \overline \nabla \times ( N \overline {\bf B} )
   \nonumber \\
   & & \qquad
       - {1 \over \sqrt{h}} ( \sqrt{h} {\bf E} )_{,0} \bigg]
       \lblue{- c g_{\phi \gamma} \left( \overline {\bf B}
       \widehat {\dot f}
       - \overline {\bf E} \times \overline \nabla f \right)},
   \label{j-MHD-ADM}
\eea
where we used Eq.\ (\ref{extrinsic-curvature-trace}). Faraday's law in Eq.\ (\ref{Maxwell-ADM-4-vector}), and Eq.\ (\ref{Maxwell-ADM-3-vector}) give
\bea
   & & {1 \over \sqrt{h}}
       ( \sqrt{h} \overline {\bf B} )_{,0}
       = - \overline \nabla \times
       ( {\bf N} \times \overline {\bf B} )
       - \overline \nabla \times ( N \overline {\bf E} ),
   \label{Faraday-MHD-ADM} \\
   & & \overline \nabla \cdot \overline {\bf B} = 0.
   \label{Maxwell-ADM-3-MHD}
\eea
Equations (\ref{E-MHD-ADM})-(\ref{Maxwell-ADM-3-MHD}) provide Maxwell equations determining the magnetic field in MHD approximation. The EM variables $\overline {\bf E}$, $\varrho_{\rm em}$ and $\overline {\bf j}$ are determined by Eqs.\ (\ref{E-MHD-ADM})-(\ref{j-MHD-ADM}). We note that $\overline {\bf E}$ and $\overline {\bf j}$ are coupled in Eqs.\ (\ref{E-MHD-ADM}) and (\ref{j-MHD-ADM}). The other equations determining the gravity, fluid and the scalar field are the {\it same} as summarized below Eq.\ (\ref{ADM-E-conserv-FL-3}).

%%%%%%%%%%%%%%%%%%%%%%%%%%%%%%%%%%%%%%%%%%%%%%%%%%%%%%%%%%%%%%%
\subsection{FNLE formulation}

Ohm's law in Eq.\ (\ref{Ohms-cov}), using Eqs.\ (\ref{j-u-n-relation}) and (\ref{b-B-relation}), gives
\bea
   & & {\bf j}
       + {\gamma^2 \over 1 + 2 \varphi}
       {1 \over c^2} {\bf v} {\bf v} \cdot {\bf j}
       - \varrho_{\rm em} \gamma^2 {\bf v}
   \nonumber \\
   & & \qquad
       = \sigma \gamma \left( {\bf E}
       + {{\bf v} \times {\bf B} \over c \sqrt{1 + 2 \varphi}}
       \right),
   \label{Ohms-FNLE-0}
\eea
where the curl operation is associated with $\eta_{ijk}$. Using the MHD condition in Eq.\ (\ref{MHD-condition}) in Gauss' law in Eq.\ (\ref{Maxwell-FNL-1}), the convective current term, $\varrho_{\rm em} \gamma^2 {\bf v}$, is negligible; we use $\nabla \cdot {\bf E} \sim {\bf E}/L$ and ${\bf v} \sim L/T$, and assume the axion contribution is of the similar order as $\varrho_{\rm em}$. Thus, Ohm's law in MHD approximation gives
\bea
   {\bf j}
       + {\gamma^2 \over 1 + 2 \varphi}
       {1 \over c^2} {\bf v} {\bf v} \cdot {\bf j}
       = \sigma \gamma \left( {\bf E}
       + {{\bf v} \times {\bf B} \over c \sqrt{1 + 2 \varphi}}
       \right).
   \label{Ohms-FNLE}
\eea
In the ideal MHD limit, we have
\bea
   {\bf E}
       = - {{\bf v} \times {\bf B} \over c \sqrt{1 + 2 \varphi}}.
   \label{E-MHD-ideal}
\eea
Now, we can set up the FNLE MHD formulation.

Ohm's law in Eq.\ (\ref{Ohms-FNLE}) determines ${\bf E}$
\bea
   {\bf E}
       = - {{\bf v} \times {\bf B} \over c \sqrt{1 + 2 \varphi}}
       + {1 \over \sigma \gamma} \left( {\bf j}
       + {\gamma^2 \over 1 + 2 \varphi}
       {1 \over c^2} {\bf v} {\bf v} \cdot {\bf j} \right).
   \label{E-MHD-FNLE}
\eea
Gauss' law in Eq.\ (\ref{Maxwell-FNL-1}) determines $\varrho_{\rm em}$
\bea
   \varrho_{\rm em}
       = {\nabla \cdot ( \sqrt{1 + 2 \varphi} {\bf E} )
       \over 4 \pi a ( 1 + 2 \varphi )^{3/2}}
       \lblue{+ g_{\phi \gamma} {{\bf B} \cdot \nabla f
       \over a (1 + 2 \varphi)}}.
   \label{varrho_em-MHD}
\eea
Amp$\grave{\rm e}$re's law in Eq.\ (\ref{Maxwell-FNL-2}) determines ${\bf j}$
\bea
   & & {\bf j}
       = {c \over 4 \pi a {\cal N} \sqrt{1 + 2 \varphi}}
       \bigg[
       \nabla \times \left( {\cal N} {\bf B} \right)
       + {1 \over a} \nabla \times
       \left( {\vec{\chi} \times {\bf E}
       \over \sqrt{1 + 2 \varphi}} \right)
   \nonumber \\
   & & \qquad
       - \vec{\chi} {1 \over a}
       {\nabla \cdot ( \sqrt{1 + 2 \varphi} {\bf E} )
       \over 1 + 2 \varphi}
       - {1 \over ac} {\partial \over \partial t}
       (a^2\sqrt{1 + 2 \varphi} {\bf E} ) \bigg]
   \nonumber \\
   & & \qquad
       \lblue{- g_{\phi \gamma} \left[ {1 \over {\cal N}} {\bf B}
       \left( \dot f + {c \vec{\chi} \cdot \nabla f
       \over a^2( 1 + 2 \varphi)} \right)
       - {c {\bf E} \times \nabla f
       \over a \sqrt{1 + 2 \varphi}} \right]},
   \label{j-MHD-FNLE}
\eea
where we used Eq.\ (\ref{varrho_em-MHD}). Faraday's law in Eq.\ (\ref{Maxwell-FNL-4}), and Eq.\ (\ref{Maxwell-FNL-3}) give
\bea
   & & {1 \over c} {\partial \over \partial t}
       \left( a^2 \sqrt{1 + 2 \varphi} {\bf B} \right)
   \nonumber \\
   & & \qquad
       = - a \nabla \times \left( {\cal N} {\bf E} \right)
       + \nabla \times \left( {\vec{\chi} \times {\bf B} \over \sqrt{1 + 2 \varphi}} \right),
   \label{Faraday-MHD-FNLE} \\
   & & \nabla \cdot \left( \sqrt{1 + 2 \varphi} {\bf B} \right)
       = 0.
   \label{no-monopole-MHD-FNLE}
\eea
Equations (\ref{E-MHD-FNLE})-(\ref{no-monopole-MHD-FNLE}) provide the Maxwell equations determining the magnetic field in MHD approximation. The EM variables ${\bf E}$, $\varrho_{\rm em}$ and ${\bf j}$ are determined by Eqs.\ (\ref{E-MHD-FNLE})-(\ref{j-MHD-FNLE}). We note that ${\bf E}$ and ${\bf j}$ are coupled in Eqs.\ (\ref{E-MHD-FNLE}) and (\ref{j-MHD-FNLE}).

In MHD limit, fundamental variables are the fluid variables ($\varrho$, ${\bf v}$, etc.) with the magnetic field (${\bf B}$); in our case we also have the equation of motion for the scalar field ($\phi$) and equations for the gravity ($\alpha$, $\varphi$, $\kappa$ and $\vec{\chi}$). The fundamental equations without imposing temporal gauge condition are the following. The equation of motion for the scalar field remains {\it the same} as in Eq.\ (\ref{EOM-FNLE}) with ${\bf E}$ determined by Eq.\ (\ref{E-MHD-FNLE}). Notice that for an ideal MHD, as ${\bf E}$ is orthogonal to ${\bf B}$, the EM contribution to the equation of motion disappears. Under our MHD condition without assuming the slow-motion, all the fluid conservation equations in Eqs.\ (\ref{eq6-FL-FNLE})-(\ref{eq6-FL-FNLE-3}) remain {\it the same}. Similarly, for the gravity part, Eqs.\ (\ref{eq1})-(\ref{eq5}) remain the same with the fluid quantities in Eqs.\ (\ref{ADM-fluid-FL-FNLE}), (\ref{ADM-fluid-SF-FNLE}) and (\ref{ADM-fluid-EM-FNLE}). The weak-gravity limit and slow-motion limit will further simplify some parts in the EM contributions as we show below.

%%%%%%%%%%%%%%%%%%%%%%%%%%%%%%%%%%%%%%%%%%%%%%%%%%%%%%%%%%%%%%%
\subsubsection{Weak-gravity limit}
                                   \label{sec:MHD-WG}

In the weak-gravity limit the complete equations are in Eqs.\ (\ref{eq6-fluid-WG})-(\ref{eq4-WG}). In MHD, the Maxwell equations in Eqs.\ (\ref{Maxwell-WG-1})-(\ref{Maxwell-WG-4}) should be replaced by Eqs.\ (\ref{E-MHD-FNLE})-(\ref{no-monopole-MHD-FNLE}). In the weak field limit, these become
\bea
   & & {\bf E}
       = - {1 \over c} {\bf v} \times {\bf B}
       + {1 \over \sigma \gamma} \left( {\bf j}
       + \gamma^2
       {1 \over c^2} {\bf v} {\bf v} \cdot {\bf j} \right),
   \label{E-MHD-WG} \\
   & & \varrho_{\rm em}
       = {1 \over 4 \pi a} \nabla \cdot {\bf E}
       \lblue{+ g_{\phi \gamma} {1 \over a}
       {\bf B} \cdot \nabla f},
   \label{varrho_em-MHD-WG} \\
   & & {\bf j}
       = {c \over 4 \pi a} \nabla \times {\bf B}
       - {1 \over 4 \pi a^2} {\partial \over \partial t}
       ( a^2 {\bf E} )
   \nonumber \\
   & & \qquad
       \lblue{- g_{\phi \gamma} \left( {\bf B} \dot f
       - {c \over a} {\bf E} \times \nabla f \right)},
   \label{j-MHD-WG} \\
   & & {1 \over c} {\partial \over \partial t}
       \left( a^2 {\bf B} \right)
       = - a \nabla \times {\bf E},
   \label{Faraday-MHD-WG} \\
   & & \nabla \cdot {\bf B} = 0.
   \label{no-monopole-MHD-WG}
\eea
The other equations in Eqs.\ (\ref{eq6-fluid-WG})-(\ref{ADM-fluid-FL-WG}) and (\ref{EOM-WG})-(\ref{eq4-WG}) remain the same.

%%%%%%%%%%%%%%%%%%%%%%%%%%%%%%%%%%%%%%%%%%%%%%%%%%%%%%%%%%%%%%%
\subsubsection{Slow-motion limit}

In the slow-motion limit with $v^2/c^2 \ll 1$, the only change in Eqs.\ (\ref{E-MHD-FNLE})-(\ref{no-monopole-MHD-FNLE}) occurs in Ohm's law in Eq.\ (\ref{E-MHD-FNLE}) which simplifies to
\bea
   {\bf E}
       = - {{\bf v} \times {\bf B} \over c \sqrt{1 + 2 \varphi}}
       + {1 \over \sigma} {\bf j}.
   \label{E-MHD-SM}
\eea
Further simplifications occur in the fluid quantities of the FL and EM field, and in the fluid momentum conservation equation. First, we can set $\gamma = 1$ and ignore $v^i v^j/c^2$ in Eq.\ (\ref{ADM-fluid-FL-FNLE}) and conservation equations. For an {\it ideal} MHD, using Eq.\ (\ref{E-MHD-ideal}), $E^2$ and $E_i E_j$ in Eq.\ (\ref{ADM-fluid-EM-FNLE}) can be ignored compared with $B^2$ and $B_i B_j$. Finally, in Eq.\ (\ref{eq7-FL-FNLE}), using Eqs.\ (\ref{E-MHD-ideal}) and (\ref{j-MHD-FNLE}), the the first $\varrho_{\rm em} E_i$ term can be ignored compared with the second Lorentz force term. The remaining equations are the same.

The weak-gravity and slow-motion limit was studied in \cite{Hwang-Noh-2022-axion-MHD-WG}.

%%%%%%%%%%%%%%%%%%%%%%%%%%%%%%%%%%%%%%%%%%%%%%%%%%%%%%%%%%%%%%%
%
%
%
%%%%%%%%%%%%%%%%%%%%%%%%%%%%%%%%%%%%%%%%%%%%%%%%%%%%%%%%%%%%%%%
\subsection{Induction equation}
                                   \label{sec:Induction-eq}

Replacing ${\bf E}$ and ${\bf j}$ using Ohm's law and Amp$\grave{\rm e}$re's law, Faraday's law gives the induction equation. In the weak-gravity approximation, using Eqs.\ (\ref{E-MHD-WG}) and (\ref{j-MHD-WG}), Eq.\ (\ref{Faraday-MHD-WG}) gives
\bea
   & & {1 \over a^2}
       \left( a^2 {\bf B} \right)^{\displaystyle{\cdot}}
       - {1 \over a} \nabla \times ( {\bf v} \times {\bf B} )
       - {c^2 \Delta {\bf B} \over 4 \pi \sigma a^2}
       = \lblue{{c g_{\phi \gamma} \over \sigma \gamma a}}
       \nabla
   \nonumber \\
   & & \qquad
       \lblue{\times \bigg[ \left( {\bf I}
       + \gamma^2
       {1 \over c^2} {\bf v} {\bf v} \right) \cdot
       \left( \dot f {\bf B}
       - {c \over a} {\bf E} \times \nabla f \right)\bigg],}
   \label{Induction-WG}
\eea
where ${\bf I}$ is $\delta^i_j$ and we assumed a constant $\sigma$. The displacement current term in Amp$\grave{\rm e}$re's law can be ignored in induction equation using the MHD condition in Eq.\ (\ref{MHD-condition}). The second term in the left-hand side is the flux conserving induction term (in the absence of the other terms the magnetic field is frozen-in with the fluid), and the third term is the diffusion. The right-hand side is a contribution from the helical coupling and can work as the scalar field source for the magnetic field;  in the slow-motion limit, the first term with $\dot f {\bf B}$ can work as the $\alpha$-dynamo \cite{Krause-Radler-1980} with $\alpha = c g_{\phi \gamma} \dot f / \sigma$ \cite{Field-Carroll-2000, Campanelli-Giannotti-2005}. The $\alpha$-dynamo term causes exponential growth of the magnetic field in the linear stage, see below. For an ideal MHD ($\sigma \rightarrow \infty$), the $g_{\phi \gamma}$-coupling disappears.

Corresponding induction equations can be derived in the three exact formulations as well. In the covariant formulation, we can use Eqs.\ (\ref{E-MHD-cov}) and (\ref{j-MHD-cov}) to remove $E_a$ and $j_a$ in Eq.\ (\ref{Faraday-MHD-cov}). In the ADM formulation, we can use Eqs.\ (\ref{E-MHD-ADM}) and (\ref{j-MHD-ADM}) to remove $\overline {\bf E}$ and $\overline {\bf j}$ in Eq.\ (\ref{Faraday-MHD-ADM}). In the FNLE formulation, we can use Eqs.\ (\ref{E-MHD-FNLE}) and (\ref{j-MHD-FNLE}) to remove ${\bf E}$ and ${\bf j}$ in Eq.\ (\ref{Faraday-MHD-FNLE}), and we have
%\begin{widetext}
\bea
   & & {1 \over a^2} \left( a^2 \sqrt{1 + 2 \varphi} {\bf B}
       \right)^{\displaystyle{\cdot}}
       - {1 \over a} \nabla \bigg( {\cal N}
       {{\bf v} \times {\bf B} \over \sqrt{1 + 2 \varphi}} \bigg)
   \nonumber \\
   & & \qquad
       + {c^2 \over 4 \pi a^2} \nabla \times
       \bigg[ {1 \over \sigma \gamma}
       \left( {\bf I} + {\gamma^2 \over 1 + 2 \varphi}
       {1 \over c^2} {\bf v} {\bf v} \right) \cdot
       {\nabla \times ({\cal N} {\bf B}) \over {\cal N}
       \sqrt{1 + 2 \varphi}} \bigg]
   \nonumber \\
   & & \qquad
       = {c \over a^2} \nabla \times \bigg(
       {\vec{\chi} \times {\bf B} \over \sqrt{1 + 2 \varphi}}
       \bigg)
   \nonumber \\
   & & \qquad
       - {c^2 \over 4 \pi a^3} \nabla \times \bigg\{
       {1 \over \sigma \gamma {\cal N} \sqrt{1 + 2 \varphi}}
       \left( {\bf I} + {\gamma^2 \over 1 + 2 \varphi}
       {1 \over c^2} {\bf v} {\bf v} \right)
   \nonumber \\
   & & \qquad
       \cdot
       \bigg[ \nabla \times \bigg(
       {\vec{\chi} \times {\bf E} \over \sqrt{1 + 2 \varphi}}
       \bigg)
       - \vec{\chi} {\nabla \cdot ( \sqrt{1 + 2 \varphi} {\bf E})
       \over 1 + 2 \varphi} \bigg] \bigg\}
   \nonumber \\
   & & \qquad
       + \lblue{{c \over a} g_{\phi \gamma} \nabla \times
       \bigg\{ {1 \over \sigma \gamma}
       \left( {\bf I} + {\gamma^2 \over 1 + 2 \varphi}
       {1 \over c^2} {\bf v} {\bf v} \right)}
   \nonumber \\
   & & \qquad
       \lblue{\cdot
       \bigg[ {1 \over {\cal N}} {\bf B}
       \bigg( \dot f + {c \vec{\chi} \cdot \nabla f
       \over a^2 (1 + 2 \varphi)} \bigg)
       - {c {\bf E} \times \nabla f \over a \sqrt{1 + 2 \varphi}}
       \bigg] \bigg\}
       }.
   \label{Induction-FNLE}
\eea
%\end{widetext}
To the linear order, this gives
\bea
   {1 \over a^2}
       \left( a^2 {\bf B} \right)^{\displaystyle{\cdot}}
       - {c^2 \Delta {\bf B} \over 4 \pi a^2 \sigma}
       = \lblue{{c g_{\phi \gamma} \over \sigma a}
       \dot f
       \nabla \times {\bf B}},
   \label{Faraday-MHD-linear}
\eea
which coincides with the one from the weak-gravity approximation in Eq.\ (\ref{Induction-WG}). In Fourier space with ${\bf B} ({\bf k}, t) = \int d^3 x e^{-i{\bf k} \cdot {\bf x}} {\bf B} ({\bf x}, t)$, and using the orthonormal helicity (circular polarization) basis \cite{Jackson-1975} $( \widehat {\bf e}_+, \widehat {\bf e}_-, \widehat {\bf e}_3 )$ with $\widehat {\bf e}_\pm \equiv ( \widehat {\bf e}_1 \pm i \widehat {\bf e}_2 ) / \sqrt{2}$, $\widehat {\bf e}_3 \equiv {{\bf k} / k}$, and ${\bf B} \equiv B_+ \widehat {\bf e}_+ + B_- \widehat {\bf e}_- + B_3 \widehat {\bf e}_3$, we have
\bea
   {1 \over a^2} ( a^2 B_\pm )^{\displaystyle{\cdot}}
       = {c^2 \over 4 \pi \sigma}
       \left( - {k^2 \over a^2}
       \lblue{\pm {4 \pi \over c}
       {k \over a} g_{\phi \gamma} \dot f}
       \right) B_\pm,
\eea
with solutions \cite{Field-Carroll-2000, Campanelli-Giannotti-2005, Long-Vachaspati-2015}
\bea
   B_\pm
       = B_{\pm i} {a_i^2 \over a^2} {\rm exp}
       \left[  \int^t_{t_i}
       {c^2 \over 4 \pi \sigma}
       \left( - {k^2 \over a^2}
       \lblue{\pm {4 \pi \over c}
       {k \over a} g_{\phi \gamma} \dot f}
       \right) dt \right],
   \label{B-MHD-solutions}
\eea
and $B_3$ pure decaying. The first term is diffusion damping. The second term causes exponential growth of the magnetic field for small enough $k$ and steady $\dot f/(\sigma a)$, with maximum growth rate for $k = {2 \pi a \over c} | g_{\phi \gamma} \dot f |$, and the system tends toward to maximal helicity state \cite{Garreston-Field-Carroll-1992, Field-Carroll-2000, Campanelli-Giannotti-2005}. The maximal helicity state can cause inverse cascade of the magnetic energy to larger scales \cite{Frisch-1975, Widrow-2002, Brandenburg-Subramanian-2005}.

%%%%%%%%%%%%%%%%%%%%%%%%%%%%%%%%%%%%%%%%%%%%%%%%%%%%%%%%%%%%%%%
\section{Discussion}

We derived three exact formulations (the covariant, the ADM and the FNLE formulations) and weak-gravity formulation of the general relativistic ED and MHD including the helically coupled scalar field. The latter two formulations are presented in the cosmological context. The complete set of equations for each formulation is summarized in the final paragraph of each section.

The covariant and the ADM formulations are reformulations of Einstein's gravity using the time-like four-vector $U_a$ (thus, $1 + 3$) and the intrinsic metric $h_{ij}$ (thus, $3 + 1$), respectively. In the covariant formulation the ED and MHD without the helical coupling was studied in \cite{Barrow-Maartens-Tsagas-2007}. The EM field in that work, however, is somehow confined to the comoving-frame ($\widetilde u_a$), thus, they only used $\widetilde e_a$, $\widetilde b_a$, etc. Whereas, the conventional EM fields in the ED and MHD are the ones measured in the normal frame ($\widetilde n_a$), thus, using $\widetilde E_a$, $\widetilde B_a$, etc. as we expressed throughout this work.

The FNLE formulation is designed as exact perturbations in the Friedmann background spacetime. The resulting FNLE equations are fully nonlinear and exact with the metric perturbation variables explicitly visible in the equations, see Eqs.\ (\ref{eq1})-(\ref{eq6-FL-FNLE-App}). Compared with the covariant and the ADM formulations where the inverse metric $g^{ab}$ and $h^{ij}$ are assumed to exist, in the FNLE formulation these are exactly derived in terms of the metric tensor $g_{ab}$ \cite{Hwang-Noh-2013, Gong-2017}.

In Minkowski background the FNLE equations are generally valid in any configuration and the energy-momentum content, just like Newtonian hydrodynamic equations are valid for any arbitrary matter configuration; still, one important restriction (not in physics though) is that the FNLE formation is based on explicitly building the general exact perturbations in the Friedmann or Minkowski background spacetime, thus in the isotropic background. The FNLE formulation is provided without fixing the temporal gauge condition. For simplicity, here we ignored the tensor-type perturbation and imposed a spatial gauge condition; formulation without these restrictions is possible \cite{Gong-2017}.

The weak-gravity approximation takes weak gravity limit in the FNLE equations while keeping the energy-momentum part including the fluid and fields fully relativistic. The formulation is consistent in uniform-expansion gauge (maximal slicing in Minkowski background). Relativistic corrections appear, in particular, in the two Poisson's equations in Eqs.\ (\ref{eq2-WG}) and (\ref{eq4-WG}). We do not address the post-Newtonian approximation \cite{Chandrasekhar-1965, Poisson-Will-2014} in this work because it demands proper Newtonian limit which is not available for the general scalar field potential we are considering. Situation changes as we consider the axion as a massive scalar field.

The axion is a case of massive scalar field with $V = {m^2 c^2 \over 2 \hbar^2} \phi^2$. The coherent oscillation stage of axion can work as the cold or fuzzy dark matter depending on the mass scale \cite{Axion-CDM, Axion-pert, Hu-Barkana-Gruzinov-2000, Axion-DM-review}. In such a field we have the Klein and the Madelung transformations which transform the SF to the Schr\"odinger equation and conservation equations with proper Newtonian (nonrelativistic) limit. In the presence of $g_{\phi \gamma}$-coupling, however, in \cite{Hwang-Noh-2022-axion-MHD-WG} we found that these two transformations are possible {\it only} for $f \propto \phi^2$ which is not a favored form in the axion physics \cite{Sikivie-1983}.

For a massive scalar field with $f = {1 \over 2} \phi^2$ coupling, under the Klein transformation \cite{Klein-1926}
\bea
   \phi ({\bf x}, t)
       \equiv {\hbar \over \sqrt{2m}}
       \left[ \psi ({\bf x}, t) e^{- i \omega_c t}
       + \psi^* ({\bf x}, t) e^{+ i \omega_c t} \right],
   \label{Klein-tr}
\eea
the equation of motion leads to the Schr\"odinger equation with $\psi$ the complex wave function; $\omega_c \equiv m c^2/\hbar$ is the Compton frequency \cite{Hwang-Noh-2022-axion}. By further applying the Madelung transformation \cite{Madelung-1927}
\bea
   \psi \equiv \sqrt{\varrho_\phi \over m}
       e^{i m u/\hbar},
   \label{Madelung-tr}
\eea
the Schr\"odinger equation leads to continuity and momentum conservation equations with $\varrho_\phi$ and $u$ interpreted as the mass density and velocity potential (${\bf u}_\phi = {1 \over a} \nabla u$), and with a characteristic quantum stress term encompassing the quantum nature \cite{Bohm-1952}. In \cite{Hwang-Noh-2022-axion-MHD-WG} we studied the weak-gravity and slow-motion limit of this case, and studied gravitational instability caused by MHD and magnetic instability caused by the helically coupled field. For the same $V$ and $f$ these two transformations can be applied in the three exact formulations as well, but we do not pursue the subject here; in the absence of the FL and EM field, see \cite{Hwang-Noh-2022-axion-PN} for the covariant formulation and the post-Newtonian approximation of the axion.

The exact equations of the helically coupled ED and MHD in the ADM and FNLE formulations will be useful to implement in numerical simulations. The ADM formulation of ED and MHD without the helical coupling was implemented in numerical relativity \cite{Baumgarte-Shapiro-2010}. The helically coupled scalar field as the (cold or fuzzy) dark matter or the dark energy will be the interesting subject to explore in the context of origin and evolution of the cosmic magnetic field.

The weak-gravity formulation in Minkowski background will be useful in many astrophysical situations where relativistic effects of fluid and fields are important while the gravity is weak. Indeed, in nearly all astrophysical situations, except for nearby compact objects like the black holes, neutron stars and white dwarfs, the gravitational fields are extremely weak with $\Phi/c^2 \sim G M / (R c^2) \lesssim 10^{-5}$ where $M$ and $R$ are characteristic mass and length scales. Even in cosmology, the gravitational field is similarly extremely weak in nearly all observable cosmic scales including the cosmic microwave background radiation. Equations in the weak-gravity approximation are similar to Newtonian hydrodynamic equations and will be much easier for numerical implementation compared with numerical relativity which is a complicated constrained system.

%%%%%%%%%%%%%%%%%%%%%%%%%%%%%%%%%%%%%%%%%%%%%%%%%%%%%%%%%%%%%%%
%
%
%
%%%%%%%%%%%%%%%%%%%%%%%%%%%%%%%%%%%%%%%%%%%%%%%%%%%%%%%%%%%%%%%
%\section*{Acknowledgments}
\vskip .5cm \centerline{\bf Acknowledgments} \vskip .5cm

We thank the referee for many valuable suggestions and questions.  H.N.\ was supported by the National Research Foundation (NRF) of Korea funded by the Korean Government (No.\ 2018R1A2B6002466 and No.\ 2021R1F1A1045515). J.H.\ was supported by IBS under the project code, IBS-R018-D1, and by the NRF of Korea funded by the Korean Government (No.\ NRF-2019R1A2C1003031).

\appendix
%%%%%%%%%%%%%%%%%%%%%%%%%%%%%%%%%%%%%%%%%%%%%%%%%%%%%%%%%%%%%%%
%
%
%
%%%%%%%%%%%%%%%%%%%%%%%%%%%%%%%%%%%%%%%%%%%%%%%%%%%%%%%%%%%%%%%
\section{Covariant (1+3) formulation}
                                  \label{Appendix-cov}

The $1+3$ covariant decomposition is based on the time-like normalized ($u^a u_a \equiv - 1$) four-vector field $u_a$ introduced in all spacetime points \cite{Ehlers-1993, Ellis-1971, Ellis-1973, Hawking-1966}. The expansion ($\theta$), the acceleration ($a_a$), the rotation ($\omega_{ab}$), and the shear ($\sigma_{ab}$) are kinematic quantities of the projected covariant derivative of flow vector $u_a$ introduced as
\bea
   & & h^c_a h^d_b u_{c;d}
       = h^c_{[a} h^d_{b]} u_{c;d}
       + h^c_{(a} h^d_{b)} u_{c;d}
       \equiv \omega_{ab} + \theta_{ab}
   \nonumber \\
   & & \qquad
       = u_{a;b} + a_a u_b, \quad
       \sigma_{ab} \equiv \theta_{ab}
       - {1 \over 3} \theta h_{ab},
   \nonumber \\
   & &
       \theta \equiv u^a_{\;\; ;a}, \quad
       a_a \equiv {\widetilde {\dot {u}}}_a
       \equiv u_{a;b} u^b,
   \label{kinematic-shear}
\eea
where $h_{ab} \equiv g_{ab} + u_a u_b$ is the projection tensor with $h_{ab} u^b = 0$ and $h^a_a = 3$; an overdot with tilde $\widetilde {\dot {}}$ indicates a covariant derivative along $u^a$; indices surrounded by $()$ and $[]$ are the symmetrization and anti-symmetrization symbols, respectively, with $A_{(ab)} \equiv {1 \over 2} (A_{ab} + A_{ba})$ and $A_{[ab]} \equiv {1 \over 2} (A_{ab} - A_{ba})$. Thus
\bea
   u_{a;b} = \omega_{ab} + \sigma_{ab}
       + {1 \over 3} \theta h_{ab} - a_a u_b.
   \label{kinematic-cov-2}
\eea
We introduce
\bea
   & & \omega^a \equiv {1 \over 2} \eta^{abcd} u_b
       \omega_{cd}, \quad
       \omega_{ab} = \eta_{abcd} \omega^c u^d,
   \nonumber \\
   & &
       \omega^2 \equiv {1 \over 2} \omega^{ab} \omega_{ab}
       = \omega^a \omega_a, \quad
       \sigma^2 \equiv {1 \over 2} \sigma^{ab} \sigma_{ab},
   \label{kinematic-cov-3}
\eea
where $\omega^a$ is a {\it vorticity vector} which has the same
information as the vorticity tensor $\omega_{ab}$. We have
\bea
   \omega_{ab} u^b = \omega_a u^a = \sigma_{ab} u^b
       = a_a u^a = 0, \quad u^b u_{b;a} = 0.
\eea

We have the antisymmetric tensor $\eta^{abcd} = \eta^{[abcd]}$ with $\eta^{0123} = 1/\sqrt{- g}$, thus
\bea
   \eta^{abcd} = {1 \over \sqrt{-g}} \epsilon^{abcd}, \quad
       \eta_{abcd} = - \sqrt{-g} \epsilon_{abcd},
\eea
where $\epsilon^{abcd}$ is an antisymmetric symbol with $\epsilon^{0123} = \epsilon_{0123} = + 1$. We have
\bea
   & &
       \eta^{abcd} \eta_{efgh}
       = - 4 ! \delta^e_{[e} \delta^b_f \delta^c_g \delta^d_{h]}
       = - \left|
       \begin{array}{llll}
       \delta^a_e & \delta^a_f & \delta^a_g & \delta^a_h \\
       \delta^b_e & \delta^b_f & \delta^b_g & \delta^b_h \\
       \delta^c_e & \delta^c_f & \delta^c_g & \delta^c_h \\
       \delta^d_e & \delta^d_f & \delta^d_g & \delta^d_h
       \end{array}
       \right|,
   \nonumber \\
   & &
       \eta^{abcd} \eta_{efgd}
       = - 3 ! \delta^a_{[e} \delta^b_f \delta^c_{g]}
       = - \left|
       \begin{array}{lll}
       \delta^a_e & \delta^a_f & \delta^a_g \\
       \delta^b_e & \delta^b_f & \delta^b_g \\
       \delta^c_e & \delta^c_f & \delta^c_g
       \end{array}
       \right|,
   \nonumber \\
   & &
       \eta^{abcd} \eta_{efcd}
       = - 4 \delta^a_{[c} \delta^b_{d]}
       = - 2 \left|
       \begin{array}{lll}
       \delta^a_e & \delta^a_f \\
       \delta^b_e & \delta^b_f
       \end{array}
       \right|
   \nonumber \\
   & & \qquad
       = - 2 \left( \delta^a_e \delta^b_f - \delta^a_f \delta^b_e
       \right),
   \nonumber \\
   & &
        \eta^{abcd} \eta_{ebcd}
       = - 6 \delta^a_e, \quad
       \eta^{abcd} \eta_{abcd} = - 24.
\eea

Our convention of the Riemann curvature and Einstein's equation are:
\bea
   & & u_{a;bc} - u_{a;cb} = u_d R^d_{\;\;abc},
   \label{Riemann-def} \\
   & & R_{ab} - {1 \over 2} R g_{ab} + \Lambda g_{ab}
       = {8 \pi G \over c^4} T_{ab}.
   \label{Einstein-eq}
\eea
The Weyl (conformal) curvature is introduced as
\bea
   & & C_{abcd} \equiv R_{abcd} - {1 \over 2} (
       g_{ac} R_{bd} + g_{bd} R_{ac}
       - g_{bc} R_{ad}
   \nonumber \\
   & & \qquad
       - g_{ad} R_{bc} )
       + {R \over 6} \left( g_{ac} g_{bd}
       - g_{ad} g_{bc} \right).
   \label{Weyl}
\eea
The electric and magnetic parts of the Weyl curvature are introduced as \cite{Bertschinger-1995}
\bea
   & & E_{ab} \equiv C_{acbd} u^c u^d, \quad
       H_{ab} \equiv C^*_{abcd} u^c u^d,
   \nonumber \\
   & & C^{abcd}
       = \left( - \eta^{ab}_{\;\;\;\; pq} \eta^{cd}_{\;\;\;\;rs}
       + g^{ab}_{\;\;\;\; pq} g^{cd}_{\;\;\;\;rs} \right)
       u^p u^r E^{qs}
   \nonumber \\
   & & \qquad
       - \left( \eta^{ab}_{\;\;\;\; pq} g^{cd}_{\;\;\;\;rs}
       + g^{ab}_{\;\;\;\; pq} \eta^{cd}_{\;\;\;\;rs} \right)
       u^p u^r H^{qs},
   \label{EM-Weyl}
\eea
where $C^*_{abcd} \equiv {1 \over 2} \eta_{ac}^{\;\;\;\;ef} C_{efbd}$. We have $E_{ab} = E_{ba}$ and $E_{ab} u^b = 0$, and similarly for $H_{ab}$. Similarly as in EM in Eq.\ (\ref{EM-invariants}), we have two invariants
\bea
   & & C^{abcd} C_{abcd}
       = 8 ( E^{ab} E_{ab} - H^{ab} H_{ab} )
       = C^{*abcd} C^*_{abcd},
   \nonumber \\
   & & C^{abcd} C^*_{abcd}
       = 16 E^{ab} H_{ab},
\eea
with $C^{abcd} C_{abcd} = R^{abcd} R_{abcd} - 2 R^{ab} R_{ab} + {1 \over 3} R^2$.

The energy-momentum tensor is decomposed into fluid quantities
based on the four-vector field $u^a$ as
\bea
   T_{ab} \equiv \mu u_a u_b
       + p \left( g_{ab} + u_a u_b \right)
       + q_a u_b + q_b u_a
       + \pi_{ab},
   \label{Tab-FL-App}
\eea
where
\bea
   u^a q_a = 0 = u^a \pi_{ab}, \quad
       \pi_{ab} = \pi_{ba}, \quad
       \pi^a_a = 0.
   \label{q-pi-relations-App}
\eea
The variables $\mu$, $p$, $q_a$ and $\pi_{ab}$ are the energy density, the isotropic pressure (including the entropic one), the energy flux and the anisotropic pressure (stress) based on $u_a$-frame, respectively. We have
\bea
   & & \mu \equiv T_{ab} u^a u^b, \quad
       p \equiv {1 \over 3} T_{ab} h^{ab}, \quad
       q_a \equiv - T_{cd} u^c h_a^d,
   \nonumber \\
   & &
       \pi_{ab} \equiv T_{cd} h_a^c h_b^d
       - p h_{ab}.
   \label{fluid-Tab-App}
\eea

The kinematic and fluid quantities and the electric and magnetic parts of Weyl tensor depend on the frame four-vector $u_a$. Although the fluid quantities of the real fluid are defined in the fluid-comoving frame where $u_a$ is the fluid four-vector, here, we may consider $u_a$ as arbitrary time-like four vector.

%%%%%%%%%%%%%%%%%%%%%%%%%%%%%%%%%%%%%%%%%%%%%%%%%%%%%%%%%%%%%%%%%
\subsection{Covariant equations}

The specific (per mass) entropy $S$ can be introduced by $T dS = c^2 d \varepsilon + p_T d v$ where $\varepsilon$ is specific internal energy with $\mu = \overline \varrho ( c^2 + \varepsilon )$, $p_T$ the thermodynamic pressure, $v \equiv 1/\overline \varrho$ the specific volume, and $T$ the temperature. We have the isotropic pressure $p = p_T + e$ where $e$ is the entropic pressure (isotropic stress). Using eqs. (\ref{cov-E-conserv-App}) and (\ref{cov-Mass-conserv-App}) below we can show
\bea
   \overline \varrho T \widetilde {\dot {S}}
       = - \left( e \theta
       + \pi^{ab} \sigma_{ab}
       + q^a_{\;\;;a} + q^a a_a \right).
   \label{entropy}
\eea
Thus, we notice that $e$, $\pi^{ab}$ and $q^a$ generate the entropy. Using a four-vector $S^a \equiv \overline \varrho u^a S + {1 \over T} q^a$ which is termed the entropy flow density \cite{Ehlers-1993} we can derive
\bea
   S^a_{\;\; ;a}
       = - {1 \over T} \left( {T_{,a} \over T}
       + a_a \right) q^a
       - {1 \over T} \left( e \theta
       + \pi^{ab} \sigma_{ab} \right).
\eea
We derive the covariant $(1+3)$ set of equations in the following, see \cite{Ehlers-1993, Ellis-1971, Ellis-1973, Hawking-1966, Bertschinger-1995}.

The energy and the momentum conservation equations follow from $u_a T^{ab}_{\;\;\;\;;b} = 0$ and $h^c_a T^{ab}_{\;\;\;\; ;b} = 0$, respectively
\bea
   & & \widetilde {\dot {\mu}}
       + \left( \mu + p \right) \theta
       + \pi^{ab} \sigma_{ab}
       + q^a_{\;\;;a} + q^a a_a = 0,
   \label{cov-E-conserv-App} \\
   & & \left( \mu + p \right) a_a
       + h^b_a \left( p_{,b} + \pi^c_{b;c}
       + \widetilde {\dot {q}}_b \right)
   \nonumber \\
   & & \qquad
       + \bigg( \omega_{ab} + \sigma_{ab}
       + {4 \over 3} \theta h_{ab} \bigg) q^b = 0.
   \label{cov-Mom-conserv-App}
\eea
The mass conservation follows from $j^a \equiv \overline \varrho u^a$ and $j^a_{\;\; ;a} = 0$ as
\bea
   \widetilde {\dot {\overline \varrho}}
       + \theta \overline \varrho = 0.
   \label{cov-Mass-conserv-App}
\eea

By applying $u^b$ on Eq.\ (\ref{Riemann-def}) we have
\bea
   \left( u_{a;c} \right)^{\widetilde \cdot}
       - a_{a;c}
       + u_{a;b} u^b_{\;\; ;c}
       + R_{abcd} u^b u^d = 0.
   \label{propagation-eq}
\eea
By applying $g^{ac}$ on Eq.\ (\ref{propagation-eq}) we have the Raychaudhuri equation
\bea
   & & \widetilde {\dot {\theta}} + {1 \over 3} \theta^2
       - a^a_{\;\; ;a}
       + 2 \left( \sigma^2 - \omega^2 \right)
       + {4 \pi G \over c^4}
       \left( \mu + 3 p \right) - \Lambda
   \nonumber \\
   & & \qquad
       = 0.
   \label{Raychaudhury-eq-cov}
\eea
By applying $\eta^{acst} u_s$ on Eq.\ (\ref{propagation-eq}) we have the vorticity propagation equation
\bea
   h^a_b \widetilde {\dot {\omega}}{}^b
       + {2 \over 3} \theta \omega^a
       = \sigma^a_b \omega^b
       + {1 \over 2} \eta^{abcd} u_b a_{c;d}.
   \label{cov-vorticity-prop}
\eea
An equivalent equation can be derived by applying $h^a_{[d} h^c_{e]}$ on Eq.\ (\ref{propagation-eq})
\bea
   h^c_a h^d_b \left( \widetilde {\dot {\omega}}_{cd}
       - a_{[c;d]} \right)
       + {2 \over 3} \theta \omega_{ab}
       = 2 \sigma^e_{[a} \omega_{b]c}.
   \label{cov-vorticity-prop-2}
\eea
By applying $h^a_{(d} h^c_{e)}$ on Eq.\ (\ref{propagation-eq}) we have the shear propagation equation
\bea
   & & h_a^c h_b^d \left( \widetilde {\dot {\sigma}}_{cd}
       - a_{(c;d)} \right)
       - a_a a_b + \omega_a \omega_b
       + \sigma_{ac} \sigma^c_b
       + {2 \over 3} \theta \sigma_{ab}
   \nonumber \\
   & & \qquad
       - {1 \over 3} h_{ab} \left( \omega^2
       + 2 \sigma^2
       - a^c_{\;\; ;c} \right) + E_{ab}
       - {4 \pi G \over c^4} \pi_{ab}
   \nonumber \\
   & & \qquad
       = 0.
   \label{shear-propagation-cov}
\eea
By applying $g^{ac} h^{be}$ on Eq.\ (\ref{Riemann-def}) we have
\bea
   & & h_{ab} \bigg( \omega^{bc}_{\;\;\; ;c}
       - \sigma^{bc}_{\;\;\; ;c}
       + {2 \over 3} \theta^{;b} \bigg)
       + \left( \omega_{ab} + \sigma_{ab} \right) a^b
   \nonumber \\
   & & \qquad
       = {8 \pi G \over c^4} q_a.
   \label{shear-constraint-cov}
\eea
From Eq.\ (\ref{Riemann-def}) we have $u_{[a;bc]} = 0$. Thus, from $\eta^{abcd} u_d u_{a;bc} = 0$ we have
\bea
   \omega^a_{\;\; ;a} = 2 \omega^b a_b.
   \label{cov-constr-2}
\eea
By applying $h^g_{(e} h^a_{f)} \eta_{gh}^{\;\;\;\; bc} u^h$ on Eq.\ (\ref{Riemann-def}) we have
\bea
   H_{ab}
       = 2 a_{(a} \omega_{b)}
       - h_a^c h_b^d \left( \omega_{(c}^{\;\;\; e;f}
       + \sigma_{(c}^{\;\;\; e;f} \right) \eta_{d)gef} u^g.
   \label{cov-constr-3}
\eea

From the Bianchi identity $R_{ab[cd;e]} = 0$ we have
\bea
   C^{abcd}_{\;\;\;\;\;\;\; ;d}
       = R^{c[a;b]}
       - {1 \over 6} g^{c[a} R^{;b]},
\eea
using $C^{abcd}$ in Eq.\ (\ref{EM-Weyl}) we can derive the four quasi-Maxwellian equations
\begin{widetext}
\bea
   & & h^a_b h^c_d E^{bd}_{\;\;\;\; ;c}
       - \eta^{abcd} u_b \sigma_c^e H_{de}
       + 3 H^a_b \omega^b
       = {4 \pi G \over c^4} \left( {2 \over 3} h^{ab} \mu_{,b}
       - h^a_b \pi^{bc}_{\;\;\;\; ;c}
       - 3 \omega^a_{\;\;b} q^b
       + \sigma^a_b q^b
       + \pi^a_b a^b
       - {2 \over 3} \theta q^a \right),
   \label{cov-Maxwell-1} \\
   & & h^a_b h^c_d H^{bd}_{\;\;\;\; ;c}
       + \eta^{abcd} u_b \sigma_c^e E_{de}
       - 3 E^a_b \omega^b
       = {4 \pi G \over c^4} \left\{ 2 \left( \mu
       + p \right) \omega^a
       + \eta^{abcd} u_b \left[ q_{c;d}
       + \pi_{ce} \left( \omega^e_{\;\; d}
       + \sigma^e_{\;\; d} \right) \right] \right\},
   \label{cov-Maxwell-2} \\
   & &  h^a_c h^b_d \widetilde {\dot {E}} {}^{cd}
        + \left( H^f_{d;e} h_f^{(a}
        - 2 a_d H_e^{(a} \right) \eta^{b)cde} u_c
        + h^{ab} \sigma^{cd} E_{cd}
        + \theta E^{ab}
        - E_c^{(a} \left( 3 \sigma^{b)c}
        + \omega^{b)c} \right)
       = {4 \pi G \over c^4} \Big[
       - \left( \mu + p \right) \sigma^{ab}
   \nonumber \\
   & & \qquad
       - 2 a^{(a} q^{b)}
       - h^{(a}_c h^{b)}_d \left( q^{c;d}
       + \widetilde {\dot {\pi}} {}^{cd} \right)
       - \left( \omega_c^{\;\;(a}
       + \sigma^{(a}_c \right) \pi^{b)c}
       - {1 \over 3} \theta \pi^{ab}
       + {1 \over 3} \left( q^c_{\;\; ;c}
       + a_c q^c
       + \pi^{cd} \sigma_{cd} \right) h^{ab} \Big],
   \label{cov-Maxwell-3} \\
   & & h^a_c h^b_d \widetilde {\dot {H}} {}^{cd}
       - \left( E^f_{d;e} h_f^{(a}
       - 2 a_d E_e^{(a} \right) \eta^{b)cde} u_c
       + h^{ab} \sigma^{cd} H_{cd}
       + \theta H^{ab}
       - H_c^{(a} \left( 3 \sigma^{b)c}
       + \omega^{b)c} \right)
   \nonumber \\
   & & \qquad
       = {4 \pi G \over c^4} \left[
       \left( q_e \sigma^{(a}_d
       - \pi^f_{d;e} h_f^{(a} \right)
       \eta^{b)cde} u_c
       + h^{ab} \omega_c q^c
       - 3 \omega^{(a} q^{b)} \right].
   \label{cov-Maxwell-4}
\eea
These follow, respectively, from
\bea
   - u_b u_c C^{abcd}_{\;\;\;\;\;\;\; ;d}, \quad
       - {1 \over 2} h^g_e u_f \eta^{ef}_{\;\;\;\; ab} u_c C^{abcd}_{\;\;\;\;\;\;\; ;d}, \quad
       - h_a^{(e} h_c^{f)} u_b C^{abcd}_{\;\;\;\;\;\;\; ;d}, \quad
       - {1 \over 2} h^{(g}_e h^{h)}_c u_f \eta^{ef}_{\;\;\;\; ab} u_c C^{abcd}_{\;\;\;\;\;\;\; ;d},
\eea
where we used the momentum conservation and the energy conservation equations in Eq.\ (\ref{cov-Maxwell-1}) and (\ref{cov-Maxwell-3}), respectively. Notice the analogy with Maxwell equations in Eqs.\ (\ref{Maxwell-cov-1})-(\ref{Maxwell-cov-4}).

We note that the time-like four-vector used above is generic one. We also note that in the presence of multiple components like several components of FL, SF, and the EM field, etc., $T_{ab}$ is the sum of each component, and consequently the fluid quantities ($\mu$, $p$, $q_a$ and $\pi_{ab}$) are also the collective ones defined in Eq.\ (\ref{fluid-Tab-App}). As we explain in Sec.\ \ref{sec:covariant-formulation}, in such a case, equations governing the individual component should be supplemented, like conservation equations for individual fluid component, the equation of motion for the SF, and the Maxwell equations for the EM field, etc.

%%%%%%%%%%%%%%%%%%%%%%%%%%%%%%%%%%%%%%%%%%%%%%%%%%%%%%%%%%%%%%%
%
%
%
%%%%%%%%%%%%%%%%%%%%%%%%%%%%%%%%%%%%%%%%%%%%%%%%%%%%%%%%%%%%%%%
\section{ADM (3+1) formulation}
                                  \label{Appendix-ADM}

In the following, we use tildes in order to clearly distinguish covariant quantities. The ADM notations \cite{ADM, Bardeen-1980} of the metric and fluid quantities are presented in Eqs.\ (\ref{ADM-metric-def}) and (\ref{ADM-fluid-def}). The extrinsic curvature is defined as
\bea
   K_{ij} \equiv {1 \over 2 N} \left( N_{i:j}
       + N_{j:i} - h_{ij,0} \right), \quad
       K \equiv h^{ij} K_{ij}, \quad
       \overline{K}_{ij} \equiv K_{ij}
       - {1\over 3} h_{ij} K,
   \label{extrinsic-curvature-def}
\eea
where the indices of $K_{ij}$ are raised and lowered by $h_{ij}$ and its inverse, and $K \equiv K^i_i$. A colon indicates a covariant derivative associated with $h_{ij}$ where the connection is
\bea
   \Gamma^{(h)i}_{\;\;\;\;\;jk} \equiv {1 \over 2}
       h^{i\ell} \left( h_{\ell j,k}
       + h_{\ell k,j} - h_{jk,\ell} \right), \quad
       \Gamma^{(h)k}_{\;\;\;\;\;ik}
       = {1\over 2} h^{k \ell} h_{k \ell,i}
       = {\sqrt{h}_{,i} \over \sqrt{h} }, \quad
       h \equiv {\rm det} (h_{ij}).
   \label{connection-h}
\eea
Thus
\bea
   K = {1\over N} \left( N^i_{\;\; :i}
       - {1\over 2} h^{ij} h_{ij,0} \right)
       = {1\over N} \bigg( N^i_{\;\; :i} - {\sqrt{h}_{,0}
       \over \sqrt{h} } \bigg).
   \label{extrinsic-curvature-trace}
\eea
We have
\bea
   & & h_{ij,0} = - 2 N K_{ij} + N_{i:j}
       + N_{j:i}, \quad
       {\sqrt{h}_{,0} \over \sqrt{h} } = - N K + N^i_{\;\; :i}, \quad
       h^{ij}_{\;\;\;,0} = - h^{ik} h^{j\ell}
       h_{k\ell,0} = 2 N K^{ij} - N^{i:j} - N^{j:i},
   \nonumber \\
   & & \Gamma^{(h)k}_{\;\;\;\;\;ij,0}
       = \left( - 2 N K^k_{(i}
       + N^k_{\;\; :(i}
       + N_{(i}^{\;\; :k} \right)_{:j)}
       + \left( N K_{ij}
       - N_{(i:j)} \right)^{:k}, \quad
       \Gamma^{(h)j}_{\;\;\;\;\;ij,0}
       = \left( - N K
       + N^j_{\;\; :j} \right)_{:i}.
\eea
The intrinsic curvature $R^{(h)i}_{\;\;\;\;\;\;\;jk\ell}$ is the Riemann curvature tensor associated with $h_{ij}$ as the metric tensor
\bea
   & & R^{(h)i}_{\;\;\;\;\;\;\;jk\ell}
       \equiv \Gamma^{(h)i}_{\;\;\;\;\;j\ell,k}
       - \Gamma^{(h)i}_{\;\;\;\;\;jk,\ell}
       + \Gamma^{(h)m}_{\;\;\;\;\;j\ell}
       \Gamma^{(h)i}_{\;\;\;\;\;km}
       - \Gamma^{(h)m}_{\;\;\;\;\;jk}
       \Gamma^{(h)i}_{\;\;\;\;\;\ell m},
   \nonumber \\
   & & R^{(h)}_{ij} \equiv
       R^{(h)k}_{\;\;\;\;\;\;\;ikj}, \quad
       R^{(h)} \equiv h^{ij} R^{(h)}_{ij}
       = R^{(h)i}_{\;\;\;\;\; i}, \quad
       \overline R^{(h)}_{ij}
       \equiv R^{(h)}_{ij} - {1 \over 3} h_{ij} R^{(h)}.
   \label{curvature-h}
\eea

We have antisymmetric tensor
\bea
   & & \widetilde \eta^{0ijk}
       = {1 \over \sqrt{- \widetilde g}}
       \epsilon^{0ijk}
       = {1 \over N \sqrt{h}} \epsilon^{0ijk}
       \equiv {1 \over N} \overline \eta^{ijk}, \quad
       \widetilde \eta_{0ijk}
       = - \sqrt{ - \widetilde g} \epsilon_{0ijk}
       = - N \sqrt{h} \epsilon_{0ijk}
       = - N \overline \eta_{ijk};
   \nonumber \\
   & &
       \overline \eta_{ijk} \equiv \widetilde \eta_{ijkd} \widetilde n^d
       = - {1 \over N} \widetilde \eta_{0ijk}, \quad
       \overline \eta^{ijk} \equiv \widetilde \eta^{ijkd} \widetilde n_d
       = N \widetilde \eta^{0ijk},
   \label{eta-ADM}
\eea
where indices of $\overline \eta_{ijk}$ are raised and lowered by $h_{ij}$ and its inverse. We have
\bea
   \overline \eta^{ijk} \overline \eta_{\ell mn}
       = 3 ! \delta^i_{[\ell} \delta^j_m \delta^k_{n]}
       = \left|
       \begin{array}{lll}
       \delta^i_\ell & \delta^i_m & \delta^i_n \\
       \delta^j_\ell & \delta^j_m & \delta^j_n \\
       \delta^k_\ell & \delta^k_m & \delta^k_n
       \end{array}
       \right|, \quad
       \overline \eta^{ijm} \overline \eta_{k\ell m}
       = 2 ! \delta^i_{[k} \delta^j_{\ell]}
       = \left|
       \begin{array}{lll}
       \delta^i_k & \delta^i_\ell \\
       \delta^j_k & \delta^j_\ell
       \end{array}
       \right|
       = \left( \delta^i_k \delta^j_\ell
       - \delta^i_\ell \delta^j_k \right), \quad
       \overline \eta^{ik\ell} \overline \eta_{jk\ell}
       = 2 \delta^i_j.
\eea

The ADM equations can be derived in a conventional way \cite{ADM, Wilson-Mathews-2003, Baumgarte-Shapiro-2010}. One other way is by using the covariant equations based on the normal-frame $\widetilde n_a$. Here we derive the set directly from Einstein's equation by brute force. We will also present the conformal tensors and the quasi-Maxwellian equations in the ADM notation.

The connection gives
\bea
   & & \widetilde \Gamma^0_{00}
       = {1\over N} \left( N_{,0} + N_{,i} N^i
       - K_{ij} N^i N^j \right), \quad
       \widetilde \Gamma^0_{0i}
       = {1\over N} \left( N_{,i}
       - K_{ij} N^j \right), \quad
       \widetilde \Gamma^0_{ij}
       = - {1\over N} K_{ij},
   \nonumber \\
   & & \widetilde \Gamma^i_{00}
       = {1\over N} N^i \left( - N_{,0}
       - N_{,j} N^j + K_{jk} N^j N^k \right)
       + N N^{:i} + N^i_{\;\; ,0} - 2 N K^{ij} N_j
       + N^{i:j} N_j,
   \nonumber \\
   & & \widetilde \Gamma^i_{0j}
       = - {1\over N} N_{,j} N^i
       - N K^i_j + N^i_{\;\; :j}
       + {1\over N} N^i N^k K_{jk}, \quad
       \widetilde \Gamma^i_{jk}
       = \Gamma^{(h)i}_{\;\;\;\;jk}
       + {1\over N} N^i K_{jk},
   \label{connection-ADM}
\eea
thus,
\bea
   \widetilde \Gamma^c_{c0} = {1\over N} N_{,0}
       - N K + N^i_{\;\; :i}
       = {1\over N} N_{,0} + {\sqrt{h}_{,0} \over \sqrt{h} }, \quad
       \widetilde \Gamma^c_{ci} = \Gamma^{(h)k}_{\;\;\;\;\; ki}
       + {1\over N} N_{,i}.
\eea
Riemann curvature tensor gives
\bea
   & & \widetilde R^0_{\;\; 00i}
       =
       - {1\over N} K_{ij,0} N^j
       - {1\over N} N_{,i:j} N^j
       + {1\over N} K_{jk:i} N^j N^k
       + {1\over N} K_{jk} N^k_{\;\; :i} N^j
       + {1\over N} K_{ik} N^{k}_{\;\; :j} N^j
       - K_{ij} K^{jk} N_k,
   \nonumber \\
   & & \widetilde R^0_{\;\; 0ij}
       =
       {1\over N} \left( K^k_{i:j} - K^k_{j:i}
       \right) N_k,
   \nonumber \\
   & & \widetilde R^0_{\;\; i0j}
       =
       - {1\over N} K_{ij,0}
       - {1\over N} N_{,i:j}
       - K^k_i K_{jk}
       + {1\over N} K^k_{i:j} N_k
       + {1\over N} K_{ik} N^k_{\;\; :j}
       + {1\over N} K_{jk} N^k_{\;\; :i}, \quad
       \widetilde R^0_{\;\; ijk}
       =
       {1\over N} \left( K_{ij:k} - K_{ik:j}
       \right),
   \nonumber \\
   & & \widetilde R^i_{\;\; 00j}
       =
       R^{(h)i}_{\;\;\;\;\;\;\; k\ell j} N^k N^\ell
       - N \left( K^i_{j,0}
       + K^i_{j:k} N^k \right)
       - N N^{:i}_{\;\;\; j}
       - K^i_j K_{k\ell} N^k N^\ell
       + N K^k_j \left( N K^i_k
       - N^i_{\;\; :k} \right)
   \nonumber \\
   & & \qquad
       + {1 \over N} N^i \left( N^k K_{jk,0}
       + N_{,k:j} N^k
       - K_{k\ell:j} N^k N^\ell
       - K_{k\ell} N^k N^\ell_{\;\; :j}
       + N K_{jk} K^{k\ell} N_\ell
       - K_{jk} N^{k:\ell} N_\ell \right)
   \nonumber \\
   & & \qquad
       + N \left( K^{ik}_{\;\;\;\; :j}
       + K_{jk}^{\;\;\;\; :i} \right) N_k
       + N K^{ik} N_{k:j}
       + K_{jk} K^i_\ell N^k N^\ell,
   \nonumber \\
   & & \widetilde R^i_{\;\; 0jk}
       = 2 N K^i_{[j:k]}
       - 2 N^i_{\;\; :[jk]}
       - 2 {1\over N} N^i N^\ell K_{\ell[j:k]}
       + 2 N^\ell K^i_{[j} K_{k]\ell},
   \nonumber \\
   & & \widetilde R^i_{\;\; j0k}
       = R^{(h)i}_{\;\;\;\;\;\; j \ell k} N^\ell
       + {1 \over N} N^i K_{jk,0}
       + K^{i\ell} N_\ell K_{jk}
       + {1 \over N} N_{,j:k} N^i
       - {1 \over N} N^i N^\ell K_{j\ell:k}
       - {1 \over N} N^i \left(
       N^\ell_{\;\; :k} K_{j\ell}
       + N^\ell_{\;\; :j} K_{k\ell} \right)
   \nonumber \\
   & & \qquad
       - K^i_k K_{j\ell} N^\ell
       + N^i K^\ell_j K_{k\ell}
       + N \left( K_{kj}^{\;\;\; :i}
       - K^i_{k:j} \right), \quad
       \widetilde R^i_{\;\; jk\ell}
       = R^{(h)i}_{\;\;\;\;\;\; jk\ell}
       + 2 K^i_{[k} K_{\ell]j}
       - 2 {1\over N} N^i K_{j[k:\ell]},
   \\
   & & \widetilde R_{0i0j}
       = R^{(h)}_{ikj\ell} N^k N^\ell
       + N \left[ K_{ij,0}
       + K_{ij:k} N^k
       + N_{,i:j}
       - \left( K_{ki:j} + K_{kj:i} \right) N^k
       - \left( K_{ki} N^k_{\;\; :j}
       + K_{kj} N^k_{\;\; :i} \right) \right]
   \nonumber \\
   & & \qquad
       + N^2 K^k_i K_{jk}
       + \left( K_{ij} K_{k\ell}
       - K_{ik} K_{j\ell} \right) N^k N^\ell,
   \nonumber \\
   & & \widetilde R_{0ijk}
       = N^\ell \left( R^{(h)}_{\ell ijk}
       - 2 K_{i[j} K_{k]\ell} \right)
       - 2 N K_{i[j:k]}, \quad
       \widetilde R_{ijk\ell}
       = R^{(h)}_{ijk\ell}
       + 2 K_{i[k} K_{\ell]j}.
\eea
Ricci tensor gives
\bea
   & & \widetilde R_{00}
       = N \left( K_{,0}
       + K_{,i} N^i
       + N^{:i}_{\;\;\; i}
       - N K^{ij} K_{ij}
       - 2 N_j K^{ij}_{\;\;\; :i} \right)
   \nonumber \\
   & & \qquad
       + {1 \over N} N^i N^j \left(
       - K_{ij,0}
       - N_{,i|j}
       + K_{jk|i} N^k
       + 2 K_{jk} N^k_{\;\;|i}
       + N K K_{ij}
       - 2 N K_i^k K_{jk}
       + N R^{(h)}_{ij} \right),
   \nonumber \\
   & & \widetilde R_{0i}
       =
       N K_{,i} - N K^j_{i:j}
       + R^{(h)}_{ij} N^j
       - {1\over N} K_{ij,0} N^j
       + {1\over N} K_{ij:k} N^j N^k
       + K_{ij} \bigg( {1\over N} N^j_{\;\;|k} N^k
       + K N^j \bigg)
   \nonumber \\
   & & \qquad
       - {1\over N} N_{,i:j} N^j
       + {1\over N} K_{jk} N^j N^k_{\;\; :i}
       - 2 K_{ij} K^j_k N^k,
   \nonumber \\
   & & \widetilde R_{ij}
       = R^{(h)}_{ij}
       - {1\over N} K_{ij,0}
       - {1\over N} N_{,i:j}
       + K K_{ij}
       - 2 K_{ik} K^k_j
       + {1\over N} \left( K_{ik} N^k_{\;\; :j}
       + K_{jk} N^k_{\;\; :i} \right)
       + {1\over N} K_{ij:k} N^k,
   \\
   & & \widetilde R^0_0
       = - {1 \over N} \left(
       K_{,0} + N^{:i}_{\;\;\; i}
       - K^j_{i:j} N^i
       - N K^{ij} K_{ij} \right), \quad
       \widetilde R^0_i
       = {1 \over N} \left( K^j_{i:j} - K_{,i} \right),
   \nonumber \\
   & & \widetilde R^i_j
       = R^{(h)i}_{\;\;\;\;\; j}
       - {1 \over N} \left( K^i_{j,0}
       - K^i_{j:k} N^k \right)
       - {1 \over N} N^{:i}_{\;\;\; j}
       + K K^i_j
       + {1 \over N} \left(
       K^i_k N^k_{\;\; :j}
       - K_j^k N^i_{\;\; :k} \right)
       + {N^i \over N} \left( K_{,j}
       - K^k_{j:k} \right).
\eea
Scalar curvature gives
\bea
   \widetilde R
       = R^{(h)} + K_{ij} K^{ij} + K^2
       - {2 \over N} \left( K_{,0} - K_{,i} N^i
       + N^{:i}_{\;\;\;i} \right).
\eea
The electric and magnetic parts of conformal tensors will be presented later in Eqs.\ (\ref{Eij-1}) and (\ref{Hij-1}).

%%%%%%%%%%%%%%%%%%%%%%%%%%%%%%%%%%%%%%%%%%%%%%%%%%%%%%%%%%%%%%%%%
\subsection{ADM equations}

The ADM equations can be derived directly from Einstein equations. The $\widetilde G^0_i$ component of Einstein's equation gives
\bea
   \overline K^j_{i:j}
       - {2 \over 3} K_{,i} = {8 \pi G \over c^4} J_i.
   \label{Mom-constraint-ADM}
\eea
The $\widetilde G^0_0$ component of Einstein's equation using Eq.\ (\ref{Mom-constraint-ADM}) gives
\bea
   R^{(h)}
       = \overline K_{ij} \overline K^{ij} - {2 \over 3} K^2
       + {16 \pi G \over c^4} E + 2 \Lambda.
   \label{E-constraint-ADM}
\eea
The trace of Einstein's equation using Eqs.\ (\ref{Mom-constraint-ADM}) and (\ref{E-constraint-ADM}) gives
\bea
   {1 \over N} \left( K_{,0} - K_{,i} N^i \right)
       + {1 \over N} N^{:i}_{\;\;\; i}
       - \overline K_{ij} \overline K^{ij}
       - {1 \over 3} K^2
       - {4 \pi G \over c^4} \left( E + S \right) + \Lambda = 0.
   \label{Raychaudhuri-eq-ADM}
\eea
A tracefree combination of Einstein'sequation  $\widetilde R^i_j - {1 \over 3} \delta^i_j \widetilde R^k_k$, using Eq.\ (\ref{Mom-constraint-ADM}) gives
\bea
   {1 \over N} \left( \overline K^i_{j,0}
       - \overline K^i_{j:k} N^k
       + \overline K_{jk} N^{i:k}
       - \overline K^i_k N^k_{\;\; :j} \right)
       = K \overline K^i_j
       - {1 \over N} \bigg( N^{:i}_{\;\;\; j}
       - {1 \over 3} \delta^i_j N^{:k}_{\;\;\; k} \bigg)
       + \overline R^{(h)i}_{\;\;\;\;\; j}
       - {8 \pi G \over c^4} \overline S^i_j.
   \label{ADM-prop-tracefree}
\eea
From $\widetilde n_a \widetilde T^{ab}_{\;\;\;\;;b} = 0$ and $\widetilde T^b_{i;b} = 0$, %using
%\bea
%   & & T^{00} = {1 \over N^2} E, \quad
%       T^{ij} = {1 \over N^2} N^i N^j E
%       - {1 \over N} (N^i J^j + N^j J^i ) + S^{ij},
%   \nonumber \\
%   & &
%       T^0_0 = - E + {1 \over N} N^i J_i, \quad
%       T^0_i = {1 \over N} J_i, \quad
%       T^i_0 = E N^i - N \left( J^i
%       + {1 \over N^2} N^i N_j J^j \right)
%       + N^j S^i_j, \quad
%       T^i_j = - {1 \over N} N^i J_j + S^i_j,
%   \nonumber \\
%   & &
%       T_{00} = N^2 E - 2 N N^i J_i + N^i N^j S_{ij}, \quad
%       T_{0i} = - N J_i + N^j S_{ij}, \quad
%       T_{ij} = S_{ij},
%   \nonumber
%\eea
%and Eq.\ (\ref{connection-ADM}),
we have the energy and momentum conservation equations
\bea
   & & {1 \over N} \left( E_{,0} - E_{,i} N^i \right)
       - K \left( E + {1\over 3} S \right)
       - \bar S^{ij} \bar K_{ij}
       + {1 \over N^2} \left( N^2 J^i \right)_{:i} = 0.
   \label{ADM-E-conserv-App} \\
   & & {1 \over N} \left( J_{i,0}
       - J_{i:j} N^j - J_j N^j_{\;\; :i} \right)
       - K J_i + {1 \over N} E N_{,i} + S^j_{i:j}
       + {1 \over N} N_{,j} S^j_i = 0.
   \label{ADM-Mom-conserv-App}
\eea
Equations (\ref{Mom-constraint-ADM})-(\ref{ADM-Mom-conserv-App}) are a complete set of ADM equations. The continuity equation is based on the fluid four vector; $(\overline \varrho \widetilde u^c)_{;c} \equiv 0$ gives
\bea
   \left( \sqrt{h} D \right)_{,0}
       + \left[ \sqrt{h} D
       \left( N V^i - N^i \right) \right]_{,i} = 0.
   \label{ADM-Mass-conserv-App}
\eea

%The ADM equations can also be derived from the covariant equations expressed in the normal-frame in Sec.\ \ref{Appendix-cov}; in this way we only need the connection in Eq.\ (\ref{connection-ADM}). The energy and momentum conservation equations in Eqs.\ (\ref{cov-E-conserv-App}) and (\ref{cov-Mom-conserv-App}) give Eqs.\ (\ref{ADM-E-conserv-App}) and (\ref{ADM-Mom-conserv-App}), respectively. Raychaudhuri equation in Eq.\ (\ref{Raychaudhury-eq-cov}) gives Eq.\ (\ref{Raychaudhuri-eq-ADM}). Vorticity propagation is identically satisfied. Shear propagation equation in Eq.\ (\ref{shear-propagation-cov}) using $E_{ij}$ in Eq.\ (\ref{Eij}) gives Eq.\ (\ref{ADM-prop-tracefree}). Equation (\ref{shear-constraint-cov}) gives Eq.\ (\ref{Mom-constraint-ADM}).

Equations (\ref{ADM-E-conserv-App})-(\ref{ADM-Mass-conserv-App}) can be written in conservative forms as
\bea
   & & \left( \sqrt{h} E \right)_{,0}
       + \left[ \sqrt{h} \left( N J^i - N^i E \right) \right]_{,i}
       = \sqrt{h} \left( N S^{ij} K_{ij}
       - N_{,i} J^i \right),
   \label{ADM-E-conserv-App-2} \\
   & & \left( \sqrt{h} J_i \right)_{,0}
       + \left[ \sqrt{h} \left( N S^j_i
       - N^j J_i \right) \right]_{,j}
       = \sqrt{h} \bigg( {1 \over 2} N S^{jk} h_{jk,i}
       + J_j N^j_{\;\;,i}
       - E N_{,i} \bigg).
   \label{ADM-Mom-conserv-App-2}
\eea
Using
\bea
   {\cal E} \equiv E - D c^2,
\eea
the energy conservation equation can be written as
\bea
   \left( \sqrt{h} {\cal E} \right)_{,0}
       + \left[ \sqrt{h} \left( N J^i
       - D c^2 N V^i
       - N^i {\cal E} \right) \right]_{,i}
       = \sqrt{h} \left( N S^{ij} K_{ij}
       - N_{,i} J^i \right).
   \label{ADM-E-conserv-App-3}
\eea

In the presence of multiple component of fluids and fields, the fluid quantities in the ADM equations are collective ones with $E = E^{\rm FL} + E^{\rm SF} + E^{\rm EM}$, etc. Instead of the conservation equations for collective component in Eqs.\ (\ref{ADM-E-conserv-App}), (\ref{ADM-Mom-conserv-App}), (\ref{ADM-E-conserv-App-2})-(\ref{ADM-E-conserv-App-3}) we can also use the ones for individual component. For the fluid, we have Eqs.\ (\ref{ADM-E-conserv-FL})-(\ref{ADM-Mom-conserv-FL-2}) and (\ref{ADM-E-conserv-FL-3}). For the EM field, we have the Maxwell equations in (\ref{Maxwell-ADM-1})-(\ref{Maxwell-ADM-4}). For the SF we have the equation of motion in Eq.\ (\ref{EOM-ADM}).

%%%%%%%%%%%%%%%%%%%%%%%%%%%%%%%%%%%%%%%%%%%%%%%%%%%%%%%%%%%%%%%%%
\subsection{Weyl equations}

The electric and magnetic parts of conformal tensors based on the normal-frame are
\bea
   & & \widetilde E^{(n)i}_{\;\;\;\;\;j}
       \equiv \overline E^i_j
       = {1 \over 2 N} \bigg( \overline K^i_{j,0}
       - \overline K^i_{j:k} N^k
       + \overline K_j^k N^i_{\;\; :k}
       - \overline K^i_k N^k_{\;\; :j}
       + N^{:i}_{\;\;\; j}
       - {1 \over 3} \delta^i_j N^{:k}_{\;\;\; k}
       \bigg)
   \nonumber \\
   & & \qquad
       - {1 \over 6} K \overline K^i_j
       - \overline K^i_k \overline K^k_j
       + {1 \over 3} \delta^i_j
       \overline K^k_\ell \overline K_k^\ell
       + {1 \over 2} \overline R^{(h)i}_{\;\;\;\;\; j},
   \label{Eij-1} \\
   & & \widetilde H^{(n)i}_{\;\;\;\;\;j}
       \equiv \overline H^i_j
       = \overline \eta^{ik\ell}
       \bigg[ K_{jk:\ell}
       + {1 \over 2} h_{j\ell}
       \left( K_{,k} - K^m_{k:m} \right) \bigg].
   \label{Hij-1}
\eea
Using Eqs.\ (\ref{E-constraint-ADM}), (\ref{ADM-prop-tracefree}) and (\ref{Mom-constraint-ADM}), these become
\bea
   & & \overline E^i_j
       = \overline R^{(h)i}_{\;\;\;\;\; j}
       + {1 \over 3} K \overline K^i_j
       - \bigg( \overline K^i_k \overline K^k_j
       - {1 \over 3} \delta^i_j
       \overline K^k_\ell \overline K^\ell_k \bigg)
       - {4 \pi G \over c^4} \overline S^i_j
   \nonumber \\
   & & \qquad
       = R^{(h)i}_{\;\;\;\;\; j}
       + K K^i_j - K^i_k K^k_j
       - \bigg( {16 \pi G \over 3 c^4} E
       + {2 \over 3} \Lambda \bigg) \delta^i_j
       - {4 \pi G \over c^4} \overline S^i_j,
   \label{Eij-2} \\
   & & \overline H^i_j
       = \overline \eta^{ik\ell}
       \bigg( K_{jk:\ell}
       - {4 \pi G \over c^4} h_{j\ell} J_k \bigg)
       = {1 \over 2} \left( \overline \eta^{ik\ell} K_{jk:\ell}
       + \overline \eta_j^{\;\;k\ell} K^i_{k:\ell} \right).
   \label{Hij-2}
\eea

Equations (\ref{cov-Maxwell-1})-(\ref{cov-Maxwell-4}) give the quasi-Maxwellian equations expressed in terms of the electric and magnetic parts of Weyl tensor in the normal-frame
\bea
   & & \overline E^j_{i:j}
       + \overline \eta_i^{\;\;jk} \overline K_j^\ell \overline H_{k\ell}
       = {8 \pi G \over c^4} \bigg[ {1 \over 3} E_{,i}
       - {1 \over 2} \left( K_{ij} - K h_{ij} \right) J^j
       - {1 \over 2} \overline S^j_{i:j} \bigg],
   \label{Eij-ADM} \\
   & & \overline H^j_{i:j}
       = \overline \eta_i^{\;\; jk}
       \bigg( K^\ell_j R^{(h)}_{k\ell}
       + {4 \pi G \over c^4} J_{j:k} \bigg),
   \label{Hij-ADM} \\
   & & {1 \over N} \left( \overline E_{ij,0}
       - \overline E_{ij:k} N^k \right)
       - 2 K \overline E_{ij}
       + \bigg( 5 K^k_{(i}
       - {2 \over N} N^k_{\;\; :(i} \bigg) \overline E_{j)k}
       - h_{ij} K^k_\ell \overline E^\ell_k
       + \overline \eta_{(i}^{\;\;\;k\ell}
       \bigg( \overline H_{j)k:\ell}
       - {2 \over N} N_{,k} \overline H_{j)\ell} \bigg)
   \nonumber \\
   & & \qquad
       = {4 \pi G \over c^4} \bigg\{
       \bigg( E + {1 \over 3} S \bigg) \overline K_{ij}
       - J_{(i:j)}
       - {2 \over N} N_{,(i} J_{j)}
       - {1 \over N} \left(
       \overline S_{ij,0}
       - \overline S_{ij:k} N^k \right)
       - \bigg( K^k_{(i}
       - {2 \over N} N^k_{\;\; :(i} \bigg)
       \overline S_{j)k}
   \nonumber \\
   & & \qquad
       + {1 \over 3} h_{ij} \bigg(
       J^k_{\;\; :k}
       + {2 \over N} N_{,k} J^k
       - K^k_\ell \overline S^\ell_k \bigg)
       \bigg\},
   \label{dot-Eij-ADM} \\
   & &
       {1 \over N} \left( \overline H_{ij,0}
       - \overline H_{ij:k} N^k \right)
       - 2 K \overline H_{ij}
       + \bigg( 5 K^k_{(i}
       - {2 \over N} N^k_{\;\;:(i} \bigg) \overline H_{j)k}
       - h_{ij} K^k_\ell \overline H^\ell_k
       - \overline \eta_{(i}^{\;\;\;k\ell}
       \bigg( \overline E_{j)k:\ell}
       - {2 \over N} N_{,k} \overline E_{j)\ell} \bigg)
   \nonumber \\
   & & \qquad
       = - {4 \pi G \over c^4} \overline \eta_{(i}^{\;\;\;k \ell}
       \left( K_{j)k} J_\ell
       + \overline S_{j)k:\ell} \right).
   \label{dot-Hij-ADM}
\eea
Using Eq.\ (\ref{Hij-2}), Eq.\ (\ref{Eij-ADM}) can be written as
\bea
   \overline E^j_{i:j}
       =
       K^{jk} \left( K_{jk:i} - K_{ij:k} \right)
       + {8 \pi G \over c^4} \left[ {1 \over 3} E_{,i}
       - \left( K_{ij} - {1 \over 3} K h_{ij} \right) J^j
       - {1 \over 2} \overline S^j_{i:j} \right].
   \label{Eij-ADM-2}
\eea
Equations (\ref{Eij-ADM}) and (\ref{Eij-ADM-2}) give the Bianchi identity $R^{(h)j}_{\;\;\;\;\; i:j} = {1 \over 2} R^{(h)}_{,i}$, and Eq.\ (\ref{Hij-ADM}) follows from Eq.\ (\ref{Hij-2}).

%%%%%%%%%%%%%%%%%%%%%%%%%%%%%%%%%%%%%%%%%%%%%%%%%%%%%%%%%%%%%%%
%
%
%
%%%%%%%%%%%%%%%%%%%%%%%%%%%%%%%%%%%%%%%%%%%%%%%%%%%%%%%%%%%%%%%
\section{FNLE formulation}
                                    \label{Appendix-FNLE}

The metric is in Eq.\ (\ref{metric-FNLE}). The FNLE fluid quantities are defined in terms of the ADM fluid quantities as
\bea
   E, \quad
       S, \quad
       J_i \equiv a c m_i, \quad
       S_{ij} \equiv a^2 (1 + 2 \varphi) m_{ij}, \quad
       \overline S_{ij} \equiv a^2 (1 + 2 \varphi)
       \overline m_{ij},
   \label{ADM-fluid-FNLE}
\eea
where indices of $m_i$ and $m_{ij}$ are raised and lowered using $\delta_{ij}$ as the metric and its inverse; we introduce $\overline m_{ij} \equiv m_{ij} - {1 \over 3} \delta_{ij} m^k_k$ with $m^k_k = S$.

We use the ADM formulation. The ADM metric quantities can be expressed in terms of the FNLE notation as
\bea
   & & N \equiv - \left( \widetilde g^{00} \right)^{-1/2}
       = a \sqrt{ 1 + 2 \alpha
       + {\chi^k \chi_k \over a^2 (1 + 2 \varphi)}}
       \equiv a {\cal N}, \quad
       N_i \equiv \widetilde g_{0i} = - a \chi_i, \quad
       N^i = - {\chi^i \over a(1 + 2 \varphi)},
   \nonumber \\
   & &
       h_{ij} \equiv \widetilde g_{ij}
       = a^2 \left( 1 + 2 \varphi \right) \delta_{ij}, \quad
       h^{ij} = {1 \over a^2 (1 + 2 \varphi)} \delta^{ij}.
   \label{N-definition}
\eea
The inverse metric, three-space connection and curvatures are given by
\bea
   & & \widetilde g^{00} = - {1 \over a^2 {\cal N}^2}, \quad
       \widetilde g^{0i}
       = - {\chi^i \over a^3 {\cal N}^2 (1 + 2 \varphi)}, \quad
       \widetilde g^{ij} = {1 \over a^2 (1 + 2 \varphi)}
       \left( \delta^{ij} - {\chi^i \chi^j \over a^2 {\cal N}^2 (1 + 2 \varphi) } \right),
   \label{metric-inverse-FNLE} \\
   & & \Gamma^{(h)i}_{\;\;\;\;\;jk}
       = {1 \over 1 + 2 \varphi} \left(
       \varphi_{,j} \delta^i_k
       + \varphi_{,k} \delta^i_j
       - \varphi^{,i} \delta_{jk} \right), \quad  \Gamma^{(h)k}_{\;\;\;\;\;ik}
       = {3 \varphi_{,i} \over 1 + 2 \varphi},
   \\
   & & R^{(h)i}_{\;\;\;\;\;\; jk\ell} = {1 \over 1 + 2 \varphi}
       \left(
       \varphi_{,jk} \delta^i_\ell -\varphi^{,i}_{\;\;k} \delta_{j\ell}
       - \varphi_{,j\ell}\delta^i_k +\varphi^{,i}_{\;\; \ell}
       \delta_{jk} \right)
   \nonumber \\
   & & \qquad
       + {1 \over (1 + 2 \varphi)^2}
       \left(
       - 3 \varphi_{,k} \varphi_{,j} \delta^i_\ell
       + 3 \varphi^{,i} \varphi_{,k} \delta_{j\ell}
       + 3 \varphi_{,j} \varphi_{,\ell} \delta^i_k
       - 3 \varphi^{,i} \varphi_{,l} \delta_{jk}
       - \varphi^{,m} \varphi_{,m} \delta^i_k \delta_{j\ell}
       + \varphi^{,m} \varphi_{,m} \delta^i_\ell \delta_{jk} \right),
   \nonumber \\
   & & R^{(h)}_{ij}
       = - {\varphi_{,ij} \over 1 + 2 \varphi}
       + 3 {\varphi_{,i} \varphi_{,j} \over (1 + 2 \varphi)^2}
       - \left( {\varphi^{,k}_{\;\;\; k} \over 1 + 2 \varphi}
       - {\varphi^{,k} \varphi_{,k} \over (1 + 2 \varphi)^2} \right) \delta_{ij},
    \nonumber \\
    & &  R^{(h)} = {2 \over a^2 (1 + 2 \varphi)^2}
       \bigg( - 2  \varphi^{,k}_{\; \; k}
       + 3 {\varphi^{,k} \varphi_{,k} \over 1 + 2 \varphi} \bigg),
   \nonumber \\
   & & \overline{R}^{(h)i}_{\;\;\;\;\;j}
       \equiv R^{(h)i}_{\;\;\;\;\;j}
       - {1 \over 3} \delta^i_j R^{(h)}
       = {1 \over a^2 (1 + 2 \varphi)^2} \left[
       - \varphi^{,i}_{\;\;j}
       + 3 {\varphi^{,i} \varphi_{,j} \over 1 + 2 \varphi}
       - {1 \over 3} \delta^i_j \left( -  \varphi^{,k}_{\;\; k}
       + 3 {\varphi^{,k} \varphi_{,k} \over 1 + 2 \varphi} \right) \right].
   \label{intrinsic-curvature}
\eea

We have the extrinsic curvature as
\bea
   & & K_{ij}
       = - {a^2 \over {\cal N}} \bigg[
       {1 \over c} \left( H + \dot \varphi + 2 H \varphi \right) \delta_{ij}
       + {1 \over 2 a^2} \left( \chi_{i,j} + \chi_{j,i} \right)
       - {1 \over a^2 (1 + 2 \varphi)} \left(
       \chi_{i} \varphi_{,j}
       + \chi_{j} \varphi_{,i}
       - \chi^{k} \varphi_{,k} \delta_{ij} \right) \bigg],
   \nonumber \\
   & & K
       = - {1 \over {\cal N} (1 + 2 \varphi)}
       \left[ {3 \over c} \left( H + \dot \varphi + 2 H \varphi \right)
       + {1 \over a^2} \chi^k_{\;\;,k}
       + {\chi^{k} \varphi_{,k} \over a^2 (1 + 2 \varphi)}
       \right]
       \equiv {1 \over c} \left( - 3 H + \kappa \right),
   \nonumber \\
   & & \overline{K}^i_j
       \equiv K^i_j - {1 \over 3} \delta^i_j K
       = - {1 \over a^2 {\cal N} (1 + 2 \varphi)} \bigg[
       {1 \over 2} \left( \chi^{i}_{\;\;,j} + \chi_j^{\;\;,i} \right)
       - {1 \over 3} \delta^i_j \chi^k_{\;\;,k}
       - {1 \over 1 + 2 \varphi} \left(
       \chi^{i} \varphi_{,j}
       + \chi_{j} \varphi^{,i}
       - {2 \over 3} \delta^i_j \chi^{k} \varphi_{,k} \right)
       \bigg],
   \nonumber \\
   & & \overline{K}^i_j \overline{K}^j_i
       = {1 \over a^4 {\cal N}^2 (1 + 2 \varphi)^2}
       \bigg\{
       {1 \over 2} \chi^{i,j} \left( \chi_{i,j} + \chi_{j,i} \right)
       - {1 \over 3} \chi^i_{\;\;,i} \chi^j_{\;\;,j}
       \nonumber \\
   & & \qquad
       - {4 \over 1 + 2 \varphi} \bigg[
       {1 \over 2} \chi^i \varphi^{,j} \left(
       \chi_{i,j} + \chi_{j,i} \right)
       - {1 \over 3} \chi^i_{\;\;,i} \chi^j \varphi_{,j} \bigg]
       + {2 \over (1 + 2 \varphi)^2} \left(
       \chi^{i} \chi_{i} \varphi^{,j} \varphi_{,j}
       + {1 \over 3} \chi^i \chi^j \varphi_{,i} \varphi_{,j} \right) \bigg\}.
   \label{extrinsic-curvature}
\eea

We have
\bea
   & &
       \widetilde \eta^{0ijk}
       =  {1 \over \sqrt{-g}} \eta^{ijk}
       = {1 \over a^4 {\cal N} (1 + 2 \varphi)^{3/2}} \eta^{ijk}, \quad
       \widetilde \eta_{0ijk}
       = - \sqrt{-g} \eta_{ijk}
       = - a^4 {\cal N} (1 + 2 \varphi)^{3/2} \eta_{ijk};
   \nonumber \\
   & &
       \overline \eta^{ijk}
       = {1 \over \sqrt{h}} \eta^{ijk}
       = {1 \over a^3 ( 1 + 2 \varphi )^{3/2}} \eta^{ijk}, \quad
       \overline \eta_{ijk}
       = \sqrt{h} \eta_{ijk}
       = a^3 ( 1 + 2 \varphi )^{3/2} \eta_{ijk};
   \nonumber \\
   & &
       \eta^{ijk} \eta_{\ell mn}
       = 6 \delta^i_{[\ell} \delta^j_{m} \delta^k_{n]}, \quad
       \eta^{ijk} \eta_{i \ell m}
       = 2 \delta^j_{[\ell} \delta^k_{m]}, \quad
       \eta^{ijk} \eta_{ij\ell} = 2 \delta^k_\ell,
   \label{eta-FNLE}
\eea
where indices of $\eta_{ijk}$ are raised and lowered by the metric $\delta_{ij}$ and its inverse.

%%%%%%%%%%%%%%%%%%%%%%%%%%%%%%%%%%%%%%%%%%%%%%%%%%%%%%%%%%%%%%%
\subsection{FNLE equations}

The ADM equations give the following.
Trace of extrinsic curvature in Eq.\ (\ref{extrinsic-curvature}):
\bea
   \kappa
       \equiv
       3 {\dot a \over a} \left( 1 - {1 \over {\cal N}} \right)
       - {1 \over {\cal N} (1 + 2 \varphi)}
       \left[ 3 \dot \varphi
       + {c \over a^2} \left( \chi^k_{\;\;,k}
       + {\chi^{k} \varphi_{,k} \over 1 + 2 \varphi} \right)
       \right].
   \label{eq1}
\eea
ADM energy constraint in Eq.\ (\ref{E-constraint-ADM}):
\bea
   - {3 \over 2} \left( {\dot a^2 \over a^2}
       - {8 \pi G \over 3 c^2} E
       - {\Lambda c^2 \over 3} \right)
       + {\dot a \over a} \kappa
       + {c^2 \Delta \varphi \over a^2 (1 + 2 \varphi)^2}
       = {1 \over 6} \kappa^2
       + {3 \over 2} {c^2 \varphi^{,i} \varphi_{,i} \over a^2 (1 + 2 \varphi)^3}
       - {c^2 \over 4} \overline{K}^i_j \overline{K}^j_i.
   \label{eq2}
\eea
ADM momentum constraint in Eq.\ (\ref{Mom-constraint-ADM}):
\bea
   & & {2 \over 3} {1 \over c} \kappa_{,i}
       + {1 \over a^2 {\cal N} ( 1 + 2 \varphi )}
       \left( {1 \over 2} \Delta \chi_i
       + {1 \over 6} \chi^k_{\;\;,ki} \right)
       =
       {1 \over a^2 {\cal N} ( 1 + 2 \varphi)}
       \bigg\{
       \left( {{\cal N}_{,j} \over {\cal N}}
       - {\varphi_{,j} \over 1 + 2 \varphi} \right)
       \left[ {1 \over 2} \left( \chi^{j}_{\;\;,i} + \chi_i^{\;,j} \right)
       - {1 \over 3} \delta^j_i \chi^k_{\;\;,k} \right]
   \nonumber \\
   & & \qquad
       - {\varphi^{,j} \over (1 + 2 \varphi)^2}
       \left( \chi_{i} \varphi_{,j}
       + {1 \over 3} \chi_{j} \varphi_{,i} \right)
       + {{\cal N} \over 1 + 2 \varphi} \nabla_j
       \left[ {1 \over {\cal N}} \left(
       \chi^{j} \varphi_{,i}
       + \chi_{i} \varphi^{,j}
       - {2 \over 3} \delta^j_i \chi^{k} \varphi_{,k} \right) \right]
       \bigg\}
       - {8 \pi G \over c^3} a m_i.
   \label{eq3}
\eea
Trace of ADM propagation in Eq.\ (\ref{Raychaudhuri-eq-ADM}):
\bea
   & & - 3 \bigg[ {1 \over {\cal N}}
       \left( {\dot a \over a} \right)^{\displaystyle\cdot}
       + {\dot a^2 \over a^2}
        + {4 \pi G \over 3 c^2} \left( E + S \right)
       - {\Lambda c^2 \over 3} \bigg]
       + {1 \over {\cal N}} \dot \kappa
       + 2 {\dot a \over a} \kappa
       + {c^2 \Delta {\cal N} \over a^2 {\cal N} (1 + 2 \varphi)}
   \nonumber \\
   & & \qquad
       = {1 \over 3} \kappa^2
       - {c \over a^2 {\cal N} (1 + 2 \varphi)} \left(
       \chi^{i} \kappa_{,i}
       + c {\varphi^{,i} {\cal N}_{,i} \over 1 + 2 \varphi} \right)
       + c^2 \overline{K}^i_j \overline{K}^j_i.
   \label{eq4}
\eea
Tracefree ADM propagation in Eq.\ (\ref{ADM-prop-tracefree}):
\bea
   & & \ddot h^i_j + 3 {\dot a \over a} \dot h^i_j
       - c^2 {\Delta \over a^2} h^i_j +
       \left( {1 \over {\cal N}} {\partial \over \partial t}
       + 3 {\dot a \over a}
       - \kappa
       + {c \chi^{k} \over a^2 {\cal N} (1 + 2 \varphi)} \nabla_k \right)
       \bigg\{ {c \over a^2 {\cal N} (1 + 2 \varphi)}
   \nonumber \\
   & & \qquad
       \times
       \left[
       {1 \over 2} \left( \chi^i_{\;\;,j} + \chi_j^{\;\;,i} \right)
       - {1 \over 3} \delta^i_j \chi^k_{\;\;,k}
       - {1 \over 1 + 2 \varphi} \left( \chi^{i} \varphi_{,j}
       + \chi_{j} \varphi^{,i}
       - {2 \over 3} \delta^i_j \chi^{k} \varphi_{,k} \right)
       \right] \bigg\}
   \nonumber \\
   & & \qquad
       - {c^2 \over a^2 ( 1 + 2 \varphi)}
       \left[ {1 \over 1 + 2 \varphi}
       \left( \nabla^i \nabla_j - {1 \over 3} \delta^i_j \Delta \right) \varphi
       + {1 \over {\cal N}}
       \left( \nabla^i \nabla_j - {1 \over 3} \delta^i_j \Delta \right) {\cal N} \right]
   \nonumber \\
   & & \qquad
       =
       - {c^2 \over a^2 (1 + 2 \varphi)^2}
       \left[ {3 \over 1 + 2 \varphi}
       \left( \varphi^{,i} \varphi_{,j}
       - {1 \over 3} \delta^i_j \varphi^{,k} \varphi_{,k} \right)
       + {1 \over {\cal N}} \left(
       \varphi^{,i} {\cal N}_{,j}
       + \varphi_{,j} {\cal N}^{,i}
       - {2 \over 3} \delta^i_j \varphi^{,k} {\cal N}_{,k} \right) \right]
   \nonumber \\
   & & \qquad
       + {c^2 \over a^4 {\cal N}^2 (1 + 2 \varphi)^2}
       \bigg[
       {1 \over 2} \left( \chi^{i,k} \chi_{j,k}
       - \chi_{k,j} \chi^{k,i} \right)
       + {1 \over 1 + 2 \varphi} \left(
       \chi^{k,i} \chi_k \varphi_{,j}
       - \chi^{i,k} \chi_j \varphi_{,k}
       + \chi_{k,j} \chi^k \varphi^{,i}
       - \chi_{j,k} \chi^i \varphi^{,k} \right)
   \nonumber \\
   & & \qquad
       + {2 \over (1 + 2 \varphi)^2} \left(
       \chi^{i} \chi_{j} \varphi^{,k} \varphi_{,k}
       - \chi^{k} \chi_{k} \varphi^{,i} \varphi_{,j} \right) \bigg]
       + {8 \pi G \over c^2} \overline m^i_j,
   \label{eq5}
\eea
where we introduced a transverse-tracefree part of the spatial metric $h_{ij}$ only to the linear order; we introduce $g_{ij} = a^2 [ ( 1 + 2 \varphi ) \delta_{ij} + 2 h_{ij} ]$ where $h^i_i = 0 = h^j_{i,j}$ with indices of $h_{ij}$ raised and lowered by $\delta_{ij}$ and its inverse. For FNLE formulation including the transverse-tracefree mode to fully nonlinear and exact order, see \cite{Gong-2017}. We have
\bea
   & & \overline{K}^i_j \overline{K}^j_i
       = {1 \over a^4 {\cal N}^2 (1 + 2 \varphi)^2}
       \bigg\{
       {1 \over 2} \chi^{i,j} \left( \chi_{i,j} + \chi_{j,i} \right)
       - {1 \over 3} \chi^i_{\;\;,i} \chi^j_{\;\;,j}
       - {4 \over 1 + 2 \varphi} \left[
       {1 \over 2} \chi^i \varphi^{,j} \left(
       \chi_{i,j} + \chi_{j,i} \right)
       - {1 \over 3} \chi^i_{\;\;,i} \chi^j \varphi_{,j} \right]
   \nonumber \\
   & & \qquad
       + {2 \over (1 + 2 \varphi)^2} \left(
       \chi^{i} \chi_{i} \varphi^{,j} \varphi_{,j}
       + {1 \over 3} \chi^i \chi^j \varphi_{,i} \varphi_{,j} \right) \bigg\}.
   \label{K-bar-eq}
\eea
The ADM energy, momentum and mass conservation equations follow from Eqs.\ (\ref{ADM-Mass-conserv-App})-(\ref{ADM-Mom-conserv-App-2})
\bea
   & &
       {1 \over a^3} \left[
       a^3 \left( 1 + 2 \varphi \right)^{3/2} E
       \right]^{\displaystyle\cdot}
       + {1 \over a} \left[
       \left( 1 + 2 \varphi \right)^{1/2}
       \left( {\cal N} c^2 m^i
       + {c \over a} \chi^i E \right) \right]_{,i}
       = - {1 \over a} \left( 1 + 2 \varphi \right)^{1/2}
       {\cal N}_{,i} c^2 m^i
   \nonumber \\
   & & \qquad
       - \left( 1 + 2 \varphi \right)^{1/2}
       \bigg[ \bigg( {\dot a \over a}
       + \dot \varphi + 2 {\dot a \over a} \varphi \bigg)
       \delta^{ij}
       + {c \over a^2} \chi^{i,j}
       - {c \over a^2 (1 + 2 \varphi)}
       \left( 2 \chi^i \varphi^{,j}
       - \delta^{ij} \chi^k \varphi_{,k} \right) \bigg]
       m_{ij},
   \label{eq6-1} \\
   & & {1 \over a^4} \left[ a^4
       \left( 1 + 2 \varphi \right)^{3/2} m_i
       \right]^{\displaystyle\cdot}
       + {1 \over a} \bigg[
       \left( 1 + 2 \varphi \right)^{3/2}
       \bigg( {\cal N} m_i^j
       + {c \chi^j m_i \over a (1 + 2 \varphi)} \bigg)
       \bigg]_{,j}
   \nonumber \\
   & & \qquad
       = {1 \over a} \left( 1 + 2 \varphi \right)^{3/2} \bigg[
       {{\cal N} \varphi_{,i} \over 1 + 2 \varphi} S
       - {\cal N}_{,i} E
       - {c \over a} \left( {\chi^j \over 1 + 2 \varphi} \right)_{,i} m_j \bigg],
   \label{eq7-1} \\
   & & {1 \over a^3} \left[ a^3
       \left( 1 + 2 \varphi \right)^{3/2}
       D \right]^{\displaystyle{\cdot}}
       + {1 \over a} \left[ \sqrt{1 + 2 \varphi}
       \left( {\cal N} v^i + {c \over a} \chi^i \right)
       D \right]_{,i} = 0,
   \label{eq0-1}
\eea
where $D \equiv \overline \varrho \gamma$. Using ${\cal E} \equiv E - D c^2$, we have
\bea
   & & {1 \over a^3} \left[
       a^3 \left( 1 + 2 \varphi \right)^{3/2} {\cal E}
       \right]^{\displaystyle\cdot}
       + {1 \over a} \left\{ \sqrt{1 + 2 \varphi}
       \left[ {\cal N} c^2 ( m^i - D v^i )
       + {c \over a} \chi^i {\cal E} \right] \right\}_{,i}
       = - {1 \over a} \sqrt{1 + 2 \varphi} {\cal N}_{,i} c^2 m^i
   \nonumber \\
   & & \qquad
       - \sqrt{1 + 2 \varphi}
       \left[ \left( H + \dot \varphi + 2 H \varphi \right) S
       - {c \over a^2}
       {\chi^i \varphi_{,i} \over 1 + 2 \varphi} S
       + {c \over a^2} m^{ij} \left( \chi_{i,j}
       - {2 \chi_i \varphi_{,j} \over 1 + 2 \varphi} \right)
       \right].
   \label{eq6-FL-FNLE-App}
\eea
The ADM related fluid quantities $E$, $S$, $m_i$ and $m_{ij}$ will be determined in terms of the fluid quantities in the comoving frame for FL, the SF and the EM field in Eqs.\ (\ref{ADM-fluid-FL-FNLE}), (\ref{ADM-fluid-SF-FNLE}) and (\ref{ADM-fluid-EM-FNLE}), respectively.

To the background order, Eqs.\ (\ref{eq2}), (\ref{eq4}), (\ref{eq6-1}) and (\ref{eq0-1}), respectively, give
\bea
   {\dot a^2 \over a^2}
       = {8 \pi G \over 3 c^2} E
       + {\Lambda c^2 \over 3}, \quad
       {\ddot a \over a}
       = - {4 \pi G \over 3 c^2} \left( E + S \right)
       + {\Lambda c^2 \over 3}, \quad
       \dot E + 3 {\dot a \over a} (E + S) = 0, \quad
       \dot {\overline \varrho}
       + 3 {\dot a \over a} \overline \varrho = 0,
\eea
where, from Eqs.\ (\ref{ADM-fluid-FL-FNLE}), (\ref{ADM-fluid-SF-FNLE}) and (\ref{ADM-fluid-EM-FNLE}), we have
\bea
   E = \mu + {1 \over 2 c^2} \dot \phi^2 + V, \quad
       S = 3 \left( p + {1 \over 2 c^2} \dot \phi^2 - V \right),
\eea
with $\mu \equiv \overline \varrho \left( c^2 + \Pi \right)$.
The EM field cannot be accommodated in the Friedmann background.

\end{widetext}
%%%%%%%%%%%%%%%%%%%%%%%%%%%%%%%%%%%%%%%%%%%%%%%%%%%%%%%%%%%%%%%
%
% Bibliography
%
%%%%%%%%%%%%%%%%%%%%%%%%%%%%%%%%%%%%%%%%%%%%%%%%%%%%%%%%%%%%%%%

%%%%%%%%%%%%%%%%%%%%%%%%%%%%%%%%%%%%%%%%%%%%%%%%%%%%%%%%%%%%%%%

%%%%%%%%%%%%%%%%%%%%%%%%%%%%%%%%%%%%%%%%%%%%%%%%%%%%%%%%%%%%%%%
\end{document}